\documentclass[12pt]{iopart}

\usepackage{citesort}
\usepackage{iopams}  
\usepackage{graphicx}
\expandafter\let\csname equation*\endcsname\relax
\expandafter\let\csname endequation*\endcsname\relax
\usepackage{amsmath}
\usepackage{amssymb}
\usepackage[pdftex]{color}
\usepackage{bigints}
\usepackage{dcolumn}
\usepackage{bm}
\usepackage{mathtools}
\usepackage{etoolbox}

\def\bra#1{\mathinner{\langle{#1}|}}
\def\ket#1{\mathinner{|{#1}\rangle}}

\makeatletter
\def\@mkboth#1#2{}
\newlength\appendixwidth
\preto\appendix{\addtocontents{toc}{\protect\patchl@section}}
\newcommand{\patchl@section}{%
  \settowidth{\appendixwidth}{\textbf{Appendix }}%
  \addtolength{\appendixwidth}{1.5em}%
  \patchcmd{\l@section}{1.5em}{\appendixwidth}{}{\ddt}%
}
\makeatother

\begin{document}

\title{Cavity-assisted mesoscopic transport of fermions: Coherent and dissipative dynamics.}

\author{David Hagenm\"{u}ller}
\address{IPCMS (UMR 7504) and ISIS (UMR 7006), University of Strasbourg and CNRS, 67000 Strasbourg, France}
\author{Stefan Sch\"utz}
\address{IPCMS (UMR 7504) and ISIS (UMR 7006), University of Strasbourg and CNRS, 67000 Strasbourg, France}
\author{Johannes Schachenmayer}
\address{IPCMS (UMR 7504) and ISIS (UMR 7006), University of Strasbourg and CNRS, 67000 Strasbourg, France}
\author{Claudiu Genes}
\address{Max Planck Institute for the Science of Light, Staudtstra{\ss}e 2, D-91058 Erlangen, Germany}
\author{Guido Pupillo}
\address{IPCMS (UMR 7504) and ISIS (UMR 7006), University of Strasbourg and CNRS, 67000 Strasbourg, France}

\begin{abstract}
We study the interplay between charge transport and light-matter interactions in a confined geometry, by considering an open, mesoscopic chain of two-orbital systems resonantly coupled to a single bosonic mode close to its vacuum state. We introduce and benchmark different methods based on self-consistent solutions of non-equilibrium Green's functions and numerical simulations of the quantum master equation, and derive both analytical and numerical results. It is shown that in the dissipative regime where the cavity photon decay rate is the largest parameter, the light-matter coupling is responsible for a steady-state current enhancement scaling with the cooperativity parameter. We further identify different regimes of interest depending on the ratio between the cavity decay rate and the electronic bandwidth. Considering the situation where the lower band has a vanishing bandwidth, we show that for a high-finesse cavity, the properties of the resonant Bloch state in the upper band are transfered to the lower one, giving rise to a delocalized state along the chain. Conversely, in the dissipative regime with low cavity quality factors, we find that the current enhancement is due to a collective decay of populations from the upper to the lower band. 
\end{abstract}


\tableofcontents

\newpage

\section{Introduction}
\label{intro}

\subsection{Context}

Investigating how the transport of excitations can be modified by the coupling to light is a topic of considerable fundamental and practical interest~\cite{plenio1,reben,marais,levi}. Recent studies have predicted drastic modifications of the transport of electron-hole pairs called excitons when interacting with photons in confined geometries such as cavities~\cite{schachenmayer2015cavity} or plasmonic resonators~\cite{feist2015extraordinary}. The modification of exciton transport in a cavity can be understood using the Tavis-Cummings model~\cite{tavis} (TC), which describes the collective behavior of $N$ dipoles (two-level systems) resonantly coupled to a single bosonic mode. As localized excitons hop toward their nearest neighboring sites, the exciton propagation from one side of the cavity to the other can be bypassed by exchanging energy with polariton modes delocalized over the entire cavity mode volume. This energy transfer can be interpreted as a long-range dipole-dipole-type interaction mediated by the cavity~\cite{taptap}.

\vspace{2mm}

Studies of charge transport modifications induced by the coupling to bosonic fields in condensed matter systems have traditionally focused on electron-phonon interactions~\cite{mahan}. In polar semiconductors, the latter provide a screening of the electron motion by the lattice polarization~\cite{0038-5670-12-1-A09,pekar,doi:10.1080/00018735400101213} (polaron), which is responsible for increasing the electron effective mass and reducing the mobility~\cite{mahan}. Electron-phonon coupling in metals is known to lead to different instabilities at sufficiently low temperature, such as BCS electron pairing leading to superconductivity~\cite{fro,PhysRev.104.1189,PhysRev.108.1175} and Peierls-type instabilities responsible for a metal-insulator phase transition in one-dimensional systems~\cite{peierls1996quantum,afre}. The crucial difference between electron-photon and electron-phonon coupling stems from the possibility of low-energy electron scattering with both vanishing and large momenta (of the order of the Fermi wavevector $k_{F}$) in the latter case. In particular, the aforementioned instabilities occur due to large-momentum ($\sim 2 k_{F}$) scattering across the Fermi surface, within a narrow energy band of the order of the Debye frequency. Conversely, light-matter coupling typically involves quasi-vertical electronic excitations across a band-gap, resulting in the absence of both low-energy and large-momentum excitations. In the macroscopic limit, this usually provides a decoupling between low-energy charge transport and light-matter coupling occuring at finite frequencies~\cite{liberato}.  

\vspace{2mm}

An emerging topic of interest is the modification of material properties using an external electromagnetic radiation~\cite{lindner}, and in particular the possibility of light-induced superconductivity in the ultraviolet~\cite{yamamoto} and terahertz portions of the spectrum~\cite{mitrano,sri,theory_mitrano,theory_mitrano2,scalp,theory_mitrano3,theory_mitrano4}, as well as the emergence of zero-resistance states in quantum Hall systems subjected to microwave radiation~\cite{mani,zudov,theory_mani,theory_mani2}. On the other hand, the study of light-matter interactions in confined geometries is attracting  increasing attention in various fields, such as in quantum optics~\cite{tsintzos,lagoudakis,kenacohen,cristofolini,schneider,javadi,sipahigil,cao,manzoni}, quantum chemistry~\cite{ebbesen,jino,herrera,baumberg,galego}, and condensed matter~\cite{tredicucci,kasprzak,amo,deng,carusotto,kavokin,ouss,Zhu}, opening the way to investigate the rich interplay between many-body physics and strong light-matter interactions~\cite{imam1,imam2}. In the case of charge transport, large conductivity enhancements ($\sim$ one order of magnitude) have been recently reported considering organic molecular semiconductors strongly coupled to a surface plasmon resonator~\cite{orgiu2015conductivity}. Inspired by these experiments, a fermionic version of the TC model has been introduced~\cite{hagenmuller}, showing that the cavity coupling can lead to very large current enhancements in the asymmetric situation where the bandwidth associated with tightly bound valence electrons is much smaller than the bandwidth of delocalised electrons in the conduction band.

\vspace{2mm}

In the present paper, we further investigate how the coupling to a cavity mode can lead to an enhancement of the steady-state current through a chain of $N$ sites with two orbitals, providing complementary and original methods to investigate this system. We characterize different regimes of transmission involving either a dissipative or a coherent dynamics, and identify the presence of collective effects and electronic correlations depending on the strength of the light-matter coupling. Our model might find direct applications in several fields, such as transport in organic semiconductors~\cite{orgiu2015conductivity}, quantum dot arrays~\cite{kagan_charge_2015,wang_fabrication_2004,frey_dipole_2012,viennot_out_equilibrium_2014,liu_semiconductor_2015,gudmundsson,moldoveanu}, and nanowires~\cite{rurali_colloquium_2010,laird_quantum_2015,vidar2}, as well as for quantum simulations using ultracold atoms~\cite{Brantut1069,laflamme} or superconducting qubits~\cite{devoret2004,clarke,koch} in the microwave domain.

\subsection{Model}
\label{phi_s}

\begin{figure}[t]
\centerline{\includegraphics[width=0.52\columnwidth]{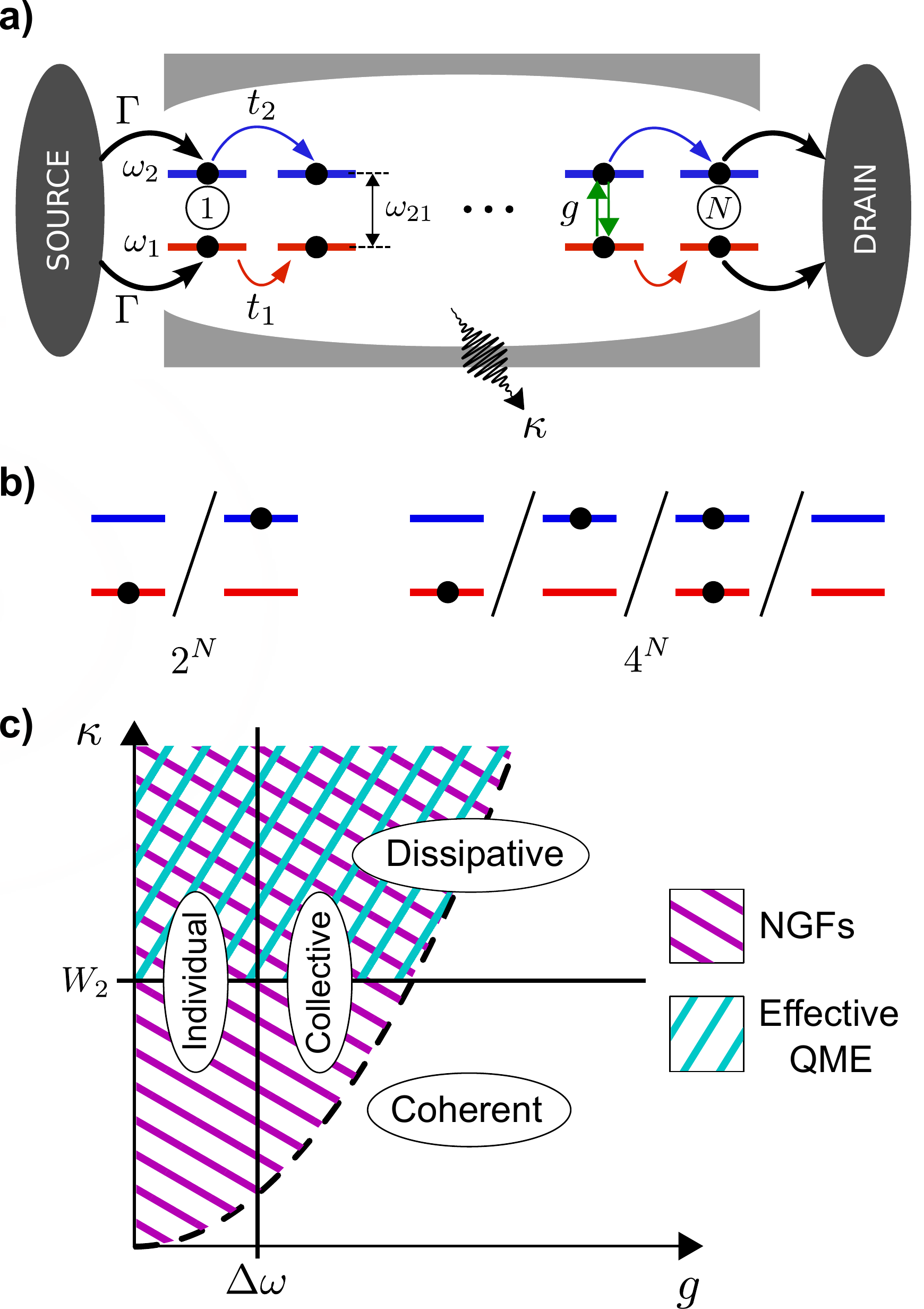}}
\caption{\textbf{a)} 1D chain of $N$ two-orbital systems, each one consisting of a lower orbital ($\alpha=1$, red) and an upper one ($\alpha=2$, blue). The first and last sites $j=1$ and $j=N$ are coupled to two leads with equal injection/extraction rate $\Gamma$. The transition with energy $\omega_{21}$ between lower and upper orbitals is resonantly coupled (with strength $g$) to a single cavity mode, with decay rate $\kappa$. $t_\alpha$ is the hopping rate between neighboring sites in the band $\alpha$ (we always consider the case $t_2 \gg t_1$). \textbf{b)} TC model: The Hilbert space associated with a given site is spanned by the two quantum states represented on the left side, providing a $2^{N}$-states basis for the whole chain. Right-side: The hopping Hamiltonian $H_{t}$ provides a coupling of these states to two new states with both orbitals either occupied or empty. The chain is thus spanned by a $4^{N}$-states basis. \textbf{c)} Sketch showing the different regimes investigated, together with the applicability domains of the different methods used in this article. $W_2$ and $\delta \omega$ denote the upper electronic bandwidth and the typical separation between two adjacent Bloch states in the upper band, respectively. NGFs stands for Non-equilibrium Green's functions, and QME for Quantum Master Equation. The full QME is valid everywhere on the diagram. The dashed line corresponds to $g^{2}/(\kappa\Gamma)=1$ (the left-hand side is the cooperativity), separating the perturbative regime (above the line) from the non-perturbative regime (below the line). The horizontal line $\kappa= W_2$ represents the separation between the dissipative regime $\kappa \gg W_2$ and the coherent regime $\kappa \ll W_2$. While for $g < \delta \omega$, the coupling to light always involves a single Bloch state (``individual dressing regime''), a collective coupling of many Bloch state arises when $g > \delta \omega$ (``collective dressing regime''). Note that since $\delta \omega \to 0$ in the macroscopic limit $N \to \infty$, the coupling is therefore always collective in this case.}
\label{fig1}
\end{figure}

We consider a 1D chain of $N$ sites with two electronic orbitals of energy $\omega_\alpha$ ($\hbar=1$), where $\alpha=1,2$ stands for lower and upper orbitals, respectively [see Fig.~\ref{fig1} \textbf{a)}]. Each orbital $\alpha$ on site $j$ is coupled to its nearest neighbors $j\pm 1$ with hopping rate $t_\alpha$, resulting in two bands in a tight-binding picture. \textit{In the following, we will always consider the situation where the upper band is much broader than the lower one ($t_{2}\gg t_{1}$).} Depending on $N$, the upper electronic bandwidth varies between $2 t_{2}$ ($N=2$) and $4t_{2}$ ($N \to \infty$), and will be denoted as $W_2$ (respectively $W_1$ for the lower band). Electrons are considered as spin-less. The edges of the chain are connected to a source and a drain (leads) with a large bias voltage across, such that the Fermi level of the source (the drain) is higher (lower) than any other energy scale in the system. This allows for injection/extraction in both orbitals at a rate $\Gamma_{\alpha}$. Although different injection/extraction rates are kept for the sake of generality, we will only discuss the results obtained for $\Gamma_{1}=\Gamma_{2}\equiv \Gamma$. All energies are in units of $\omega_{21}$ (set to $1$), which is assumed to be the largest parameter. The on-site transition between lower and upper orbitals with energy $\omega_{21}=\omega_2-\omega_1$ is resonantly coupled (with a coupling strength $g$) to a single cavity mode with decay rate $\kappa$. Letting the contributions from the leads and the extra-cavity photonic environment aside for now, the 1D chain Hamiltonian can be written as $H_{S}=H_{e}+H_{c} + H_{t}+H_{I}$, where:
 
\begin{align*}
H_{e}&= \sum_{\alpha}\sum_{j=1}^{N} \omega_{\alpha} c^{\dagger}_{\alpha,j} c_{\alpha,j} \nonumber \\
H_{c}&=\omega_{c} a^\dagger a,
\end{align*}
describe the free orbitals and free cavity mode contributions, respectively. The fermionic operators $c_{\alpha,j}$ and $c^{\dagger}_{\alpha,j}$ respectively annihilate and create an electron in the orbital $\alpha$ on site $j$, and satisfy the anti-commutation relations $\{c_{\alpha,i},c^{\dagger}_{\alpha',j}\}=\delta_{\alpha,\alpha'}\delta_{i,j}$. On the other hand, $a$ and $a^{\dagger}$ denote the bosonic annihilation and creation operators of a photon in the cavity mode with energy $\omega_{c}$, and satisfy the commutation relation $[a,a^{\dagger}]=1$. The nearest-neighbor hopping in both orbitals is described by the contribution: 

\begin{align}
H_{t} = - \sum_{\alpha} t_{\alpha} \left(\sum_{j=1}^{N-1} c^{\dagger}_{\alpha,j+1} c_{\alpha,j} + \sum_{j=2}^{N} c^{\dagger}_{\alpha,j-1} c_{\alpha,j} \right), 
\label{H_kinetic2}
\end{align}
and the light-matter coupling by the term:

\begin{align}
H_{I} = g \sum_{j=1}^{N} \left(c^{\dagger}_{2,j} c_{1,j} + c^{\dagger}_{1,j} c_{2,j} \right) A,
\label{H_intera2} 
\end{align}
with $A=a+a^{\dagger}$. In the absence of Eq.~(\ref{H_kinetic2}), and if one restricts the orbital occupation to one per site, $H_{S}$ corresponds to the TC Hamiltonian~\cite{tavis} with counter-rotating terms, where the bosonic field $A$ is coupled to the collective pseudo-spin operator:

\begin{align*}
S_{x}=\frac{1}{2\sqrt{N}}\sum_{j=1}^{N} \left(c^{\dagger}_{2,j} c_{1,j} + c^{\dagger}_{1,j} c_{2,j} \right).
\end{align*}

In the case of the TC model, the size of the electronic part of the Hilbert space is $2^N$ [see Fig.~\ref{fig1} \textbf{b)}], and one can use boson mapping techniques~\cite{klein} to find the spectrum of $H_{S}$. The cavity field thus interacts with a collective mode formed of a coherent superposition of $N$ single-spin excitations, with an enhanced coupling strength $\Omega=g\sqrt{N}$ called vacuum Rabi frequency~\cite{raimond}. In particular, the strong coupling regime of cavity QED~\cite{raimond} is achieved when $\Omega >\kappa$, allowing a quasi-reversible energy transfer between the collective dipole and the cavity field, and providing vacuum Rabi oscillations at a frequency $\Omega$. Instead of the bare cavity resonance, the cavity spectrum features two polariton resonances separated by a splitting $2\Omega$.

\vspace{2mm}

In the presence of $H_{t}$, however, charge transport can occur due to the coupling between the two quantum states associated with each local pseudo-spin and the two new states with both orbitals either occupied or empty [see Fig.~\ref{fig1} \textbf{b)}]. Our model thus exhibits a larger Hilbert space ($4^N$) compared to the TC model, and features a more complex physics. Introducing the electron density operator in the orbital $\alpha$ as $\hat{n}_{\alpha j}=c^{\dagger}_{\alpha,j} c_{\alpha,j}$, one realizes that the total density at a given site $j$ is conserved by the light-matter coupling Hamiltonian, namely $[H_{I},\sum_{\alpha}\hat{n}_{\alpha j}]=0$. This means that in contrast to exciton transport, the cavity-induced modification of charge transport can only occur through the interplay between $H_{I}$ and $H_{t}$.  

\vspace{2mm}

\subsection{Main results}

The main results of the paper are summarized in the following: 

\begin{itemize}
\item In Sec.~\ref{ss_cucu23_part}, we explain how to compute the relevant physical observables (current, populations, electron and photon density of states) using different theoretical methods that are presented in detail. In Sec.~\ref{gf_self_ener_part}, we introduce a frequency-domain method based on the self-consistent solutions of Non-equilibrium Green's Functions (NGFs), valid in the perturbative regime where the cooperativity $g^{2}/(\kappa \Gamma) < 1$ [above the dashed line on Fig.~\ref{fig1} \textbf{c)}]. The results obtained with this method are benchmarked with a suitable Quantum Master Equation (QME) presented in Sec.~\ref{qme_eff_part}, exact in the rotating-wave approximation~\cite{scully} and as long as the Markovian approximation for the system-baths coupling holds true, but nevertheless limited to a small number of sites. In Sec.~\ref{sec:eff_master}, we show that in the dissipative regime $\kappa/W_2 \gg 1$ and for small coupling strengths, an effective QME can be derived, in which light-matter interactions are entirely cast into a dissipator $\propto g^{2}/\kappa$.

\item We present our results in Sec.~\ref{res_part}, by first discussing the physical properties of the system in the absence of light-matter coupling (Sec.~\ref{no_coupl_part}), and explaining how the electron density of states (DOS) is broadened by light-matter interactions in the perturbative regime (Sec.~\ref{spec_lan_jd}). We further explain how polariton modes arise from the dressing of the photon GF by the electron-hole polarization. In the asymmetric situation where $t_2\gg t_1$, we show that the light-matter coupling is responsible for opening a new transmission channel in the lower band, which leads to an enhancement of the steady-state current. In Sec.~\ref{sec:comparison}, we compare the current enhancement predicted by the different numerical methods, identifying the regimes of interest based on the ratio between the upper electronic bandwidth $W_2\sim t_{2}$ and the cavity photon decay rate $\kappa$. In the dissipative regime $\kappa/W_2 \gg 1$, we find that the current enhancement scales with the cooperativity.  

\item We further investigate the dissipative regime in Sec.~\ref{diss_reg_sec} [upper part on Fig.~\ref{fig1} \textbf{c)}]. In particular, we derive an analytical formula for the current enhancement valid for small coupling strengths (Sec.~\ref{anal_app_part}), and characterize the presence of collective effects by computing the different observables numerically in Sec.~\ref{coll_dress_part}. We show that a collective coupling of many Bloch states to the cavity mode occurs when $g >\delta \omega$, namely when the coupling strength is larger than the typical energy separation between two adjacent Bloch states in the upper band. In this dissipative, collective ``dressing'' regime, the current enhancement stems from a global transfer of populations from the upper to the lower band, with only marginal propagation through the lower band. For large coupling strengths (Sec.~\ref{non_local_coco2}), we show that the current enhancement saturates to about twice its value for $g=0$, and that the collective coupling is responsible for the existence of non-local electronic correlations.

\item The ``coherent'' regime obtained for $\kappa/W_2 \ll 1$ is studied in Sec.~\ref{ind_dress_part}. When $g <\delta \omega$, only one given resonant Bloch state is individually coupled to the cavity mode [left bottom part on Fig.~\ref{fig1} \textbf{c)}], which is refered to as ``individual dressing regime''. After having characterized the latter by computing the different observables in Sec.~\ref{ttdress_part43}, we show that a transfer of spectral weight $\sim 10\%$ occurs from the upper to the lower band, resulting in a new state with energy $\sim \omega_1$ delocalized across the whole chain (Sec.~\ref{ttdress_part666}). In this case, the current enhancement can be interpreted as stemming from coherent dynamics sustained by the absorption and emission of cavity photons. Ultimately, for $N \gg 1$, or when the coupling strength becomes larger than the upper electronic bandwidth, we expect to recover a collective coupling of the Bloch states to the cavity mode [right bottom part on Fig.~\ref{fig1} \textbf{c)}]. 

\item Concluding remarks concerning the cavity photons population and the scaling of the current with the chain length $N$ are presented in Sec.~\ref{comp_part}, and perspectives are drawn in Sec.~\ref{conclll}.
\end{itemize}

\section{Methods}
\label{method_part}

This section is structured as follows. In Sec.~\ref{ss_cucu23_part}, we introduce the steady-state current flowing through the chain in the presence of light-matter coupling, showing that this observable can be calculated by using the QME and NGFs formalisms, depending on whether the problem is formulated in real time or in the frequency domain, respectively. In Sec.~\ref{gf_self_ener_part}, we focus on the NGFs method, introducing the total Hamiltonian including the contributions from the environment, and explain how to compute the current by solving a set of self-consistent equations for electron and photon Green's functions (GFs). In Sec.~\ref{qme_eff_part}, we introduce a suitable QME to compute the steady-state current, exact but limited to a small number of sites. In Sec.~\ref{sec:eff_master}, we introduce an effective master equation valid in the dissipative regime where the fast cavity field evolution can be adiabatically eliminated, resulting in an effective QME involving only electronic degrees of freedom. 

\subsection{Steady-state current}
\label{ss_cucu23_part}

In the frequency domain, the steady-state current can be put in a form reminiscent of the Lan\-dau\-er formula~\cite{landauer} for equilibrium mesoscopic systems~\cite{haug}:

\begin{equation}
J = \frac{J_{s}-J_{d}}{2} =\sum_{\alpha} \frac{e \Gamma_{\alpha}}{2} \int \!\! \frac{d\omega}{2\pi} T_{\alpha}(\omega),
\label{transsm}
\end{equation}
where $J_{s}$ ($J_{d}$) is the steady-state current flowing through the source (drain), $\omega$ the frequency, and $e$ the electron charge. In the high-bias regime, the transmission spectrum $T_{\alpha}(\omega)$ is expressed in terms of the electron GFs $\underline{G}^{r}_{\alpha}$ and $\underline{G}^{<}_{\alpha}$ defined in Sec.~\ref{caca_ngfs}:

\begin{align}
T_{\alpha}(\omega) &= \textbf{Tr} \left[-2\underline{\sigma}^{1} \circ \Im \underline{G}^{r}_{\alpha} (\omega) + \left( \underline{\sigma}^{N} -\underline{\sigma}^{1}\right) \circ \Im  \underline{G}^{<}_{\alpha} (\omega) \right],
\label{trans_long_coco2}
\end{align}
where underlined quantities denote $N\times N$ matrices, $\circ$ is the element-wise Hadamard product, $\Im$ stands for imaginary part, and $\textbf{Tr}$ denotes the sum over all matrix elements. In Sec.~\ref{self_self22}, we explain in detail how the current can be computed by solving numerically a set of coupled self-consistent equations for electron GFs.

\vspace{2mm}

On the other hand, the steady-state current can be directly obtained from the electron populations at the first and last site:

\begin{align}
J = \frac{J_{s}-J_{d}}{2} = \sum_{\alpha} \frac{e \Gamma_{\alpha}}{2} \left(\langle 1 - \hat{n}_{\alpha 1} \rangle + \langle \hat{n}_{\alpha N} \rangle \right), 
\label{Jinout}
\end{align}
and can be directly computed by simulating the time-evolution of the joint density operator for the chain and the cavity field. In Sec.~\ref{qme_eff_part}, we explain how this can be done by using a suitable QME. We also show that the cavity mode can be adiabatically eliminated in the dissipative regime (lossy cavity), resulting in an effective QME involving only electronic degrees of freedom (Sec.~\ref{sec:eff_master}).    

\subsection{Non-equilibrium Green's functions}
\label{gf_self_ener_part}

In this section, we focus on the NGFs method. In Sec.~\ref{tt_hamil}, we introduce the total Hamiltonian including the contributions from the environment. In Sec.~\ref{caca_ngfs}, we present the derivation of the steady-state current written in terms of electron GFs, and explain how to compute this current by solving a set of self-consistent equations for electron and photon Green's functions in Sec.~\ref{self_self22}. This method is based on a generalization of the model presented in the Chapter 12 of ~\cite{haug}, with a treatment of light-matter coupling similar to the one presented in ~\cite{engelsberg1963coupled} for electron-phonon interactions. 

\subsubsection{Total Hamiltonian.}
\label{tt_hamil}

In the framework of the NGFs formalism, the environment is described by Hamiltonian terms. In total, one can write $H=H_{S}+H_{L} +H_{P}$, where the chain Hamiltonian $H_{S}$ is given in Sec.~\ref{phi_s}. 

The two leads injecting and extracting electrons are described by the contribution:  

\begin{align*}
H_{L} =\sum_{\alpha}\sum_{\eta=s,d} \sum_{\bf q} \omega_{\bf q} b^{\dagger}_{\alpha,{\bf q},\eta} b_{\alpha,{\bf q},\eta} + \sum_{\alpha} \sum_{\eta=s,d} \sum_{j,{\bf q}}  \lambda_{\alpha,j,{\bf q},\eta} \left( c_{\alpha,j} b^{\dagger}_{\alpha,{\bf q},\eta} + b_{\alpha,{\bf q},\eta} c^{\dagger}_{\alpha,j}\right),
\end{align*}
with coupling constants 
\begin{equation}
\lambda_{\alpha,j,{\bf q},s} = \begin{cases}  \lambda_{\alpha,{\bf q}} \; \textrm{for} \; j=1 \\
0 \quad \;\; \textrm{for} \; j\neq 1 \end{cases} \lambda_{\alpha,j,{\bf q},d} = \begin{cases} \lambda_{\alpha,{\bf q}} \; \textrm{for} \; j=N \\
0 \quad \;\; \textrm{for} \; j\neq N.  \end{cases}
\label{cou_le_ref}
\end{equation}    

The operators $b^{\dagger}_{\alpha,{\bf q},\eta}$ ($b_{\alpha,{\bf q},\eta}$) create (annihilate) a fermion in the state $(\alpha,{\bf q})$ with energy $\omega_{\bf q}$ in the lead $\eta$, and obey fermionic commutation relations. The photonic bath responsible for cavity photon losses is described by the Hamiltonian:

\begin{equation}
H_{P} = \sum_{\bf p} \omega_{\bf p} a^{\dagger}_{\bf p} a_{\bf p} + \sum_{\bf p} \mu_{\bf p} A_{\bf p} A,
\end{equation}
where $a^{\dagger}_{\bf p}$ and $a_{\bf p}$ denote the extra-cavity photon operators (obeying bosonic commutation rules) with corresponding energy $\omega_{\bf p}$, and $A_{\bf p}=a_{\bf p}+a^{\dagger}_{\bf p}$. The cavity photons-bath coupling strength is denoted as $\mu_{\bf p}$. The continous variables ${\bf q}$ and ${\bf p}$ are arbitrary quantities respectively associated with the electronic (leads) and photonic baths. 

The chain operators can be expanded in the Bloch states basis as $c_{\alpha,j}=\sum_{k=1}^{N} \varphi^{j}_{k} \tilde{c}_{\alpha,k}$, with 

\begin{equation}
\varphi^{j}_{k}=\sqrt{\frac{2}{N+1}} \sin \left(\frac{\pi j k}{N+1} \right),
\label{bloch_sfou}
\end{equation}
such that the contribution $H_{e}+H_{t}$ takes the diagonal form $\sum_{\alpha,k} \omega_{\alpha,k} \tilde{c}^{\dagger}_{\alpha,k} \tilde{c}_{\alpha,k}$ with $\omega_{\alpha,k}=\omega_{\alpha}-2t_{\alpha}\cos(\pi k/(N+1))$. The Hamiltonian $H_{S}$ can thus be partitioned into a diagonal part $H_{0}=H_{e}+H_{t}+H_{c}$ with known eigenstates, and the light-matter interaction Eq.~(\ref{H_intera2}) treated perturbatively. 

\subsubsection{Steady-state current.}
\label{caca_ngfs}

In the steady-state, the charge current $J_{\eta}$ flowing through the lead $\eta$ is given by the continuity equation $J_{\eta} = -e \partial_t \langle N_{\eta} \rangle = - i e \langle [H,N_{\eta}] \rangle$, with $J_{s}=-J_{d}$. Here, $\langle \cdots \rangle$ denotes the statistical average with respect to the density operator $\varrho$ of the whole system (chain+environment), whose evolution is governed by the total Hamiltonian $H$. $N_{\eta}=\sum_{\alpha,{\bf q}} b^{\dagger}_{\alpha,{\bf q},\eta} b_{\alpha,{\bf q},\eta}$ is the number of electrons in the lead $\eta$. As detailed in \ref{negfs_form}, the steady-state current can be put in the form given by Eqs.~(\ref{transsm}) and (\ref{trans_long_coco2}). The matrix elements of $\underline{\sigma}^{j}$ entering Eq.~(\ref{trans_long_coco2}) are given by $\sigma^{j}_{k,k'}=\varphi^{j}_{k}\varphi^{j}_{k'}$, and the matrix elements of the so-called retarded and ``lesser'' electron GFs are respectively defined (in the frequency domain) as:

\begin{align*}
G^{r}_{\alpha,k,k'} (\omega) & =-i \int_{0}^{+\infty} \!\! d\tau e^{i \omega \tau} \langle \{ \tilde{c}_{\alpha,k} (\tau), \tilde{c}^{\dagger}_{\alpha,k'} (0) \} \rangle \nonumber \\
G^{<}_{\alpha,k,k'} (\omega) & =i \int_{-\infty}^{+\infty} \!\! d\tau e^{i \omega \tau} \langle \tilde{c}^{\dagger}_{\alpha,k'} (0) \tilde{c}_{\alpha,k} (\tau) \rangle,
\end{align*}
where $\{\cdots \}$ denotes the anticommutator. On the other hand, the time-ordered electron GF is given by the expression:

\begin{equation}
G_{\alpha,k,k'} (\tau-\tau') =-i \langle \mathcal{T} \tilde{c}_{\alpha,k} (\tau) \tilde{c}^{\dagger}_{\alpha,k'} (\tau') \rangle= -i \frac{\langle \mathcal{T} \tilde{c}_{\alpha,k} (\tau) \tilde{c}^{\dagger}_{\alpha,k'} (\tau') e^{-i \int \! d \tau_{1} H (\tau_{1})}\rangle_{0}}{\langle e^{-i \int \! d \tau_{1} H (\tau_{1})} \rangle_{0}}, 
\label{GF_def_ufc}
\end{equation}
where $\mathcal{T}$ denotes the time-ordered product for fermions, and $\langle \cdots \rangle_{0}$ refers to the statistical average with respect to the density operator $\varrho$ of the whole system (chain+environment), whose evolution is governed by the free Hamiltonian ($H$ without $H_{I}$ and the interaction terms entering $H_{L}$ and $H_{P}$). The first contribution of Eq.~(\ref{trans_long_coco2}) involves the trace of the electron spectral function $\underline{A}^{(\alpha)} (\omega)=- 2 \Im \underline{G}^{r}_{\alpha}(\omega)$ in the band $\alpha$. Physically, the quantity $\sum_{k,k'} {A}^{(\alpha)}_{k,k'} (\omega)$  corresponds to the normalized electron DOS in the band $\alpha$. The spectral function normalization $\int \! d\omega A^{(\alpha)}_{k,k'}(\omega) =2\pi \delta_{k,k'}$ implies that the effect of light-matter interactions on the steady-state current is entirely determined by the second term in Eq.~(\ref{trans_long_coco2}), which is proportional to the trace of the ``lesser'' electron GF. The latter can be used to compute the steady-state electron population in real space as~\cite{pourfath}:

\begin{align*}
n_{\alpha j} = \langle \hat{n}_{\alpha,j} \rangle=\sum_{k,k'} \varphi^{j}_{k} \varphi^{j}_{k'} \int \! \frac{d\omega}{2\pi} \Im G^{<}_{\alpha,k,k'} (\omega).
\end{align*} 

Using the spectral function normalization and inverting the previous equation, the steady-state current Eq.~(\ref{transsm}) takes the form given in Eq.~(\ref{Jinout}):

\begin{align}
J =\sum_{\alpha} \frac{e\Gamma_{\alpha}}{2} \left(1- n_{\alpha 1} + n_{\alpha N} \right) \left(=\sum_{\alpha} e\Gamma_{\alpha} n_{\alpha N} \right),
\label{current_ss_pop}
\end{align} 
showing that the latter only depends on the electron populations at the edges of the chain. 

\subsubsection{Self-consistent equations for electrons and photons.}
\label{self_self22}

We now explain in detail the procedure to compute the electron GFs entering the expression of the transmission function Eq.~(\ref{trans_long_coco2}). It can be shown (see \ref{negfs_form}) that retarded and advanced electron GFs obey a Dyson equation of the form: 

\begin{align}
\underline{G}^{\beta}_{\alpha} (\omega) &= \left((\underline{G}^{0\beta}_{\alpha} (\omega))^{-1} - \underline{\Sigma}^{\beta}_{\alpha} (\omega) \right)^{-1},
\label{dyson_aspi}
\end{align}
with $\beta=r,a$, while ``lesser'' and ``greater'' GFs are obtained from the Keldysh equation:

\begin{align}
\underline{G}^{\gamma}_{\alpha} (\omega) = \underline{G}^{r}_{\alpha}(\omega) \underline{\Sigma}^{\gamma} (\omega)  \underline{G}^{a}_{\alpha}(\omega),
\label{gless_ff_mat}
\end{align}
with $\gamma=<,>$ for lesser and greater. The matrix elements of the unperturbed GFs $\underline{G}^{0}_{\alpha} (\omega)$ (evaluated in the absence of light-matter coupling and interactions with the leads) are all proportional to $\delta_{k,k'}$:

\begin{align}
G^{0<}_{\alpha,k,k'} (\omega)&= -2i\pi \delta_{k,k'} \delta (\omega - \omega_{\alpha,k} ) n^{0}_{\alpha,k}  \nonumber \\
G^{0>}_{\alpha,k,k'} (\omega)&= 2i\pi \delta_{k,k'} \delta (\omega - \omega_{\alpha,k} ) \left(1-n^{0}_{\alpha,k}\right)  \nonumber \\
G^{0a}_{\alpha,k,k'} (\omega) &= \frac{\delta_{k,k'}}{\omega - \omega_{\alpha,k} - i 0^{+}},
\label{gg_elec32}
\end{align}
and $G^{0r}_{\alpha,k,k'}=(G^{0a}_{\alpha,k,k'})^{*}$, where $n^{0}_{\alpha,k}=\langle \tilde{c}^{\dagger}_{\alpha,k} \tilde{c}_{\alpha,k} \rangle_{0}$ is the population of the Bloch states $(\alpha,k)$ in the initial, non interacting ground state. 

\begin{figure}[t]
\centerline{\includegraphics[width=0.7\columnwidth]{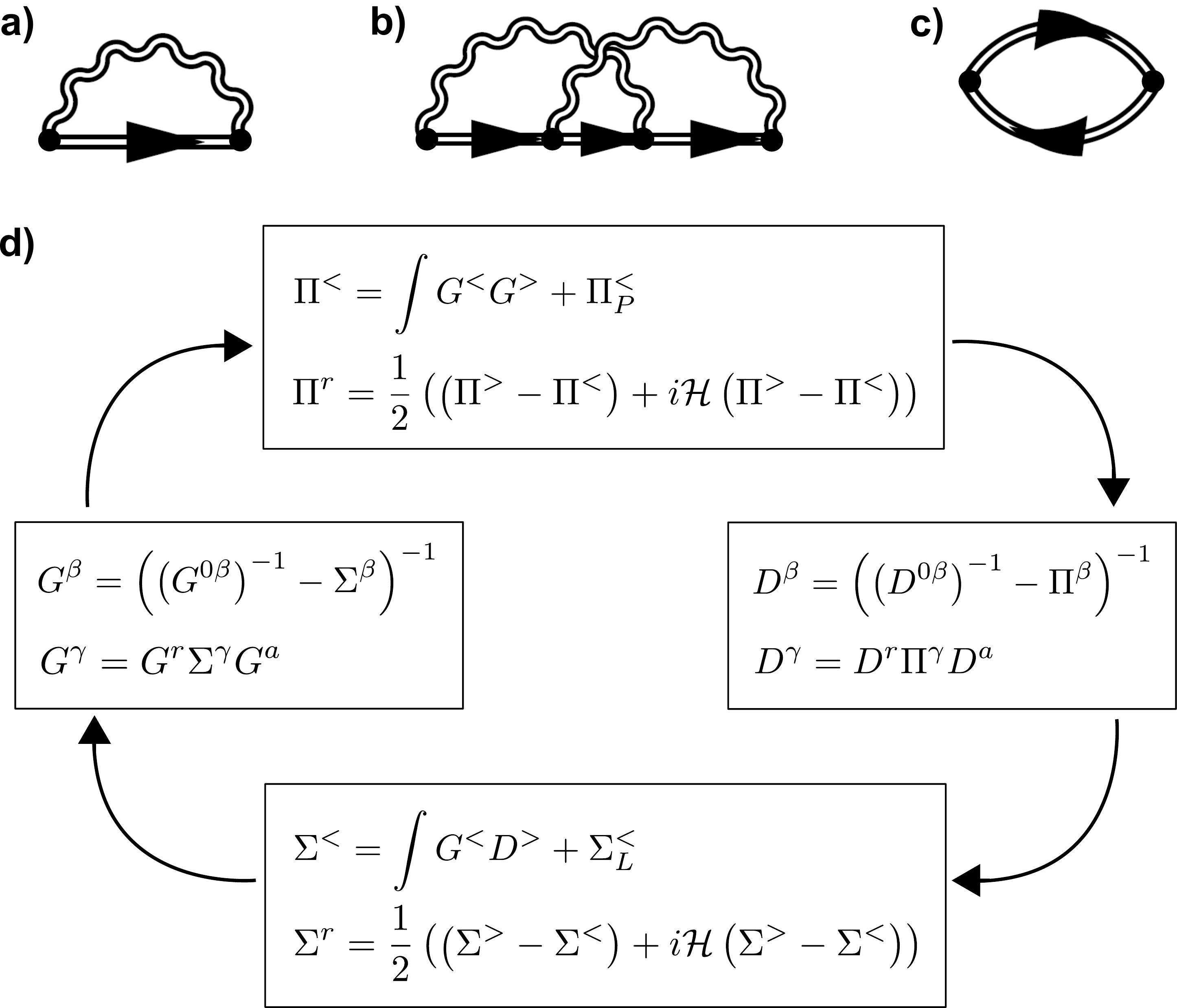}}
\caption{\textbf{a)} Electron SE diagram $\sim g^{2}$ corresponding to the SCBA. \textbf{b)} Example of vertex corrections diagram $\sim g^{4}$ where different photon lines cross eachother, that are not taken into account in the SCBA. \textbf{c)} Bubble diagram for the photon SE $\sim g^{2}$ in the SCBA. Electron GFs are represented as double straight lines while photon GFs are represented as double wiggly lines. \textbf{d)} Self-consistent algorithm used to compute electron and photon GFs. We proceed by successive iterations starting from the non-interacting electron GFs (left box with $\Sigma=0$) until convergence is reached. $\gamma=<,>$ stands for ``lesser'' and ``greater'' GFs, respectively.}
\label{fig_self}
\end{figure}

In the framework of the Self-Consistent Born Approximation (SCBA), the ``lesser'' and ``greater'' electron Self-Energies (SEs) can be decomposed as $\underline{\Sigma}^{\lessgtr}_{\alpha} (\omega) = \underline{\Sigma}^{\lessgtr}_{I,\alpha} (\omega)+ \underline{\Sigma}^{\lessgtr}_{L,\alpha}$, where

\begin{align}
\underline{\Sigma}^{<}_{I,\alpha} (\omega)&= i g^{2} \sum_{\alpha'} \left(1-\delta_{\alpha,\alpha'} \right)\int \!\! \frac{d\omega'}{2\pi} \underline{G}^{<}_{\alpha'} (\omega+\omega') D^{>} (\omega')  \nonumber \\
\underline{\Sigma}^{>}_{I,\alpha} (\omega)&= i g^{2} \sum_{\alpha'} \left(1-\delta_{\alpha,\alpha'} \right)\int \!\! \frac{d\omega'}{2\pi} \underline{G}^{>}_{\alpha'} (\omega+\omega') D^{<} (\omega')
\label{lm_se_lala}
\end{align}
represent the electron SE corrections [represented by the diagram in Fig.~\ref{fig_self} \textbf{a)}] due to the light-matter coupling, stemming from the emission/absorption of cavity excitations [poles of $D(\omega)$] when electrons undergo optical transitions between the two bands. The contributions:

\begin{align}
&\underline{\Sigma}^{<}_{L,\alpha} =i\Gamma_{\alpha} \underline{\sigma}^{1}, \qquad \underline{\Sigma}^{>}_{L,\alpha} =- i\Gamma_{\alpha} \underline{\sigma}^{N},&
\label{leads_se_true}
\end{align}
exact as long as the Markovian approximation for the system-baths coupling holds true,  represent the broadening of electron states due to the coupling between the chain and the leads. $D^{>}(\omega)$ and $D^{<}(\omega)$ respectively denote the ``greater'' and ``lesser'' photon GFs defined in the following. 

Importantly, we find that the SCBA is exact in the framework of the rotating wave-approximation~\cite{scully}, i.e. when one neglects the counter-rotating terms $\propto (c^{\dagger}_{2,j} c_{1,j} a^{\dagger}+\textrm{h.c.})$ in Eq.~(\ref{H_intera2}). Indeed, by doing so, the property\footnote{This property holds true as long as the photonic part of the non-interacting ground state is not a squeezed state.} $\langle a a \rangle_{0}=\langle a^{\dagger} a^{\dagger}\rangle_{0}=0$ implies that the crossed diagram $\sim g^{4}$ represented on Fig.~\ref{fig_self} \textbf{b)} is absent from the contribution $\propto \langle \mathcal{T} H_{I}^{4} c_{\alpha,k}(\tau) c^{\dagger}_{\alpha,k} (\tau')\rangle_{0}$ in the perturbative expansion Eq.~(\ref{GF_def_ufc}). One can generalize this result to all orders regarding diagrams where different photon lines cross each other. Moreover, it is easy to check that vertex corrections, which are neglected in the SCBA, precisely provide this type of diagrams. Note that the presence of counter-rotating terms in Eq.~(\ref{H_intera2}) is known to lead to squeezing effects associated with the so-called ultrastrong coupling regime~\cite{PhysRevB.72.115303}. In this case, an additional contribution proportional to the squared cavity vector potential generally has to be included in the Hamiltonian~\cite{pol1,yamanoi,knight,lagaf,gaston,keeling1,nataf,viehmann,hagenmuller2,PhysRevLett.110.133603,vukics2,bamba}.

The retarded and advanced electron SEs can then be efficiently calculated using the real-time equality~\cite{pourfath} $\underline{\Sigma}^{r}_{\alpha} (t)=\theta(t) \left(\underline{\Sigma}^{>}_{\alpha} (t) - \underline{\Sigma}^{<}_{\alpha} (t) \right)$, where $\theta$ is the Heaviside function. Introducing the broadening function $\underline{\chi}_{\alpha} (\omega)=i\left(\underline{\Sigma}^{>}_{\alpha} (\omega) - \underline{\Sigma}^{<}_{\alpha} (\omega) \right)$, the previous equality can be written in the frequency domain as:

\begin{align} 
\underline{\Sigma}^{r}_{\alpha} (\omega)= \frac{1}{2} \left(-i \underline{\chi}_{\alpha} (\omega) + \mathcal{H} [\underline{\chi}_{\alpha}] (\omega) \right),
\label{rere_self_causal}
\end{align} 
where 

\begin{align*} 
\mathcal{H} [\underline{\chi}_{\alpha} ] (\omega)= \frac{1}{\pi} \textrm{p.v} \int \! d\omega'  \frac{\underline{\chi}_{\alpha} (\omega')}{\omega-\omega'}
\end{align*} 
denotes the Hilbert transform, and p.v the Cauchy principal value. As a causal function, the real and imaginary parts of $\underline{\Sigma}^{r}_{\alpha} (\omega)$ are related to each other by Kramers-Kronig relations, as it can be checked directly from Eq.~(\ref{rere_self_causal}). The advanced SE is given by $\underline{\Sigma}^{a}_{\alpha} (\omega)=\left(\underline{\Sigma}^{r}_{\alpha} (\omega)\right)^{\dagger}$. \textit{The function $\underline{\chi}_{\alpha} (\omega)$ describes the broadening of Bloch states induced by the coupling to the leads and to the cavity mode}, while the real part of $\underline{\Sigma}^{r}_{\alpha} (\omega)$ provides a shift of the Bloch state energies $\omega_{\alpha,k}$. The retarded and ``lesser'' photon GFs are defined as:
\begin{align}
D^{r} (\omega) & =-i \int_{0}^{+\infty} \!\! dt e^{i \omega t} \langle \left[ A (t), A (0) \right] \rangle. \nonumber \\
D^{<} (\omega) & =-i \int_{-\infty}^{+\infty} \!\! dt e^{i \omega t} \langle A (t) A (0) \rangle, 
\label{DD_less}
\end{align} 
with similar definitions for $D^{a} (\omega)$ and $D^{>} (\omega)$. As for electrons, one can show (see \ref{negfs_form}) that $D^{r}$ and $D^{a}$ satisfy the Dyson equation:

\begin{align}
D^{\beta} (\omega)&=\left(\left(D^{\beta}_{0} (\omega) \right)^{-1} - \Pi^{\beta} (\omega) \right)^{-1}, 
\label{dydy_ph1}
\end{align}  
while $D^{>}$ and $D^{<}$ are obtained from the Keldysh equation:

\begin{align}
D^{\gamma} (\omega)&=D^{r}(\omega) \Pi^{\gamma} (\omega) D^{a}(\omega). 
\label{kehl_ph1}
\end{align} 

The expressions of the non-interacting (in the cavity vacuum state) photon GFs $D_{0} (\omega)$ are given by:

\begin{align}
D^{0<} (\omega)&=-2i\pi \delta(\omega+\omega_{c}) \nonumber \\
D^{0>} (\omega)&=-2i\pi \delta(\omega-\omega_{c}) \nonumber \\
D^{0a} (\omega)&=\frac{2\omega_{c}}{(\omega - i 0^{+})^{2} - \omega^{2}_{c}},
\label{ph_less_cav_nonint}
\end{align} 
and $D^{0r} =(D^{0a})^{*}$. 

In the SCBA, the ``lesser'' and ``greater'' photon SEs can again be decomposed as $\Pi^{\lessgtr} (\omega)=\Pi^{\lessgtr}_{I} (\omega)+\Pi^{\lessgtr}_{P} (\omega)$, where the light-matter contribution 

\begin{align}
\Pi^{<}_{I} (\omega)&=-i g^{2}\sum_{\alpha,\alpha'}\left(1-\delta_{\alpha,\alpha'} \right) \textbf{Tr} \!\! \int \!\! \frac{d\omega'}{2\pi} \underline{G}^{<}_{\alpha} (\omega+\omega')\underline{G}^{>}_{\alpha'} (\omega') \nonumber \\
\Pi^{>}_{I} (\omega)&=-i g^{2}\sum_{\alpha,\alpha'}\left(1-\delta_{\alpha,\alpha'} \right) \textbf{Tr} \!\! \int \!\! \frac{d\omega'}{2\pi} \underline{G}^{>}_{\alpha} (\omega+\omega')\underline{G}^{<}_{\alpha'} (\omega') 
\label{popo_ph_less}
\end{align}	
can be identified with the polarization function associated with the transition dipole moments, which provides a dressing of the bare cavity photon GF $D_{0}$. The polarization is represented by the bubble diagram shown on Fig.~\ref{fig_self} \textbf{c)}. On the other hand, the coupling between the cavity mode and the photon bath is described by the ``exact'' SE contribution: 

\begin{align}
&\Pi^{<}_{P} (\omega)=-i\kappa \theta (-\omega), \qquad \Pi^{>}_{P} (\omega)=-i\kappa \theta (\omega).&
\label{popo_ph_less22}
\end{align}
	
Here, we have assumed a vanishing mean population of extra-cavity photons, namely $\langle a^{\dagger}_{\bf p} a_{\bf p}\rangle \approx 0$. Similarly to electrons, the retarded and advanced photon SEs can be computed from the equation $\Pi^{r} (t)=\theta(t) \left(\Pi^{>} (t) - \Pi^{<} (t) \right)$, by introducing a photonic broadening function similar to Eq.~(\ref{rere_self_causal}). The retarded photon GF can be used to define the normalized cavity photon DOS as: 

\begin{align}
A_{c}(\omega)=-2\Im D^{r}(\omega),
\label{CAS_ff}
\end{align}	
which can be directly accessed experimentally by measuring the cavity excitation spectrum. Note that the photon GF $D^{<} (\omega)$ is related to the mean cavity photon number in the steady state (up to small squeezing terms) as $\bar{n} \equiv \langle a^{\dagger} a\rangle = -\frac{1}{2} \left(\int \frac{d\omega}{2\pi} \Im D^{<}(\omega) +1 \right)$.   

\vspace{2mm}

Above, we have shown that electron/photon SEs and GFs are related to each other by a closed set of integro-differential equations. The numerical procedure to solve these equations self-consistently is sketched on Fig.~\ref{fig_self} \textbf{d)}: One substitutes the fully interacting electron GFs in Eq.~(\ref{popo_ph_less}) with the non-interacting ones Eq.~(\ref{gg_elec32}), to compute the first-order ``lesser'' and ``greater'' photon SEs. From the latter, one deduces the retarded and advanced photon SEs and then computes the first-order photon GFs using Eqs.~(\ref{dydy_ph1}), (\ref{kehl_ph1}), and (\ref{ph_less_cav_nonint}). These photon GFs combined with the non-interacting electron GFs Eq.~(\ref{gg_elec32}) are then used to compute the first-order electron SEs from Eqs.~(\ref{lm_se_lala}) and (\ref{rere_self_causal}), which in turn can be substituted in the Dyson and Keldysh equations (\ref{dyson_aspi}) and (\ref{gless_ff_mat}) to obtain the first-order electron GFs. The whole cycle is repeated until convergence.  

\subsection{Quantum master equation formalism}
\label{qme_eff_part}

In this section, we introduce the full QME relevant to investigate the system (Sec.~\ref{QME_eeqq}), and show how it can be used to compute the steady-state current in Sec.~\ref{QME_steadtyy}. 

\subsubsection{Full quantum master equation.}
\label{QME_eeqq}

The time evolution of the joint density operator $\rho$ for the 1D chain and the cavity mode is given by the QME:
 
\begin{align}
\partial_\tau \rho &= - i [H_{S}, \rho] + \mathcal{L}_{1} \rho +\mathcal{L}_{N}\rho +\mathcal{L}_{\rm ph} \rho. 
\label{full_mast}
\end{align}

Here, the commutator $- i[H_{S}, \rho]$ describes the coherent dynamics due to the Hamiltonian $H_{S} = H_{e} + H_{t} + H_{I} + H_{c}$ introduced in Sec.~\ref{phi_s}. In the previous section, we have seen that while counter-rotating terms are formally included in the coupling Hamiltonian $H_{I}$, they do not play any role in the absence of vertex corrections. Here, we directly use the rotating-wave approximation and consider the coupling Hamiltonian:

\begin{align}
H_{I} = g \sum_{j=1}^{N} \left(c^{\dagger}_{2,j} c_{1,j} a + a^{\dagger} c^{\dagger}_{1,j} c_{2,j} \right),
\label{ha_full_path}
\end{align}
instead of Eq.~(\ref{H_intera2}). The additional terms in the right-hand side of Eq.~(\ref{full_mast}) are due to the coupling of the chain to the external degrees of freedom. The injection of electrons at the first site is described by the term~\cite{Medvedyeva2013}:
 
\begin{align*}
\mathcal{L}_{1} \rho = \sum_{\alpha} \frac{\Gamma_{\alpha}}{2} \mathcal{D} [c^{\dagger}_{\alpha,1}] \rho.
\end{align*}

Similarly, the extraction of electrons at the last site is given by:

\begin{align*}
\mathcal{L}_{N} \rho = \sum_{\alpha} \frac{\Gamma_{\alpha}}{2} \mathcal{D} [c_{\alpha,N}] \rho, 
\end{align*}
while the action of $\mathcal{D} [A]$ on $\rho$ is defined by the Lindblad superoperator~\cite{PhysRevB.80.035110,PhysRevB.91.205416}:
 
\begin{align*}
\mathcal{D} [A] \rho = - \{A^{\dagger} A, \rho\} + 2 A \rho A^{\dagger}.
\end{align*}

Assuming the extra-cavity photon bath close to its vacuum state, the cavity photon decay is described by the term: 

\begin{align*}
\mathcal{L}_{P} \rho =  \frac{\kappa}{2} \mathcal{D} [a] \rho.
\end{align*}

\subsubsection{Steady-state current.}
\label{QME_steadtyy}

Since we are interested in the steady-state current flowing through the chain, we now explain how the latter can be computed from the QME (\ref{full_mast}). The time evolution of the expectation value of a generic observable $A$ is given by the equation:

\begin{align}
\partial_\tau \langle A \rangle = {\rm Tr} (A \partial_{\tau} \rho),
\label{eq:expec_t}
\end{align} 
where the trace ${\rm Tr}$ denotes the sum over the diagonal elements in matrix representation. Using Eq.~(\ref{full_mast}), one can show that the expectation value of the total charge operator $Q_S = e \sum_{\alpha,j} \hat{n}_{\alpha j}$ evolves according to:

\begin{align}
\partial_{\tau} \langle Q_S \rangle = J_{s} + J_{d},
\label{eq:tot_cur}
\end{align} 	
where the currents flowing through the source and the drain (leads) are respectively expressed as: 

\begin{align}
J_{s} &= \sum_{\alpha} e \Gamma_{\alpha} \langle 1 - \hat{n}_{\alpha 1} \rangle = \sum_{\alpha} \frac{e \Gamma_{\alpha}}{2} {\rm Tr} \big( \hat{n}_{\alpha 1} \mathcal{D} [c^{\dagger}_{\alpha,1}] \rho \big) \nonumber \\
J_{d} &= - \sum_{\alpha} e \Gamma_{\alpha} \langle \hat{n}_{\alpha N} \rangle =\sum_{\alpha} \frac{e \Gamma_{\alpha}}{2} {\rm Tr} \big( \hat{n}_{\alpha N} \mathcal{D} [c_{\alpha,N}] \rho \big). 
\label{Jinout2}
\end{align}

The last equalities in the right-hand side of both lines can be derived by using the cyclic properties of the trace and fermionic commutation relations. In the steady-state, since $\partial_{\tau} \rho = 0$, we have $\partial_{\tau} \langle Q_S \rangle = 0$ and from Eq.~(\ref{eq:tot_cur}), $J_{s} = - J_{d}$. It is straightforward to check that the steady-state current calculated from Eq.~(\ref{Jinout2}) corresponds to Eq.~(\ref{current_ss_pop}) of Sec.~\ref{gf_self_ener_part}. We numerically solve for the steady-state, either by computing the time evolution of Eq.~(\ref{full_mast}) using the Runge-Kutta method (fourth order), or by looking for the null eigenvector of the Liouvillian $\mathcal{L}$ (with $\partial_{\tau} \rho= \mathcal{L} \rho$) written in a matrix form~\cite{carlos}. Details on how to implement fermionic operators in matrix representation can be found in ~\cite{moritz}.

\subsection{Effective quantum master equation: Pure electron dynamics}
\label{sec:eff_master}

In this section, we consider the dissipative regime obtained when the photon decay rate $\kappa$ is larger than any other energy scale except $\omega_{21}$. We show that the fast cavity field evolution can be adiabatically eliminated in this regime, resulting in an effective QME involving only electronic degrees of freedom. 

\vspace{2mm}

It is convenient to introduce the density operator $\tilde{\rho} = U \rho U^{\dagger}$ in the rotating frame defined by the unitary operator:

\begin{align*}
U (\tau) = \exp\big[ i (H_{e} + \omega_{21}  a^{\dagger} a) \tau\big].
\end{align*}

The time evolution of the operator $\tilde{\rho}$ is then derived as:

\begin{align}
\partial_\tau \tilde{\rho} &= - i [\widetilde{H}, \tilde{\rho}] + \mathcal{L}_{1} \tilde{\rho} +\mathcal{L}_{N} \tilde{\rho} +\mathcal{L}_{P} \tilde{\rho},
\label{eq:rho_tilde}
\end{align}
with the Hamiltonian

\begin{align*}
\widetilde{H} = H_{t} + \widetilde{H}_{c} + H_{I}.
\end{align*}

The (rescaled) cavity Hamiltonian $\widetilde{H}_c = - \Delta a^{\dagger} a$ contains the detuning $\Delta = \omega_{21} - \omega_c$ between the transition and the cavity mode frequencies $\omega_{21}$ and $\omega_c$. Despite the fact that we will only discuss results obtained in the resonant case $\Delta = 0$, we perform the adiabatic elimination in the general situation for the sake of completeness. The adiabatic elimination procedure using projectors ~\cite{Bonifacio1971,Schuetz2013} is detailed in \ref{master_form1} and outlined in the following. We first recast the right-hand side of Eq.~(\ref{eq:rho_tilde}) as:

\begin{align*}
\partial_{\tau} \tilde{\rho} = \mathcal{L}_{e} \tilde{\rho} + \widetilde{\mathcal{L}}_{c} \tilde{\rho} + \mathcal{L}_I \tilde{\rho},
\end{align*}
in terms of the purely electronic part $\mathcal{L}_{e} \tilde{\rho} = - i [H_t , \tilde{\rho}] + \mathcal{L}_{1} \tilde{\rho}  + \mathcal{L}_{N} \tilde{\rho}$, the photonic part $\widetilde{\mathcal{L}}_{c} \tilde{\rho} = \mathcal{L}_c \tilde{\rho} + \kappa a \tilde{\rho} a^{\dagger}$, as well as the interaction part $\mathcal{L}_{I} \tilde{\rho} = - i [H_I , \tilde{\rho}]$. The photonic part $\widetilde{\mathcal{L}}_{c} \tilde{\rho}$ contains the contribution:

\begin{align*}
\mathcal{L}_{c} \tilde{\rho} &= \left(i \Delta - \frac{\kappa}{2}\right) a^{\dagger} a \tilde{\rho} + \left(- i \Delta - \frac{\kappa}{2} \right) \tilde{\rho} a^{\dagger} a,
\end{align*}
which generally gives rise to damped oscillations for the relaxation of the cavity field. When the cavity decay rate $\kappa$ is much larger than the rates governing the electron dynamics (i.e. $t_{\alpha}$ and $\Gamma_{\alpha}$), one can separate the fast cavity dynamics from the electronic one occuring on a comparably long time-scale. In the presence of light-matter interactions, such a separation is still possible whenever the light-matter coupling strength $g$ is sufficiently weak. In \ref{master_form1}, we present a detailed derivation of the equation of motion for the electronic dynamics only, in which the cavity field has been adiabatically eliminated (retardation effects scaling with $t_{\alpha}/\kappa$ and/or $\Gamma_{\alpha}/\kappa$ are neglected), and where the light-matter interaction is treated to second order. Moreover, we restrict our discussion to the case where the cavity field remains close to its vacuum state, which is consistent with the large damping rate $\kappa$. In this limit, the time evolution of $\hat{\rho}$, the full density operator projected onto the cavity vacuum, is governed by the effective QME:

\begin{align}
\partial_\tau \hat{\rho} =\mathcal{L}_{e} \hat{\rho} 
- i  \big[ \frac{g^{2}\Delta}{\Delta^{2}+(\kappa/2)^2} S^{+} S^{-} , \hat{\rho} \big] - \frac{g^{2}\kappa/2}{\Delta^{2}+(\kappa/2)^2} \big(  S^{+} S^{-} \hat{\rho} + \hat{\rho} S^{+} S^{-}  - 2 S^{-} \hat{\rho} S^{+} \big). 
\label{eff_master_gene}
\end{align}

Here, $S^{+}=\sum_{j} c^{\dagger}_{2,j} c_{1,j}$ ($S^{-} = (S^{+})^{\dagger}$) denotes a collective raising (lowering) operator for the electrons from the lower (upper) to the upper (lower) band. In the resonant case ($\omega_{21} = \omega_c$), the time evolution of $\hat{\rho}$ can be simplified:

\begin{align}
\partial_\tau \hat{\rho} = \mathcal{L}_{e} \hat{\rho} +\mathcal{L}_{\Gamma_{c}} \hat{\rho}, 
\label{full_eff_m}
\end{align}
\textit{where light-induced interactions between electrons are entirely cast into the dissipator:}

\begin{align}
\mathcal{L}_{\Gamma_{c}} \hat{\rho} = - 2\Gamma_{c} \left( S^{+} S^{-}\hat{\rho} + \hat{\rho} S^{+} S^{-} - 2 S^{-} \hat{\rho} S^{+} \right),
\label{dissi_Gc_m} 
\end{align}
with $\Gamma_c = g^2/\kappa$. \textit{This shows that in the dissipative regime where $\kappa$ is the largest parameter, light-matter interactions are governed by the parameter $\Gamma_{c}$}. We remark that such a term also appears in the case of pseudo-spins (e.g. in a two-level atomic description) coupled to a cavity mode with strong dissipation~\cite{Bonifacio1971,Bullough1987,Garraway2011}. Introducing the local raising operators $s^{+}_{j}=c^{\dagger}_{2,j}c_{1,j}$ and the corresponding lowering operators $s^{-}_{j}=c^{\dagger}_{1,j}c_{2,j}$, the dissipator Eq.~(\ref{dissi_Gc_m}) can be rewritten as:

\begin{align}
\mathcal{L}_{\Gamma_{c}} \hat{\rho} = - 2 \Gamma_{c} \sum_{j=1}^{N} \left( s^{+}_{j} s^{-}_{j}\hat{\rho} + \hat{\rho} s^{+}_{j} s^{-}_{j} - 2 s^{-}_{j} \hat{\rho} s^{+}_{j} \right) - 2\Gamma_{c} \sum_{\substack{i,j \\ i\neq j}}^{N} \left( s^{+}_{j} s^{-}_{i}\hat{\rho} + \hat{\rho} s^{+}_{j} s^{-}_{i} - 2 s^{-}_{i} \hat{\rho} s^{+}_{j} \right). 
\label{dissi_Gspil} 
\end{align}
Here, both local and non-local coupling terms can be identified, and correspond to the first and second terms in the right-hand side of Eq.~(\ref{dissi_Gspil}), respectively. In spin-cavity setups, non-local terms can induce spin-spin correlations and ultimately lead to synchronization and superradiance~\cite{Meiser2010,Jaeger2017,Bonifacio1971}. In our situation, they give rise to non-local exchange of interband excitations, which are partly taken into account in the SCBA, as discussed in Sec.~\ref{self_self22}. Regarding charge transport, the dissipator Eq.~(\ref{dissi_Gc_m}) induces a global (collective) population transfer of electrons from the upper to the lower band. Indeed, denoting the total electron population in the band $\alpha$ by $N_{\alpha} = \sum_{j}  \hat{n}_{\alpha j}$, its time evolution due to light-matter interactions in the dissipative regime is:

\begin{align*}
\partial_{\tau} \langle N_{1} \rangle &= {\rm Tr} (N_1 \mathcal{L}_{{\Gamma}_c} \hat{\rho})
= 4 \Gamma_c \langle S^+ S^- \rangle \\
\partial_{\tau} \langle N_{2} \rangle &= {\rm Tr} (N_2 \mathcal{L}_{{\Gamma}_c} \hat{\rho})
= - 4 \Gamma_c \langle S^+ S^- \rangle,   
\end{align*}
which provides $\partial_{\tau} \langle N_1 \rangle = - \partial_{\tau} \langle N_2 \rangle$, and demonstrates the population exchange between the two bands. Moreover, the rate associated with this population transfer can be related to the mean intra-cavity photon number, approximated as $\langle a^{\dagger} a \rangle \simeq (4 \Gamma_{c} / \kappa) \langle S^+ S^- \rangle$ in the adiabatic limit~\cite{Meiser2010,Bonifacio1971}. The change of the first band population thus takes the simple form $\partial_{\tau} \langle N_1 \rangle = \kappa \langle a^{\dagger} a\rangle$: The population transfer from the upper to the lower band is accompanied by the creation of photons which are then dissipated with the rate $\kappa$. In Sec.~\ref{anal_app_part}, we derive an analytical estimate for the current enhancement in the dissipative regime, by calculating the time evolution of expectation values $\langle c^{\dagger}_{\alpha,i} c_{\alpha,j}\rangle$ with suitable approximations.

\section{Results}
\label{res_part}

In this section, we present both analytical and numerical results using the QME and NGFs methods. In Sec.~\ref{no_coupl_part}, we first discuss the situation without light-matter coupling, by computing the steady-state current, the electron density profile in both bands, as well as the time evolution of the electron spectral function. In Sec.~\ref{spec_lan_jd}, we explain how light-matter interactions lead to a broadening of the electron DOS, and show that the latter scales with the cooperativity. We further explain how polariton modes arise from the dressing of the photon GF by the electron-hole polarization. In Sec.~\ref{sec:comparison}, we compare numerical results for the steady-state current obtained with the different methods, and distinguish between two regimes characterized by the ratio between the cavity photon decay rate $\kappa$ and the upper electronic bandwidth $W_2$. In Sec.~\ref{diss_reg_sec}, we investigate the dissipative regime $\kappa/W_2\gg 1$ using both analytical and numerical calculations, and demonstrate the existence of a collective coupling to light when $g$ exceeds the energy spacing between adjacent Bloch states in the upper band. First, an analytical expression of the steady-state current valid for small coupling strength is given in Sec.~\ref{anal_app_part}, while numerical calculations using both NGFs and QMEs are presented in Sec.~\ref{coll_dress_part}. In Sec.~\ref{non_local_coco2}, we show that non-local electronic correlations occur for large coupling strengths in this regime. The ``coherent'' regime $\kappa /W_2 \ll 1$ is investigated in Sec.~\ref{ind_dress_part}. In particular, numerical results obtained with the NGFs method for the transmission spectrum and the cavity photon DOS are presented in Sec.~\ref{ttdress_part43}. In Sec.~\ref{ttdress_part666}, we compute the time evolution of the electron spectral function in the lower orbital states, and show that a coherent dynamics involving delocalized states takes place when $g$ is smaller than the energy spacing between adjacent Bloch states in the upper band. Concluding remarks concerning the cavity photon population and the chain length are given in Sec.~\ref{comp_part}.

\subsection{Absence of light-matter coupling ($g=0$)}
\label{no_coupl_part}

Here, we discuss the system properties in the absence of light-matter coupling, by computing the steady-state current, the electron density spatial profile, as well as the time evolution of the electron spectral function.

\vspace{2mm}

In the absence of light-matter coupling ($g=0$), the two bands are independent and the eigenstates of the chain consist of two identical sets of the $N$ Bloch states defined in Eq.~(\ref{bloch_sfou}). The only finite SE contribution Eq.~(\ref{leads_se_true}) is due to the coupling to the leads, and is proportional to the decay rate $\Gamma_{\alpha}$ of the Bloch states in the band $\alpha$. The transport properties of the chain are only driven by the ratio between $\Gamma_{\alpha}$ and $t_{\alpha}$, and the steady-state current does not depend on the chain length $N$. Using the spectral function sum rule $\int \! d\omega A_{\alpha,k,k'}(\omega) =2\pi \delta_{k,k'}$ in Eq.~(\ref{transsm}), the steady-state current $J^{(0)}_{\alpha}$ flowing through the band $\alpha$ can be written as: 

\begin{align}
J^{(0)}_{\alpha} = \frac{e \Gamma_{\alpha}}{2} \, \left( 1 + \int \frac{d \omega}{2 \pi} \text{\bf Tr} \left[ \left( \underline{\sigma}^N - \underline{\sigma}^1\right) \circ \Im \underline{G}^{<}_{\alpha} (\omega)\right]  \right).
\label{jj_0_causal}
\end{align}

One can then use Eqs.~(\ref{dyson_aspi}), (\ref{gless_ff_mat}), (\ref{leads_se_true}), and (\ref{rere_self_causal}) with e.g. $N=2$, and obtain the current as:

\begin{align}
J^{(0)}_{\alpha} &= \frac{e \Gamma_{\alpha}/2}{1 + \left(\frac{\Gamma_{\alpha}}{2t_{\alpha}}\right)^2},
\label{jj_1_causal}
\end{align}
in agreement with the results of ~\cite{Medvedyeva2013}. The electron populations at the edges of the chain follow from Eq.~(\ref{current_ss_pop}):

\begin{align*}
n_{\alpha N} = 1-n_{\alpha 1}=\frac{1}{2 + \frac{\Gamma^{2}_{\alpha}}{2t^{2}_{\alpha}}}.
\end{align*}

Two different regimes of transport can be distinguished. When $t_{\alpha} \ll \Gamma_{\alpha}$, transport is inhibited due to the small lifetime of Bloch states compared to the typical hopping time. Bloch states are thus not well resolved and the steady-state current is given by $J^{(0)}_{\alpha} \sim 2 e t^{2}_{\alpha}/\Gamma_{\alpha} \approx 0$. In this situation, the first and last sites are respectively fully occupied and completely empty, i.e. $n_{\alpha 1}\approx 1$ and $n_{\alpha N} \approx 0$. The opposite regime $t_{\alpha} \gg \Gamma_{\alpha}$ features single-electron transport through well-resolved Bloch states. In this regime, the current $J^{(0)}_{\alpha} \approx e\Gamma_{\alpha}/2$ is only limited by the rate $\Gamma_{\alpha}$, and the first and last sites are half-filled, namely $n_{\alpha 1}=n_{\alpha N}=0.5$. 

\begin{figure}[ht]
\centerline{\includegraphics[width=0.7\columnwidth]{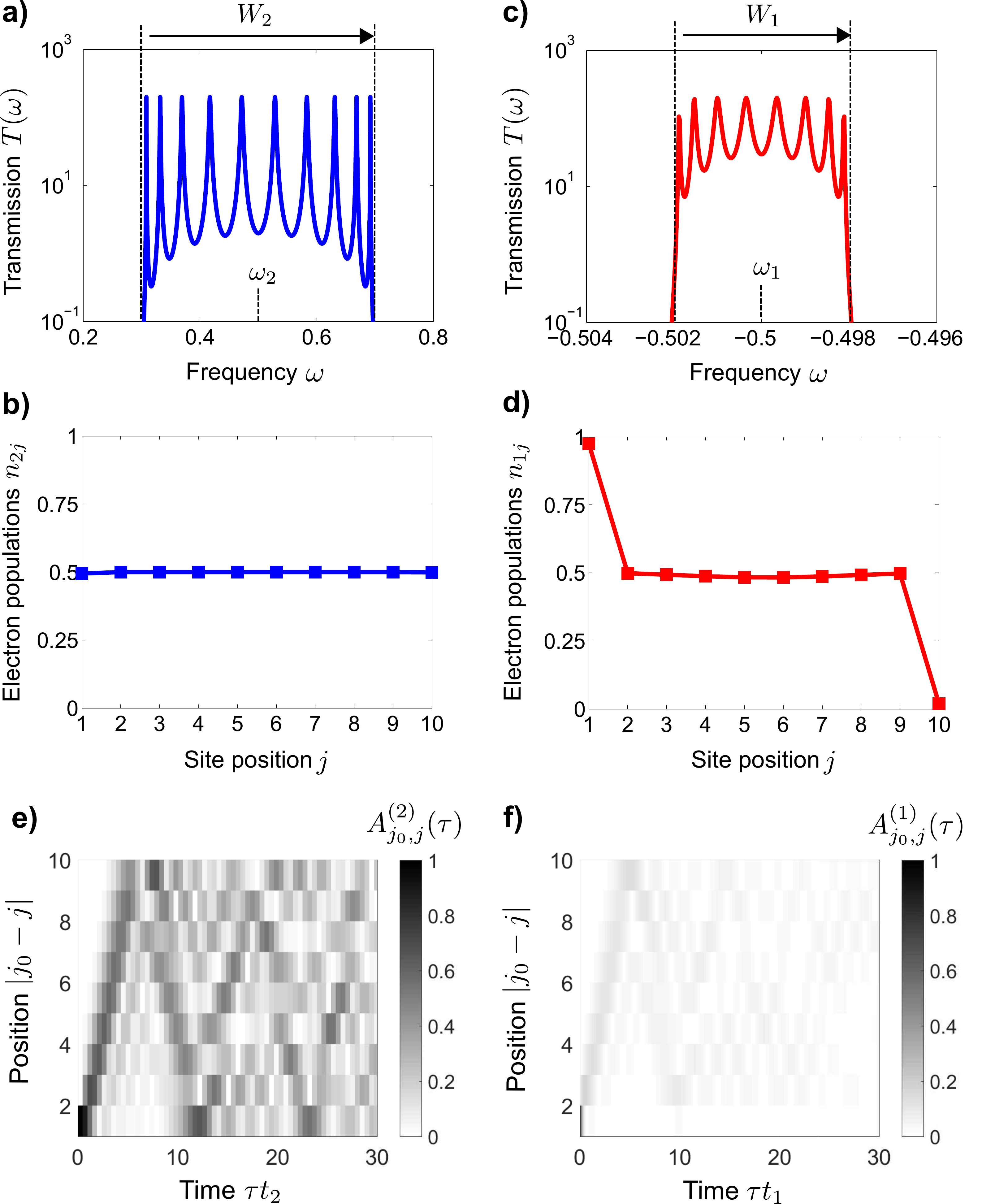}}
\caption{\textbf{a)}-\textbf{c)} (Log-scale) Transmission spectrum $T(\omega)$ versus frequency for $g=0$, in the vicinity of \textbf{a)} the upper orbital energy $\omega_{2}=0.5$ (blue line), and \textbf{c)} the lower orbital energy $\omega_{1}=-0.5$ (red line). \textbf{b)}-\textbf{d)} Spatial profiles of the electron density $n_{\alpha j}$ for $g=0$, in \textbf{b)} the upper orbitals (blue squares), and \textbf{d)} the lower orbitals (red squares). \textbf{e)}-\textbf{f)} Spectral function $A^{(\alpha)}_{j_0,j}(\tau)$ for $g=0$ as a function of position and time, obtained after injection of a particle at site $j_0=1$ and time $\tau=0$ in \textbf{e)} the upper orbital, and \textbf{f)} the lower orbital. Time is in units of the hopping rates in the lower and upper bands, respectively. Parameters are $N=10$, $t_{1}=10^{-3}$, $\Gamma= 10^{-2}$, and $t_{2}=0.1$.}
\label{fig3}
\end{figure}

\vspace{2mm}

\textit{As already mentioned, we only consider the situation where $\Gamma_1=\Gamma_2 \equiv \Gamma$ and $t_{2} \gg \Gamma \gg t_{1}$.} The different transport regimes can be identified in the transmission spectrum $T(\omega)$, which is represented on Fig.~\ref{fig3} for $g=0$. In the vicinity of the upper orbital $\omega \approx \omega_{2}$, the relation $t_2 \gg \Gamma$ leads to $N$ well-resolved peaks (Bloch states) of width $\sim \Gamma/N$, distributed over the bandwidth $W_{2}$ [Fig.~\ref{fig3} \textbf{a)}]. Moreover, all sites are half-filled [Fig.~\ref{fig3} \textbf{b)}], and the partial current obtained by integrating $T(\omega)$ in the vicinity of $\omega_{2}$ is $J^{(0)}_{2} \approx e\Gamma/2$. In the vicinity of the lower orbital $\omega \approx \omega_{1}$, however, the dynamics does not involve well-resolved Bloch states since $t_{1}\ll \Gamma$. This results in a number of peaks smaller than $N$ within the bandwidth $W_{1}$ [Fig.~\ref{fig3} \textbf{c)}], a half-filling of all sites except for the first and last ones [Fig.~\ref{fig3} \textbf{d)}], and therefore a very small current $J^{(0)}_{1}/e\Gamma \approx 2 (t_{1}/\Gamma)^2 \ll 1$.  

\newpage

It is interesting to analyze the propagation of excitations in the steady-state, by considering the electron spectral function $A^{(\alpha)}_{j_0,j}(\tau)$ defined as the Fourier transform (over both space and time variables) of the function $A^{(\alpha)}_{k,k'}(\omega)$ introduced in Sec.~\ref{caca_ngfs}:

\begin{align*}
A^{(\alpha)}_{j_0,j}(\tau) = 2 \Re \langle \{ c_{\alpha,j} (\tau), c^{\dagger}_{\alpha,j_0} (0) \} \rangle.
\end{align*}  

Physically, this function can be interpreted as follows: Considering an electron injected in the steady-state at site $j_0$ and time $\tau=0$ in the level $\alpha$, its wavefunction will be decomposed over the different sites under the time evolution governed by the total Hamiltonian (including interactions with the leads). The function $A^{(\alpha)}_{j_0,j}(\tau)$ corresponds to the overlap between this wavefunction at later time $\tau >0$ and that of an electron injected at time $\tau$ at an other site $j$, and provides information on what the wavefunction of an electron (or a hole) injected at a given site at $\tau=0$ looks like after a certain time $\tau$. This function is represented on Fig.~\ref{fig3} \textbf{e)} and \textbf{f)}, considering excitations propagating in the upper and the lower orbitals, respectively. In the former case, the dynamics of a particle injected in the upper level of site $j_0=1$ at $\tau=0$ involves a decomposition over the different well-resolved Bloch states of the upper band, resulting in the propagation of this particle throughout the chain [Fig.~\ref{fig3} \textbf{e)}]. On the other hand, since $t_{1}\ll \Gamma$, propagation in the lower band is hampered and most of the spectral weight stays localized at the injection site before being damped after a typical time $\sim 1/\Gamma$ [Fig.~\ref{fig3} \textbf{f)}]. In the following, we investigate how this physical picture is modified when the coupling to the cavity mode is turned on. 

\subsection{Spectral broadening and polaritons}
\label{spec_lan_jd}

In this section, we compute the first-order GFs and SEs, and explain how the electron DOS is broadened by the presence of light-matter interactions. In particular, we show that this broadening scales with the cooperativity in the dissipative regime, and explain how the dressing of the photon GF by the electron-hole polarization results in the appearance of polariton states.   

\vspace{2mm}

When $g\neq 0$, electrons can undergo interband transitions concurrently with the aborption/emission of cavity photons with energy $\omega \in [\omega_{21}-2 t_2, \omega_{21}+2 t_{2}]$ (for $t_1 \ll t_2$). This leads to the hybridization of the two bands, and provides a modification of the electron DOS and the transmission spectrum. In Sec.~\ref{gf_self_ener_part}, we have seen that the coupling between the two electronic bands and the cavity field is a self-consistent problem. The electron dynamics is affected by the electromagnetic field through emission/absorption of cavity photons, and the cavity field is in turn dressed by its interactions with the electron-hole polarization. The simplest approximation consists in neglecting the self-consistency\footnote{We find that this approximation gives a correct description only for very small values of the ratio $\Gamma_{c}/\Gamma$. If it is not the case, neglecting the self-consistency leads to spurious limiting behaviors as a result of the breaking of conservation laws such as the continuity equation for the current.} and calculate the first-order ($\propto g^{2}$) electron SE induced by the coupling to the leaky cavity mode [or equivalently the broadening function entering Eq.~(\ref{rere_self_causal})]. This self-energy is obtained by substituting the fully interacting electron and photon GFs in Eq.~(\ref{lm_se_lala}) with the non-interacting electron GFs Eq.~(\ref{gg_elec32}), and the photon GFs calculated without the interband contribution Eq.~(\ref{popo_ph_less}):

\begin{align} 
D^{r} (\omega) = \frac{2\omega_{c}}{\omega^{2}-\omega^{2}_{c}+ i \kappa \omega_{c} \textrm{sgn} (\omega)},
\label{dd_ret_fun12} 
\end{align}
where $\textrm{sgn}$ denotes the sign function. We point out that considering only the coupling of cavity photons to the external electromagnetic environment when calculating the photon SE is expected to be valid in the dissipative regime, where $\kappa$ is the largest energy scale. At resonance $\omega_{c}=\omega_{21}$, the first-order broadening function defined in Sec.~\ref{self_self22} is calculated as:

\begin{align} 
\chi_{\alpha,k,k'} (\omega)= \sum_{\alpha'} \frac{4\kappa g^{2} \omega^{2}_{21} (1-\delta_{\alpha,\alpha'})\delta_{k,k'}}{\left((\omega-\omega_{\alpha',k})^{2}-\omega^{2}_{21}\right)^{2}+(\kappa \omega_{21})^{2}} \Big( (1- n^{0}_{\alpha' k} ) \theta (\omega-\omega_{\alpha' k}) + n^{0}_{\alpha' k}  \theta (\omega_{\alpha',k}-\omega)\Big),
\label{broad_causal} 
\end{align}
where $n^{0}_{\alpha k}$ is the population of the Bloch state $(\alpha,k)$ in the initial ground state, without any interactions. This broadening function is diagonal with respect to $k$. Considering a Bloch state $k$ in the lower band $\alpha=1$, its light-induced broadening depends on the filling of the state $k$ in the upper band $\alpha'=2$. When $n^{0}_{2 k}=1$, the associated electron can undergo a transition from the upper to the lower band by emitting a photon with energy $\omega_{2,k}-\omega_{1,k}$. For $\omega = \omega_{1,k}$, and $\kappa/W_{2} \gg 1$ (dissipative regime), one has $\omega_{2,k} - \omega \approx \omega_{21}$, which simply yields: 

\begin{align} 
\chi_{1,k} (\omega) \approx 4\Gamma_{c},
\label{broad_causal2} 
\end{align}
where $\Gamma_{c}=g^{2}/\kappa$ has been introduced in Sec.~\ref{sec:eff_master}. Since the SE broadening due to the coupling to the leads is $\propto \Gamma$, one finds that \textit{in the dissipative regime, the light-induced relative broadening of the electron DOS is driven by the ratio $\Gamma_{c}/\Gamma$, which plays the role of a cooperativity parameter}. Moreover, the validity domain of the NGFs method is limited to the perturbative (``quasiparticle'') regime with $\Gamma_{c}/\Gamma\lesssim 1$. 

\vspace{2mm}

In the regime $t_{2}\gg t_{1}$, when the coupling strength $g$ becomes eventually larger than the typical energy spacing between two adjacent Bloch states in the upper band, a collective coupling of the different Bloch states to the cavity mode arises, and the electron-hole polarization given by Eq.~(\ref{popo_ph_less}) can no longer be neglected. In order to see how this collective coupling is related to the polarization dressing of the photon GF, one can compute the first-order retarded photon GF in the absence of cavity losses. We thus proceed in an opposite way to the one used previously, by neglecting the contribution due to the coupling to extra-cavity photons Eq.~(\ref{popo_ph_less22}), and replacing the fully interacting electron GFs in Eq.~(\ref{popo_ph_less}) with the non-interacting ones Eq.~(\ref{gg_elec32}). The photon SE is derived as:

\begin{align} 
\Pi^{r} (\omega) = \sum_{k} \frac{2g^{2} \left(n^{0}_{1 k}-n^{0}_{2 k}\right) \left(\omega_{2,k}-\omega_{1}\right)}{\left(\omega + i 0^{+}\right)^{2}-\left(\omega_{2,k}-\omega_{1}\right)^{2}}.
\label{selfener_ph_fo} 
\end{align}   

Furthermore, if we also assume $t_{2} \ll \omega_{21}$, namely neglecting the upper bandwidth with respect to the transition frequency, Eq.~(\ref{selfener_ph_fo}) takes the form of the usual interband polarization~\cite{wendler,lee,todorov} (which enters the definition of the dielectric permittivity~\cite{hopfield}) involving a collective response of the electron states:

\begin{align} 
\Pi^{r} (\omega) = \frac{2\Omega_{n_{0}}^{2} \omega_{21}}{\left(\omega + i 0^{+}\right)^{2}-\omega^{2}_{21}}.
\label{selfener_ph_fo22} 
\end{align}        

At this level of approximation, the collective vacuum Rabi frequency is defined as $\Omega_{n_{0}}=g\sqrt{\sum_{k} n^{0}_{1 k}-n^{0}_{2 k}}$, and depends on the initial population imbalance between the two bands. Replacing Eq.~(\ref{selfener_ph_fo22}) in the Dyson equation (\ref{dydy_ph1}), the first-order retarded photon GF can be written as:

\begin{align} 
\widetilde{D}^{r} (\omega) = \frac{2 \omega_{21}\left(\omega^{2}-\omega^{2}_{21}\right)}{\left[\left(\omega + i0^{+}\right)^{2}-\omega^{2}_{+}\right]\left[\left(\omega + i0^{+}\right)^{2}-\omega^{2}_{-}\right]},
\label{GF_ph_fo22} 
\end{align}  
at resonance ($\omega_{c}=\omega_{21}$). This function exhibits poles at the polariton frequencies $\omega_{\pm}=\sqrt{\omega^{2}_{21}+2\omega_{21}\Omega_{n_{0}}}$. Note that taking the cavity decay rate $\kappa$ into account would turn the latter into quasi-modes with imaginary frequency. The effect of this collective dressing of the photon GF on the electron spectral broadening can then be studied by computing the electron SE Eq.~(\ref{lm_se_lala}) together with Eqs.~(\ref{GF_ph_fo22}) and (\ref{gg_elec32}). While the result depends on the initial populations $n^{0}_{\alpha k}$ at this level of approximation, it is not the case when the self-consistency is taken into account, namely when using the fully interacting electron GFs in the photon SE Eq.~(\ref{popo_ph_less}). This will be studied numerically in Sec.~\ref{coll_dress_part}. 

\subsection{Comparison between the different methods}
\label{sec:comparison}

In this section, we benchmark the different methods used to compute the steady-state current, and show that the light-matter coupling is responsible for a current enhancement driven by the cooperativity parameter in the dissipative regime. We also compute numerically the broadening function introduced in Sec.~\ref{gf_self_ener_part} using self-consistent NGFs. 

\vspace{2mm}

\begin{figure}[ht]
\centerline{\includegraphics[width=1\columnwidth]{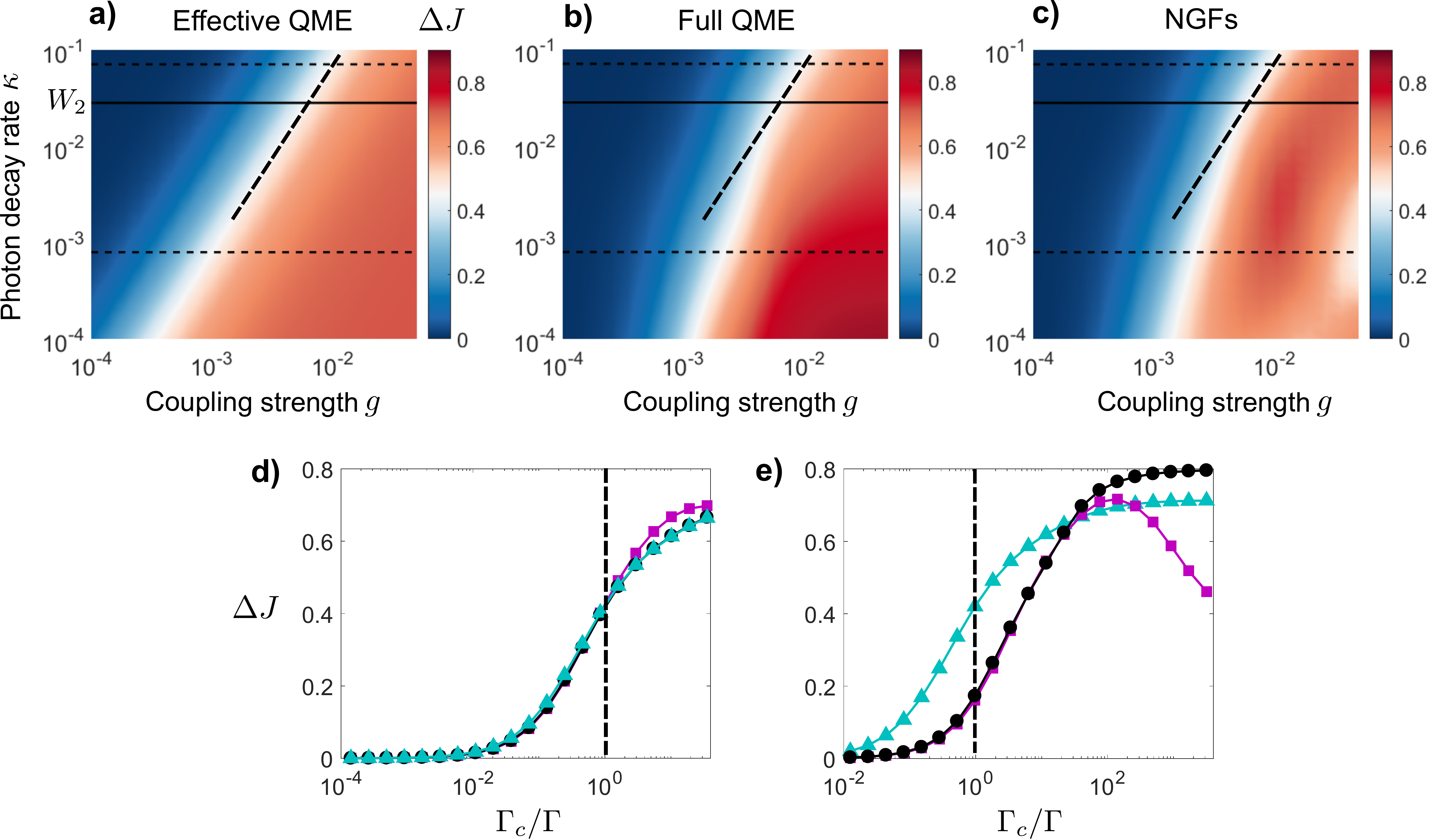}}
\caption{Relative current enhancement $\Delta J$ (see text) versus coupling strength $g$ and photon decay rate $\kappa$ (log-scale), obtained from \textbf{a)} the effective QME (\ref{full_eff_m}), \textbf{b)} the full QME (\ref{full_mast}), and \textbf{c)} the NGFs method. The diagonal dashed line $g^{2}/\kappa=\textrm{cst}$ is a guide to the eye, and the horizontal solid line corresponds to $\kappa=W_2$. \textbf{d)}-\textbf{e)} Relative current enhancement $\Delta J$ versus cooperativity $\Gamma_{c}/\Gamma$, for two different values of $\kappa$ represented by the horizontal dashed lines in the upper panels. \textbf{d)} $\kappa=0.07$ ($\kappa/W_2\approx 5$). \textbf{e)} $\kappa=8 \times 10^{-4}$ ($\kappa/W_2\approx 0.05$). The results are shown for the effective QME (blue triangles), the full QME (black circles), and the NGFs method (magenta squares). The vertical lines correspond to $\Gamma_{c}/\Gamma=1$. Parameters are $N=3$, $t_{1}=10^{-4}$, $\Gamma= 10^{-3}$, $t_{2}=10^{-2}$. For the full QME method, the maximum number of photons in the Hilbert space is set to $3$.}
\label{plot2d_ed}
\end{figure}

Introducing $J^{(0)}=J^{(0)}_{1}+J^{(0)}_{2}$ the overall steady-state current in the absence of light-matter coupling ($g=0$) [see Eq.~(\ref{jj_1_causal})], we now study numerically the relative current enhancement $\Delta J=(J/J^{(0)})-1$ as a function of the coupling parameters $g$ and $\kappa$. This is shown on Fig.~\ref{plot2d_ed}, for an example in the regime $t_{2}\gg \Gamma \gg t_{1}$ with $N=3$, $t_{1}=10^{-4}$, $\Gamma= 10^{-3}$, and $t_{2}=10^{-2}$ ($W_2\approx 0.03$). 

Panels \textbf{a)}, \textbf{b)}, and \textbf{c)} correspond respectively to the results obtained from the effective QME (\ref{full_eff_m}), the full QME (\ref{full_mast}), and the NGFs method. \textit{We observe an enhancement of the steady-state current with respect to the non-interacting case $g=0$, as the coupling strength is increased for a given decay rate $\kappa$}. Furthermore, this enhancement is substantially larger in the high-finesse cavity regime with small $\kappa$. Note that the full QME result is exact (assuming the Markovian approximation for the system-lead coupling) as long as counter-rotating terms can be neglected in the coupling Hamiltonian Eq.~(\ref{H_intera2}), which is here assumed in all cases. 

As already discussed in Sec.~\ref{sec:eff_master}, the effective QME result only depends on the parameter $\Gamma_{c}=g^{2}/\kappa$, which explains that the lines with constant current enhancement on panel \textbf{a)} scale linearly with $\log g$ and $\log \kappa$ over the whole range of parameters. Nevertheless, we point out that this result is only valid in the dissipative regime where $\kappa/W_2 \gg 1$. This is shown on panels \textbf{b)} and \textbf{c)}, where the full QME and NGFs results feature the same scaling law as the effective QME result for $\kappa/W_2 \gg 1$. However, a different scaling law is observed for $\kappa/W_2 \ll 1$, indicating the emergence of a new regime with different physical properties than the ones discussed in Secs.~\ref{sec:eff_master} and \ref{spec_lan_jd}. We will show later on that this regime can be characterized by a coherent dynamics stemming from the hybridization of only one Bloch state in the upper band with the states of the lower band.  

\vspace{2mm}

The relative current enhancement $\Delta J$ is represented on Figs.~\ref{plot2d_ed} \textbf{d)} and \textbf{e)} as a function of the cooperativity $\Gamma_{c}/\Gamma$, for two different values of $\kappa$ (horizontal dashed lines in the upper panels). The results obtained with the effective QME, the full QME, and the NGFs method are represented as light-blue triangles, black circles, and magenta squares, respectively. In the dissipative regime [Fig.~\ref{plot2d_ed} \textbf{d)}], all methods coincide for $\Gamma_{c}/\Gamma \ll 1$ (perturbative regime). As $\Gamma_{c}/\Gamma$ becomes larger than $1$, discrepancies between the NGFs and the full QME results increase, while the effective QME and full QME results are still in a surprisingly good agreement, given that the former is expected to be valid only for small coupling strengths. In the ``coherent'' regime with $\kappa/W_2 \ll 1$ [Fig.~\ref{plot2d_ed} \textbf{e)}], while the effective QME fails to reproduce the full QME result even in the perturbative regime, NGFs provide a surprisingly good approximation of the current even far away from the perturbative regime $\Gamma_{c}/\Gamma \gg 1$. However, while the current enhancement obtained from the two master equation methods always increases with $g$, this qualitative trend is not reproduced by the NGFs method when $\Gamma_{c}/\Gamma \gtrsim 100$. 

\begin{figure}[ht]
\centerline{\includegraphics[width=0.4\columnwidth]{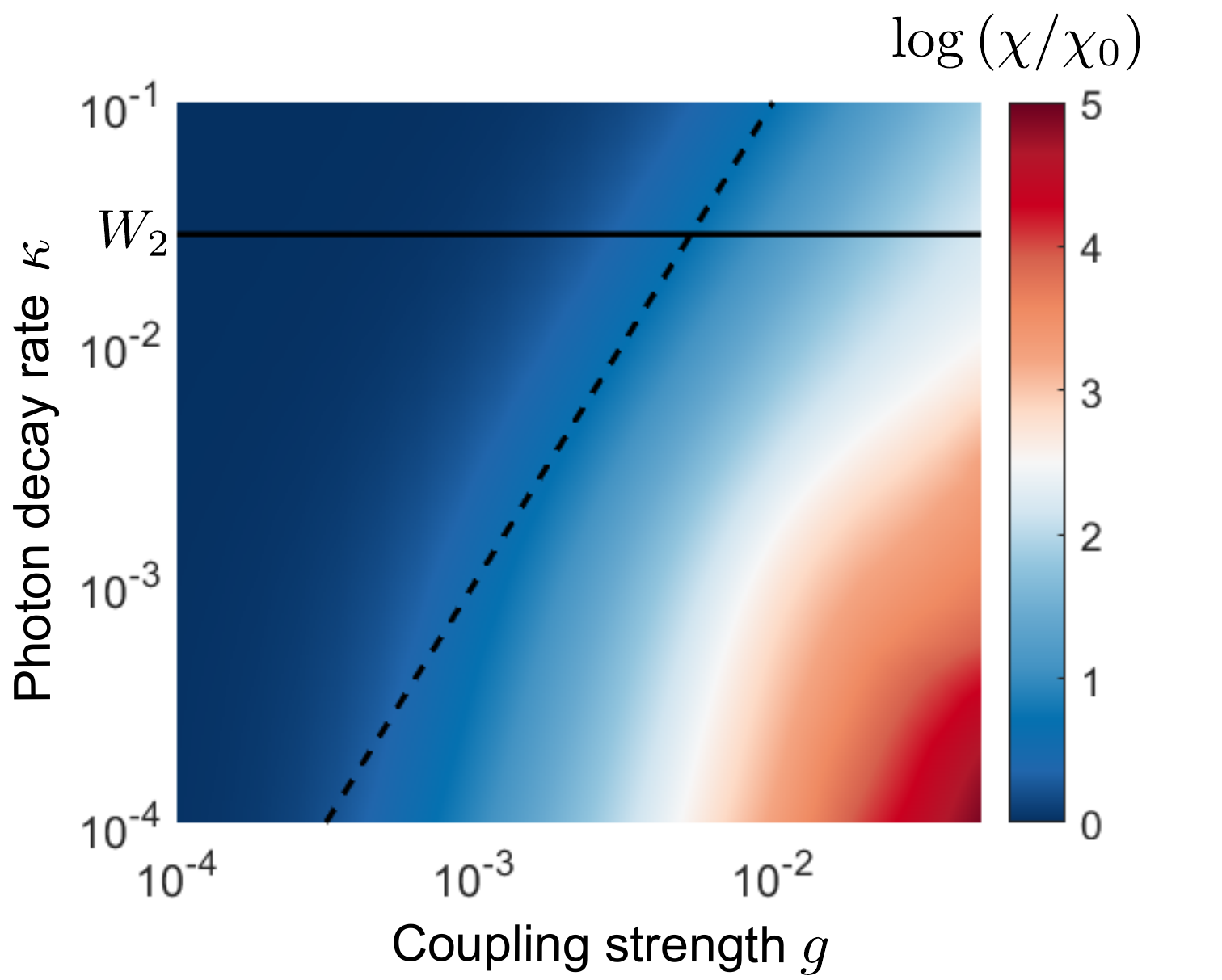}}
\caption{Relative broadening of the electron DOS $\log (\chi/\chi_{0})$ (see text) versus coupling strength $g$ and photon decay rate $\kappa$ (log-scale), obtained from the NGFs method (self-consistent calculation). The dashed line represents the equation $\Gamma_{c}/\Gamma=1$. Parameters are identical to that of Fig.~\ref{plot2d_ed}.}
\label{plotbroad_ed}
\end{figure}

Furthermore, it is interesting to compare the current enhancement represented on Fig.~\ref{plot2d_ed} \textbf{c)} with the cavity-induced broadening of the electron DOS [see Secs.~\ref{self_self22} and \ref{spec_lan_jd}] computed self-consistently with the NGFs method. We denote by $\chi \equiv \chi_{1,k_{0}} (\omega_{1,k_{0}})$ the broadening function of the resonant Bloch states with quasi-momentum $k_{0}=(N+1)/2$ (center of the bands), evaluated at the energy $\omega=\omega_{1,k_{0}}$. This quantity represents the linewidth of the electron DOS (with a Lorentzian lineshape) in the lower band for $g\neq 0$. For $g=0$, the linewidth $\chi_{0}$ is only determined by the retarded SE due to the coupling to the leads $\underline{\Sigma}^{r}_{L,1} \propto \Gamma$. The relative broadening $\log (\chi/\chi_{0})$ is shown on Fig.~\ref{plotbroad_ed}, as a function of $g$ and $\kappa$ (log-scale). As for the current enhancement, we observe that the lines with constant relative broadening scale with the cooperativity $\Gamma_{c}/\Gamma$ for $\kappa/W_2 \gg 1$, consistently with the results Eq.~(\ref{broad_causal2}) of the previous section. However, the coherent regime $\kappa/W_2 \ll 1$ features a different scaling law, qualitatively similar to that of the current enhancement.

\subsection{Dissipative regime $\kappa/W_2 \gg 1$}
\label{diss_reg_sec}

In order to further investigate the physics in the dissipative regime $\kappa/W_2 \gg 1$, we now present numerical calculations using both NGFs in the frequency domain, and QME methods in the time domain. In Sec.~\ref{anal_app_part}, we show that an analytical expression of the current valid for small coupling strengths can be derived starting from the effective QME, confirming the scaling of the current enhancement with the cooperativity. In Sec.~\ref{coll_dress_part}, we present numerical results for the transmission spectrum and the cavity photon DOS, evidencing the presence of a collective coupling to light when the coupling strength is larger than the typical separation between two adjacent Bloch states. For large coupling strengths, we show that the current enhancement saturates to about twice its value for $g=0$, and that the system features non-local electron-electron correlations when one goes beyond the perturbative regime $\Gamma_{c}/\Gamma > 1$ [Sec~.\ref{non_local_coco2}].

\subsubsection{Analytical approach with rate equations.}
\label{anal_app_part}

In Sec.~\ref{ss_cucu23_part}, we have seen that the steady-state current is directly related to the populations of the first/last site in both orbitals. Hence, the former can be obtained by computing the expectation values $\langle c^{\dagger}_{\alpha, i} c_{\alpha, j} \rangle$, with $i=j = 1$ or $i=j = N$. In the absence of light-matter interactions ($g=0$), the expectation values $\langle c^{\dagger}_{\alpha, i} c_{\alpha, j} \rangle$ evolve as:

\begin{align}
\partial_t \langle c^{\dagger}_{\alpha, i} c_{\alpha,j} \rangle &= i t_{\alpha} \sum_{\ell=1}^{N-1} \langle c_{\alpha,i}^{\dagger} \big( \delta_{\ell+1,j} c_{\alpha,\ell} + \delta_{\ell,j} c_{\alpha,\ell+1} \big) \rangle  
+ \textrm{h.c.} \nonumber \\
&- \frac{\Gamma_{\alpha}}{2} \Big( \delta_{i1} + \delta_{j1} + \delta_{iN} +  \delta_{jN} \Big) \langle c^{\dagger}_{\alpha,i} c_{\alpha,j} \rangle
+ \Gamma_{\alpha} \delta_{i1} \delta_{j1},
\label{eq:diff_unc}
\end{align}
with h.c.~the hermitian conjugate, and where we have used $\partial_t \langle c^{\dagger}_{\alpha, i} c_{\alpha,j} \rangle={\rm Tr} (c^{\dagger}_{\alpha,i} c_{\alpha,j} \mathcal{L}_{e} \hat{\rho} )$ according to Eq.~(\ref{eq:expec_t}). Equation (\ref{eq:diff_unc}) forms a closed set of linear differential equations. When solving these equations in the case of uncoupled bands, and plugging the solution in Eq.~(\ref{Jinout}), one recovers the overall steady-state current which is the sum of the individual currents [see Eq.~(\ref{jj_1_causal})] flowing through the two bands. 

\vspace{2mm}

We now explain how to modify the set Eq.~(\ref{eq:diff_unc}) in the presence of light-matter interactions in the dissipative regime. In the resonant case $\Delta=0$, we see from Eq.~(\ref{full_eff_m}) that we need to compute the additional contribution:  

\begin{align*}
\partial_t \langle c^{\dagger}_{\alpha, i} c_{\alpha, j} \rangle={\rm Tr} (c^{\dagger}_{\alpha, i} c_{\alpha, j} \mathcal{L}_{\Gamma_c} \hat{\rho}),
\end{align*}
with $\mathcal{L}_{\Gamma_c} \hat{\rho}$ given by Eq.~(\ref{dissi_Gc_m}). We obtain:

\begin{align}
\partial_t \langle c^{\dagger}_{1,i} c_{1,j} \rangle
&= 2\Gamma_{c}  \langle c^{\dagger}_{2,i} c_{1,j} S^- + S^+ c^{\dagger}_{1,i} c_{2,j} \rangle \nonumber \\
\partial_t \langle c^{\dagger}_{2,i} c_{2,j} \rangle &= - 2\Gamma_{c} \langle c^{\dagger}_{2,i} c_{1,j} S^- + S^+ c^{\dagger}_{1,i} c_{2,j} \rangle.
\label{eq:cij}
\end{align}

The differential equations (\ref{eq:cij}) now contain four-operator products, in contrast to the non-interacting case with only quadratic operators. Full computation of the expectation values thus involves higher-order correlation functions, and the NGFs method can be efficiently used in this case (see Sec.~\ref{self_self22}). In order to get a first estimate of the current enhancement, starting from Eq.~\eqref{eq:cij} for the time evolution of $\langle c^{\dagger}_{\alpha,i} c_{\alpha,j} \rangle$, we only focus on the populations ($i=j$) and discard non-local contributions arising from the second term in the right-hand-side of Eq.~(\ref{dissi_Gspil}). We have verified numerically that the effective QME with and without non-local coupling terms in the dissipator gives comparable results for small $g$, as we will discuss in more detail in Sec.~\ref{non_local_coco2}. If, in addition, we factorize the expectation value of four-operator products as:

\begin{align}
\langle \hat{n}_{1 i} \hat{n}_{2 i} \rangle \simeq \langle \hat{n}_{1 i} \rangle \langle \hat{n}_{2 i} \rangle,
\label{factorrrr}
\end{align} 
Eq.~(\ref{eq:cij}) provides: 

\begin{align}
\partial_{\tau} \langle \hat{n}_{1 i} \rangle &= 4 \Gamma_c \langle \hat{n}_{2 i} \rangle \big( 1 - \langle \hat{n}_{1 i} \rangle \big) \nonumber \\
\partial_{\tau} \langle \hat{n}_{2 i} \rangle &= -4 \Gamma_c \langle \hat{n}_{2 i} \rangle \big( 1 - \langle \hat{n}_{1 i} \rangle \big). 
\label{eq:cij_approx}
\end{align}

At this level of approximation, the cavity mode induces a local population transfer from the upper to the lower orbitals at each site. Solving the differential equations (\ref{eq:diff_unc}) together with the contribution Eq.~(\ref{eq:cij_approx}) stemming from the light-matter coupling, one can check numerically that the steady-state current is nearly independent of the chain length $N$ as long as $t_{1} \ll \Gamma$. We therefore restrict the calculation to the case $N=2$. For the sake of simplicity, we limit the derivation to the case $t_1 = 0$ and $\Gamma_1 = \Gamma_2 \equiv \Gamma$. The time-evolution of the mean population $n_{11}$ in the lower band at the first site is obtained as:

\begin{align*}
\partial_\tau n_{11} = - \Gamma n_{11} + \Gamma + 4 \Gamma_c n_{21} \left( 1 - n_{11} \right).
\end{align*}  

In the steady-state, $\partial_{\tau} n_{11}=0$, which provides the solution $n_{11}=1$. Furthermore, as a solution of the equations $\partial_\tau \langle c^{\dagger}_{11} c_{12} \rangle = - \Gamma \langle c^{\dagger}_{11} c_{12} \rangle$ and $\partial_\tau \langle c^{\dagger}_{12} c_{11} \rangle = - \Gamma \langle c^{\dagger}_{12} c_{11} \rangle$, the lower orbital coherence $\langle c^{\dagger}_{12} c_{11}\rangle =\langle c^{\dagger}_{11} c_{12}\rangle$ vanishes in the steady-state. The remaining set of equations can be rewritten as:
 
\begin{align}
\partial_\tau n_{21} &= - \Gamma n_{21} - 2 t_{2} \mathcal{C}  + \Gamma \nonumber \\
\partial_\tau n_{22} &= - \Gamma n_{22} + 2 t_{2} \mathcal{C} - 4 \Gamma_c n_{22} \left( 1 - n_{12} \right) \nonumber \\
\partial_\tau n_{12} &= - \Gamma n_{12} + 4 \Gamma_c n_{22} \left( 1 - n_{12}  \right) 
\nonumber \\
\partial_\tau \mathcal{C} &= - \Gamma \mathcal{C} + t_{2} \left( n_{21} - n_{22} \right), 
\label{eq:diffc} 
\end{align}
where we have introduced the imaginary part of the upper orbital coherence $\mathcal{C}=\Im \langle c^{\dagger}_{21} c_{22} \rangle$. The latter is related to the local current in the upper band between the first and the second site. Setting the left-hand side of Eq.~(\ref{eq:diffc}) to zero, one can compute the overall steady-state current $J=e\Gamma (n_{21}+n_{22})$ as:

\begin{align}
J=\frac{e\Gamma t^{2}_{2}/2}{t^{2}_{2} (1+ \phi)/2 + \Gamma^{2}/4}, 
\label{eq:pop_22}
\end{align}
with 

\begin{align*}
\phi=\frac{4\Gamma_{c} n_{22} + \Gamma}{4 \Gamma_c (n_{22}+1)+ \Gamma}.
\end{align*}

Since the population $n_{22} >0$, the function $\phi$ is positive and has an upper bound $1$. This value is reached, for instance, in the absence of light-matter coupling $\Gamma_{c}=0$. In this case, one can verify that the current coincides with Eq.~(\ref{jj_1_causal}). \textit{Further inspection shows that whenever $\Gamma_c \neq 0$, $\phi < 1$, resulting in an enhancement of the steady-state current.} Nevertheless, we expect the result Eq.~(\ref{eq:pop_22}) to be a reasonable approximation only for small $\Gamma_c$ (small coupling strength), as pointed out before. It is therefore convenient to expand Eq.~(\ref{eq:pop_22}) to the lowest non-vanishing order in $\Gamma_c$, which provides:

\begin{align*}
J=J^{(0)} \left(1+\Delta J \right) + \mathcal{O}(\Gamma^{2}_{c}), 
\end{align*}
where $J^{(0)}$ is the overall steady-state current for $g=0$, and the relative current enhancement introduced in Sec.~\ref{sec:comparison}:

\begin{align}
\Delta J=2\frac{t^{2}_{2}}{t^{2}_{2}+ \Gamma^{2}/4} \left(\frac{\Gamma_{c}}{\Gamma} \right).
\label{eq:pop_258}
\end{align}

In this regime, the current enhancement is induced by a population transfer from the upper to the lower band. Indeed, the upper band population at the last site

\begin{align*}
n_{22}=n^{(0)}_{22} \left(1- 2\frac{t^{2}_{2}+\Gamma^{2}/2}{t^{2}_{2}+ \Gamma^{2}/4} \left(\frac{\Gamma_{c}}{\Gamma} \right) \right) + \mathcal{O}(\Gamma^{2}_{c})
\end{align*}
is a decreasing function of $\Gamma_c$. Here, $n^{(0)}_{22} = \frac{t^{2}_{2}/2}{t^{2}_{2}+\Gamma^{2}/4}$ denotes the population in the upper band at the last site for $g=0$. As $\Gamma_c$ increases, the population in the lower band at the last site increases as:

\begin{align*}
n_{12}= 2\frac{t^{2}_{2}}{t^{2}_{2}+\Gamma^{2}/4}  \frac{\Gamma_{c}}{\Gamma},
\end{align*}
and vanishes for $g=0$ (as long as $t_1=0$). Importantly, the overall population at the last site increases with $\Gamma_{c}$, which explains the observed current enhancement~\cite{hagenmuller}. In the previous derivation, we only considered the local terms in the dissipator Eq.~(\ref{dissi_Gc_m}), which is valid for small coupling strengths, and further discarded the contributions of these terms to the time-evolution of the intraband coherence $\partial \mathcal{C}/\partial t$. Taking them explicitely into account, the last equation of motion in Eq.~(\ref{eq:diffc}) is modified as:     

\begin{align}
\partial_\tau \mathcal{C} = - \Gamma \mathcal{C} + t_{2} \left( n_{21} - n_{22} \right) - 2\Gamma_{c} \left(2-n_{11}-n_{12}\right)\mathcal{C}, 
\label{eq:diffc_final} 
\end{align}
where we have used a factorization procedure similar to Eq.~(\ref{factorrrr}) for the four-operator products entering the last term in the right-hand-side of Eq.~(\ref{eq:diffc_final}). This term describes an additional damping of the intraband coherence due to the light-matter coupling. Since the intraband coherence is proportional to the local current, we therefore expect this correction to lead to a smaller current enhancement. Moreover, we have checked numerically that it can even lead to a reduction of the overall current when $t_2 \ll \Gamma$. In this article, we only focus on the regime $t_2 \gg \Gamma$, where one can show that the effect of the additional term in Eq.~(\ref{eq:diffc_final}) becomes negligible for the relative current enhancement. In this case, Eq.~(\ref{eq:pop_258}) simply reduces to 

\begin{align*}
\Delta J = 2\Gamma_c /\Gamma.
\end{align*}

Here, we clearly confirm the relevance of the cooperativity $\Gamma_{c}/\Gamma$, as found in Secs.~\ref{spec_lan_jd} and \ref{sec:comparison}. Nevertheless, \textit{we point out that non-local contributions entering the dissipator Eq.~(\ref{dissi_Gspil}) have not been taken into account in this derivation. Therefore, this analytical estimation is unable to describe any collective effects arising from these non-local terms, namely long-range electronic correlations due to the collective coupling to the cavity mode~\cite{Meiser2010,Jaeger2017}}. As we will see in the next section, these terms play a crucial role beyond the perturbative regime $\Gamma_c/\Gamma >1$.

\newpage

\subsubsection{Transmission spectrum and cavity DOS.}
\label{coll_dress_part}

\begin{figure}[ht]
\centerline{\includegraphics[width=0.7\columnwidth]{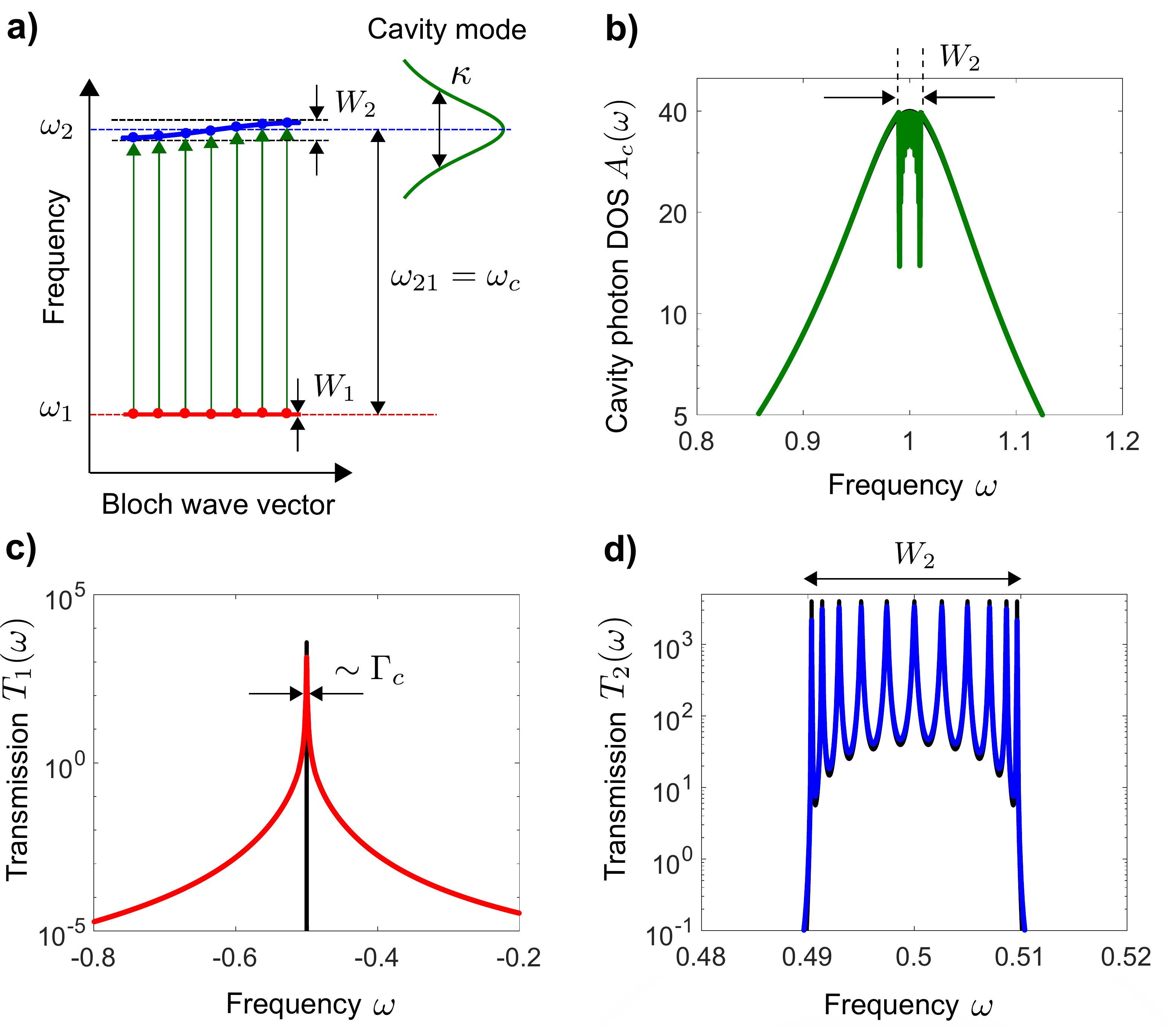}}
\caption{\textbf{a)} Sketch of the energy bands in the dissipative regime with $t_{2}=5 \times 10^{-3}$, and $\kappa=0.1$. \textbf{b)} (Log-scale) Cavity photon DOS $A_{c}(\omega)$. \textbf{c)} (Log-scale) Transmission spectrum $T_{1}(\omega)$ in the vicinity of the lower orbital energy $\omega_{1}=-0.5$. \textbf{d)} (Log-scale) Transmission spectrum $T_{2}(\omega)$ in the vicinity of the upper orbital energy $\omega_{2}=0.5$. The black lines correspond to $g=0$, while the red, blue, and green lines correspond to $g=2.2 \times 10^{-3}$. The chain length is $N=11$, and the other parameters are $t_{1}=5 \times 10^{-5}$ and $\Gamma=5 \times 10^{-4}$.}
\label{fig9}
\end{figure}

In the dissipative regime $\kappa/W_2 \gg 1$, all Bloch states are comprised within the cavity linewidth, which allows a collective coupling to arise when the coupling strength is larger than the typical separation between two adjacent Bloch states, namely $g > \delta \omega = \omega_{2,k+ 1}-\omega_{2,k}$\footnote{This feature can be qualitatively understood from the spectrum of an effective, bosonic TC hamiltonian (obtained from the TC model, by considering the leading-order of the Holstein-Primakoff~\cite{holstein} expansion of spin operators in terms of bosons) $H_{\rm TC}=\omega_{c}a^{\dagger} a+ \sum_{k=1}^{N} (\omega_{2,k}-\omega_{1}) b^{\dagger}_{k} b_{k} + g \sum_{k=1}^{N} (b^{\dagger}_{k} a+ b_{k} a^{\dagger})$, where $b_{k},b^{\dagger}_{k}$ are bosonic operators associated with the $N$ interband transitions.}. This regime is refered to as ``collective dressing regime''~\cite{hagenmuller}. Frequency domain calculations using the NGFs method are shown on Fig.~\ref{fig9}, for an example with $N=11$, $t_{1}=5 \times 10^{-5}$, $\Gamma=5 \times 10^{-4}$, $t_{2}=5 \times 10^{-3}$, $\kappa=0.1$ ($\kappa/W_2 \approx 20$), and $g=2.2 \times 10^{-3}$ ($\Gamma_{c}/\Gamma \approx 0.1$). 

In this regime with large photon damping, the interband transitions with frequencies $\omega_{2,k}-\omega_{1}$ between the states of the quasi-flat lower band and the upper band Bloch states [Fig.~\ref{fig9} \textbf{a)}] are all quasi-resonant to the broad bare cavity mode of width $\kappa$ [thin black line on Fig.~\ref{fig9} \textbf{b)}, hardly visible], resulting in a collective coupling of the Bloch states to the cavity mode when $g > \delta \omega$ (in this case $\delta \omega \lesssim 2.5 \times 10^{-3}$). The photon DOS [Eq.~(\ref{CAS_ff})] is shown as a thick green line on Fig.~\ref{fig9} \textbf{b)} for $g=2.2 \times 10^{-3}$. The central region of width $W_2 \approx 4t_{2}$ and centered at $\omega_{21}$ consists of $N-1$ peaks originating from the individual dressing of interband transitions. As explained in more details in the following, the finite photon spectral weight present in this region is crucial to the existence of the current enhancement, as it connects the two bands through aborption and emission of cavity photons. 

\vspace{2mm}

Figure~\ref{fig9} \textbf{c)} and \textbf{d)} display the transmission spectrum given by Eq.~(\ref{trans_long_coco2}) for $g=0$ (black line), $g=2.2 \times 10^{-3}$ (colored lines), in the vicinity of the lower and upper orbital energies $\omega_{1}$ and $\omega_{2}$, respectively. The key feature is that the narrow transmission associated with the partial current $J_{1} \sim 2 e t^{2}_{1}/\Gamma$ flowing through the lower band for $g=0$ is broadened by a quantity $\sim \Gamma_{c}$, giving rise to the current enhancement ($\Delta J \approx 0.1$). On the other hand, the transmission in the vicinity of $\omega_2$ is only slightly reduced with respect to the case $g=0$. 

\vspace{2mm}

\begin{figure}[ht]
\centerline{\includegraphics[width=1\columnwidth]{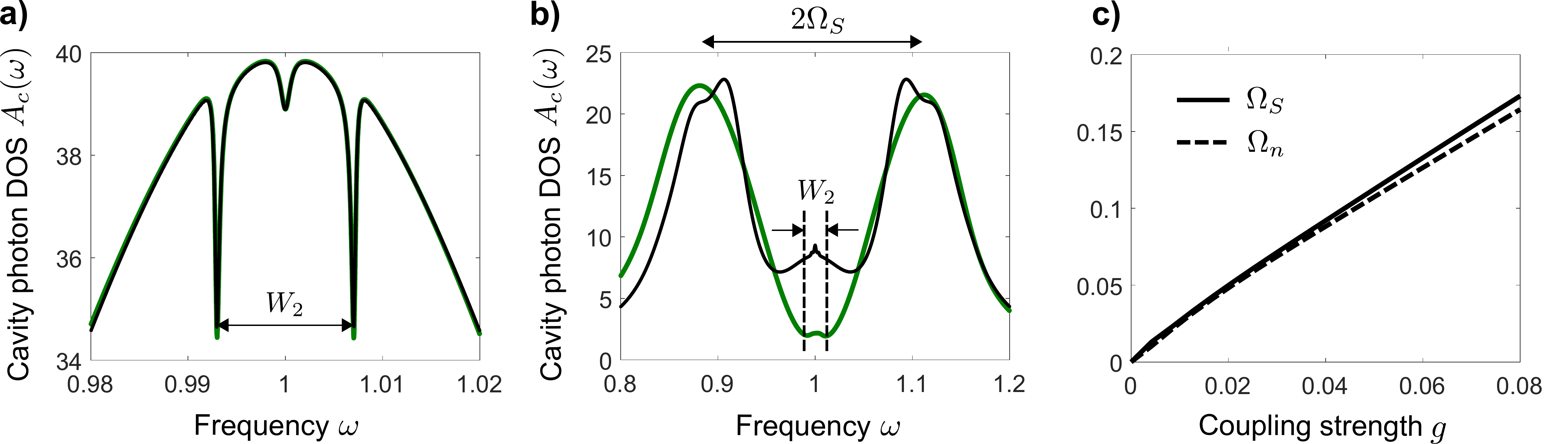}}
\caption{\textbf{a)}-\textbf{b)} (Log-scale) Cavity photon DOS $A_{c}(\omega)$ for $N=3$, computed with the full QME method using the quantum regression theorem (thin black line) and the NGFs method (thick green line). \textbf{a)} Small coupling strength $g=2.2 \times 10^{-3}$ ($\Gamma_{c}/\Gamma\approx 0.1$). \textbf{b)} Large coupling strength $g=8.5 \times 10^{-2}$ ($\Gamma_{c}/\Gamma\approx 145$). \textbf{c)} Polariton half-splitting $\Omega_S$ (solid line) and vacuum Rabi frequency $\Omega_n$ (dashed line) as a function of $g$ for $N=11$. Other parameters are the same as in Fig.~\ref{fig9}, and the maximum number of photons in the calculation using the full QME is set to $2$.}
\label{fig6}
\end{figure}

The cavity photon DOS calculated with the full QME and the NGFs methods is represented on Fig.~\ref{fig6} \textbf{a)}-\textbf{b)}, in the perturbative regime ($\Gamma_{c}/\Gamma =0.1$) and at large coupling strength ($\Gamma_{c}/\Gamma=145$), respectively. The former case [Fig.~\ref{fig6} \textbf{a)}] corresponds to Fig.~\ref{fig9} \textbf{b)} for $N=3$ instead of $N=11$, but with the same other parameters. Here, the two methods are in good agreement, showing the validity of the NGFs method in the perturbative regime $\Gamma_{c}/\Gamma\ll 1$. On the other hand, when $g > \delta \omega$ ($\delta \omega \lesssim 0.03$ in this case), the collective coupling gives rise to two polariton peaks separated by a splitting which we define as $\Omega_{S} > g$ [Fig.~\ref{fig6} \textbf{b)}]. 

Importantly, the result obtained with the NGFs method features a small photon spectral weight in the central region of width $W_2$, and inaccurately predicts that this photon spectral weight vanishes in the limit of large coupling strengths. This explains why the current enhancement computed with NGFs decreases in this region, as observed in ~\cite{hagenmuller}. Indeed, when no photon weight is present in the range $[\omega_{21}-2 t_{2},\omega_{21}+2t_{2}]$, photon absorption or emission can not take place between the two bands, resulting in a vanishing current enhancement. In contrast, as predicted by the full QME, the photonic weight in the central region saturates to a finite value as $g$ is increased while the two polariton peaks further split. This is consistent with the observation of Sec.~\ref{sec:comparison} that the (exact) current enhancement computed with the full QME always increases with $g$. The relative current enhancements obtained in the case of small and large coupling strengths correspond respectively to: $\Delta J\approx 0.13$ for both methods [Fig.~\ref{fig6} \textbf{a)}], and $\Delta J\approx 0.56$ for NGFs and $\Delta J\approx 0.7$ for the full QME [Fig.~\ref{fig6} \textbf{b)}]. Note that the asymmetry of the polariton peaks computed with NGFs is inherited from the asymmetry of the first-order photon GF Eq.~(\ref{dd_ret_fun12}), only valid in the perturbative regime, namely when the SE due to the coupling to extra-cavity photons is much smaller than the bare cavity photon energy $\kappa \ll \omega_{c}$. 

On Fig.~\ref{fig6} \textbf{c)}, we compare the polariton half-splitting $\Omega_S$ with the vacuum Rabi frequency defined as $\Omega_{n}=g\sqrt{N_1 - N_2}$ (see Sec.~\ref{spec_lan_jd}), where $N_\alpha = \sum_{j} n_{\alpha,j}$ is obtained from the steady-state population imbalance between the two bands for $g\neq 0$~\cite{todorov}. As already mentioned in ~\cite{hagenmuller}, we observe that these two quantities coincide in the dissipative regime, thereby connecting to the physics of the TC model~\cite{tavis} where the relevant coupling strength is not $g$ but the collective coupling constant $\Omega_{n}$. Importantly, since sites with both orbitals occupied (or empty) are not effectively coupled to light, we always find $\Omega_n < g\sqrt{N}$, in contrast to the TC model [see Sec.~\ref{phi_s}].

\vspace{2mm}

As a side comment, we have already mentioned that the full QME method predicts that the total current admits an upper limit $< e \Gamma$ as $g \to \infty$. We find numerically that for $t_2 \gg \Gamma \gg t_1$, and for the two values $\kappa=8 \times 10^{-4}$ (coherent regime) and $\kappa=0.07$ (dissipative regime) [see Fig.~\ref{plot2d_ed}], this upper bound becomes closer to $e \Gamma$ (twice the current for $g=0$) when counter-rotating terms are included in the coupling Hamiltonian Eq.~(\ref{ha_full_path}). This shows that higher-order correlations such as the one depicted on Fig.~\ref{fig_self} \textbf{b)} are important to determine the full current enhancement in the limit of large coupling strengths. These effects will be further investigated in a future work. 

\subsubsection{Non-local correlations.}
\label{non_local_coco2}

\begin{figure}[ht]
\centerline{\includegraphics[width=0.7\columnwidth]{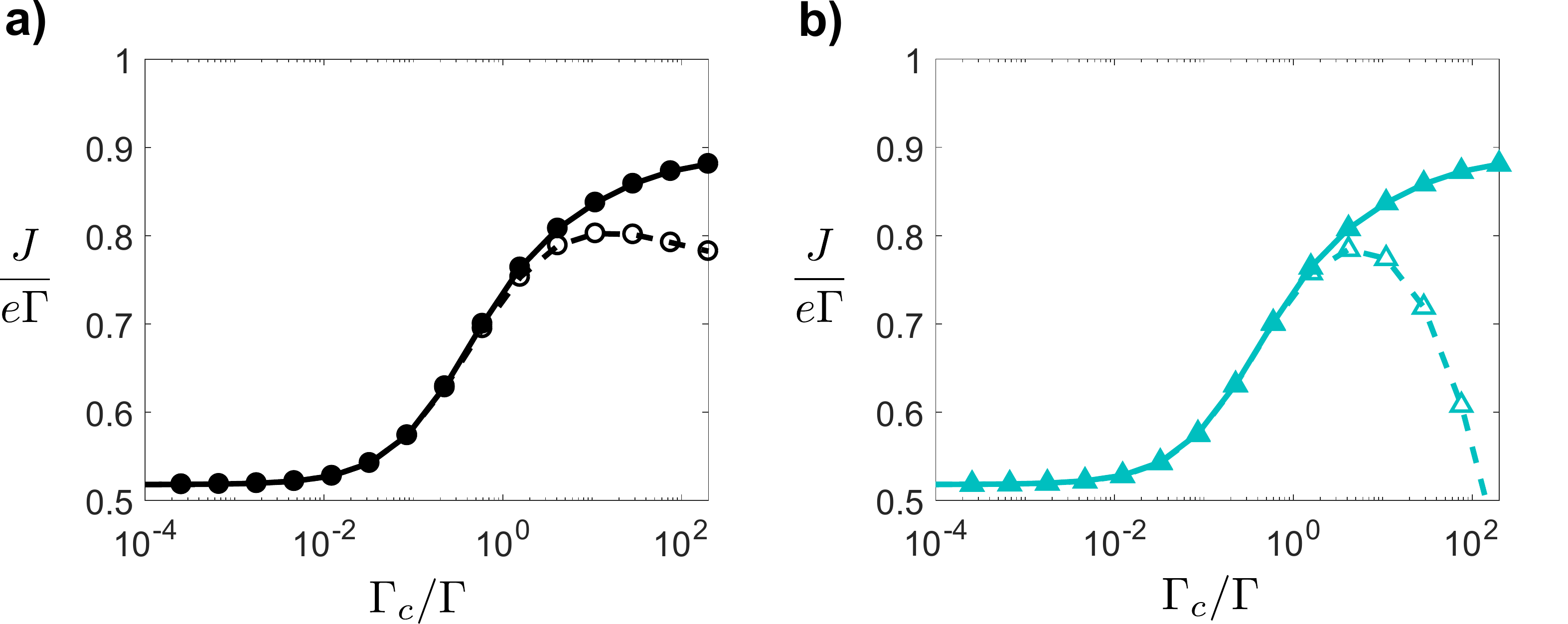}}
\caption{Steady-state current $J/e\Gamma$ as a function of the cooperativity $\Gamma_{c}/\Gamma$ (log-scale) for $N=3$. \textbf{a)} Calculation using the full QME with fermions (filled circles) and hard-core bosons (empty circles). \textbf{b)} Same quantity computed with the effective QME, using the full dissipator Eq.~(\ref{dissi_Gspil}) (filled triangles), or only the local terms (empty triangles). Other parameters are identical to that of Fig.~\ref{fig9}, and the maximum number of photons in the calculation using the full QME is set to $2$.} 
\label{fig7}
\end{figure}

As already mentioned in Sec.~\ref{sec:eff_master}, an interesting point is the existence of non-local electron-electron correlations for large coupling strengths. This can be seen on Fig.~\ref{fig7}, where we have represented the steady-state current computed with different approximations as a function of the cooperativity. Fig.~\ref{fig7} \textbf{a)} displays a comparison between the full QME results with fermions (filled circles) and hard-core bosons (empty circles). The discrepancy observed for $\Gamma_{c}/\Gamma > 1$ points to the existence of fermionic correlations that can not be reproduced with hard-core bosons. On Fig.~\ref{fig7} \textbf{b)}, we have represented the results obtained with the effective QME, using either the full dissipator of Eq.~(\ref{dissi_Gspil}) (filled triangles), or only the local terms in the right-hand side of the same equation (empty triangles). We remark that these local terms would correspond to a situation in which each site is individually coupled to its own lossy cavity (with decay rate $\kappa$ and coupling strength $g$).    

Moreover, we observe on Fig.~\ref{fig7} \textbf{b)} that non-local terms play an important role as one moves away from the perturbative regime. This points to the existence of non-local electronic correlations for $\Gamma_{c}/\Gamma > 1$, that are obtained only when considering that all sites are coupled to the same cavity mode. These correlations can be understood by transforming the light-matter coupling Hamiltonian Eq.~(\ref{H_intera2}) in terms of a two-body retarded interaction between electrons, as in the case of electron-phonon interactions for BCS superconductivity~\cite{bruus}. This is done by rearranging the real-time electron GF given by Eq.~(\ref{GF_def_ufc}), after having replaced $H$ by the light-matter coupling Hamiltonian Eq.~(\ref{H_intera2}) without counter-rotating terms. The latter is then rewritten exactly as:

\begin{align*}
H_{I} (\tau) = \frac{1}{2}\sum_{i \neq j} g^{2} \int\!\! d\tau' D^{r}(\tau-\tau') c^{\dagger}_{2,i} (\tau') c^{\dagger}_{1,j} (\tau) c_{2,j} (\tau) c_{1,i} (\tau') + {\rm h.c.},
\end{align*}
where $D^{r}(\tau-\tau')$ is the real-time retarded photon GF entering Eq.~(\ref{DD_less}). In this form, the light-matter coupling can be interpreted as a retarded dipole-dipole interaction, where an interband excitation is created on site $i$ and destroyed at later time on site $j$. This plays the role of a retarded, long-wavelength interaction mediated by the cavity mode, which induces long-range correlations $\propto g^{4}$. Note that the overall correlations can be described by including the static Coulomb repulsion, which will not be addressed in this paper.      
 
\subsection{Coherent regime $\kappa/W_2 \ll 1$}
\label{ind_dress_part}

We now focus on the ``coherent'' regime where $\kappa /W_2 \ll 1$, and in particular on the ``individual dressing regime'' occuring when the coupling strength $g$ is smaller than the typical separation between two adjacent Bloch states in the upper band~\cite{hagenmuller}. After having characterized the transmission spectrum and the cavity photon DOS using the NGFs method in Sec.~\ref{ttdress_part43}, we compare the electron density profiles along the chain obtained in the dissipative and in the coherent regime. Furthermore, by computing the time evolution of the electron spectral function, we show that in contrast to the dissipative regime where the electronic excitations stay essentially localized, a small transfer of electron spectral weight occurs between the two bands in the individual dressing regime, resulting in the emergence of a delocalized state in the lower band.   

\subsubsection{Transmission spectrum and cavity DOS.}
\label{ttdress_part43}

\begin{figure}[ht]
\centerline{\includegraphics[width=0.7\columnwidth]{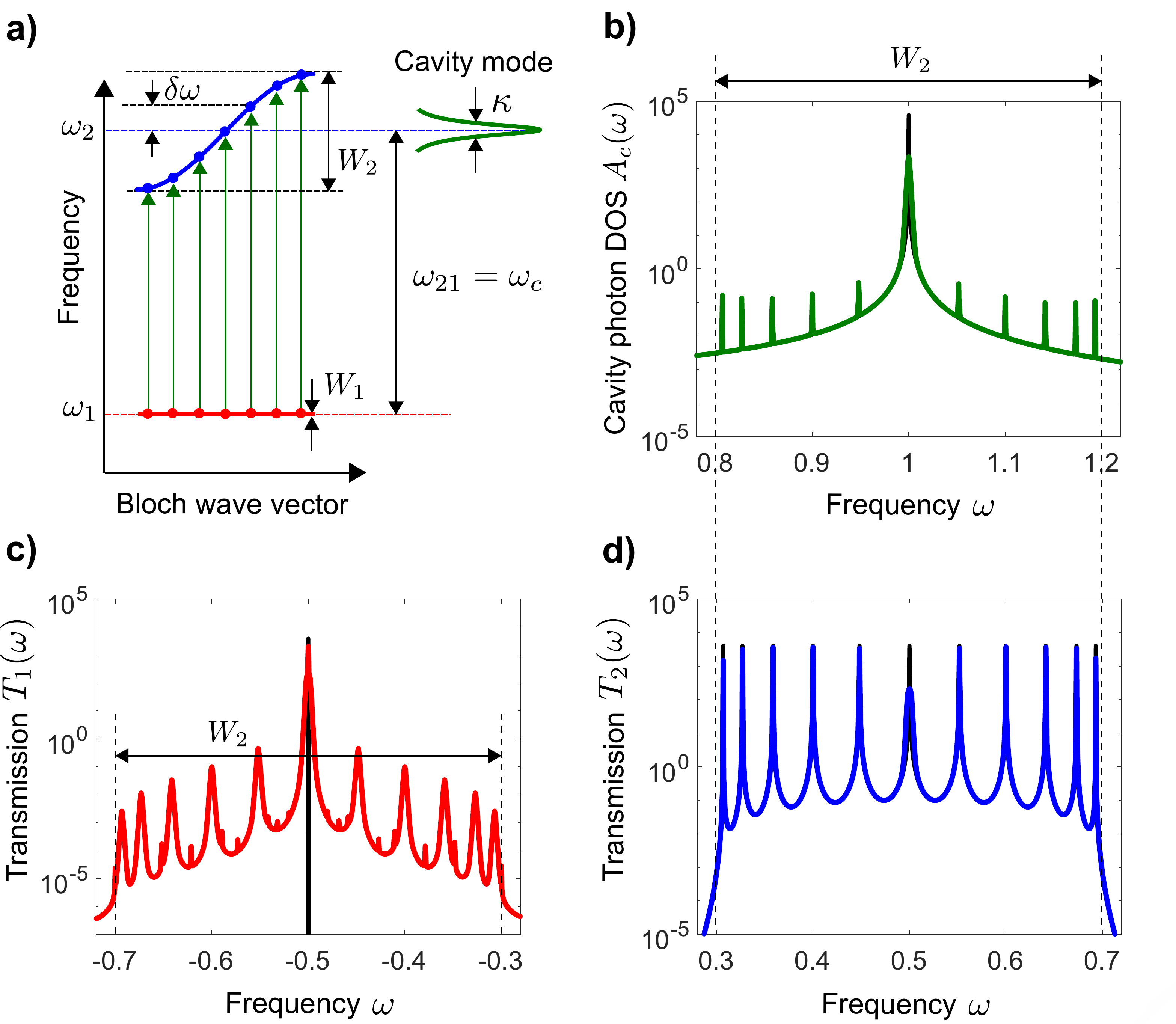}}
\caption{\textbf{a)} Sketch of the energy bands in the individual dressing regime with $t_{2}=0.1$ and $\kappa=10^{-4}$. $\delta \omega$ denotes the typical energy spacing between adjacent Bloch states in the upper band. \textbf{b)} (Log-scale) Cavity photon DOS $A_{c}(\omega)$. \textbf{c)} (Log-scale) Transmission spectrum $T_{1}(\omega)$ in the vicinity of the lower orbital energy $\omega_{1}=-0.5$. \textbf{d)} (Log-scale) Transmission spectrum $T_{2}(\omega)$ in the vicinity of the upper orbital energy $\omega_{2}=0.5$. The black lines correspond to $g=0$, while the red, blue, and green lines correspond to $g=2.2 \times 10^{-3}$. The other parameters are identical to that of Fig.~\ref{fig9} ($N=11$, $t_{1}=5\times 10^{-5}$, and $\Gamma= 5\times 10^{-4}$).}
\label{fig10}
\end{figure}

Frequency domain calculations using the NGFs method are presented on Fig.~\ref{fig10}, for an example with $N=11$, $t_{1}=5 \times 10^{-5}$, $\Gamma=5 \times 10^{-4}$, $t_{2}=0.1$, $\kappa=10^{-4}$ ($\kappa/W_2 \approx 2.5\times 10^{-4}$), and $g=2.2 \times 10^{-3}$ ($\Gamma_{c}/\Gamma \approx 100$). In this regime, $\delta \omega \lesssim 0.05$, and each transition between the states of the lower band (not resolved) and the different Bloch states of the upper band can therefore be addressed individually by the narrow cavity mode. \textit{As already mentioned, we focus on the situation where the cavity mode is resonant with the transition between the flat lower band and the Bloch state lying in the center of the upper band ($k_{0}=(N+1)/2$ for $N$ odd).} The latter corresponds to a spatial half-period of two sites with the maximum Bloch velocity $2 t_{2}$ [Fig.~\ref{fig10} \textbf{a)}].    

\vspace{2mm}

The cavity photon DOS given by Eq.~(\ref{CAS_ff}) is represented on Fig.~\ref{fig10} \textbf{b)} in this regime. The bare cavity mode centered at $\omega_{21}$ (black line) is dressed by the resonant interband transition resulting in a broadened cavity resonance, as well as small satellite peaks originating from the dressing of the detuned interband transitions (green line). The transmission spectrum $T_{\alpha}(\omega)$ is shown on Fig.~\ref{fig10} \textbf{c)} and \textbf{d)}, for $g=0$ (thin black line), $g=2.2 \times 10^{-3}$ (colored lines), in the vicinity of $\omega_{1}$ and $\omega_{2}$, respectively. Similarly as in the dissipative regime, we observe a broad peak centered at $\omega_1$ [Fig.~\ref{fig10} \textbf{c)}], responsible for the current enhancement ($\Delta J \approx 0.16$). In the vicinity of $\omega_{2}$, the peak corresponding to the resonant Bloch state in the upper band is reduced compared to the case $g=0$ [hardly visible on Fig.~\ref{fig10} \textbf{d)}]. In this regime, the light-matter coupling is clearly dominated by this resonant Bloch state (note the log-scale). The other small peaks are reminiscent of the off-resonant Bloch states that are only weakly coupled to the cavity field. 

\subsubsection{Coherent dynamics and spectral weight transfer.}
\label{ttdress_part666}

\begin{figure}[ht]
\centerline{\includegraphics[width=0.7\columnwidth]{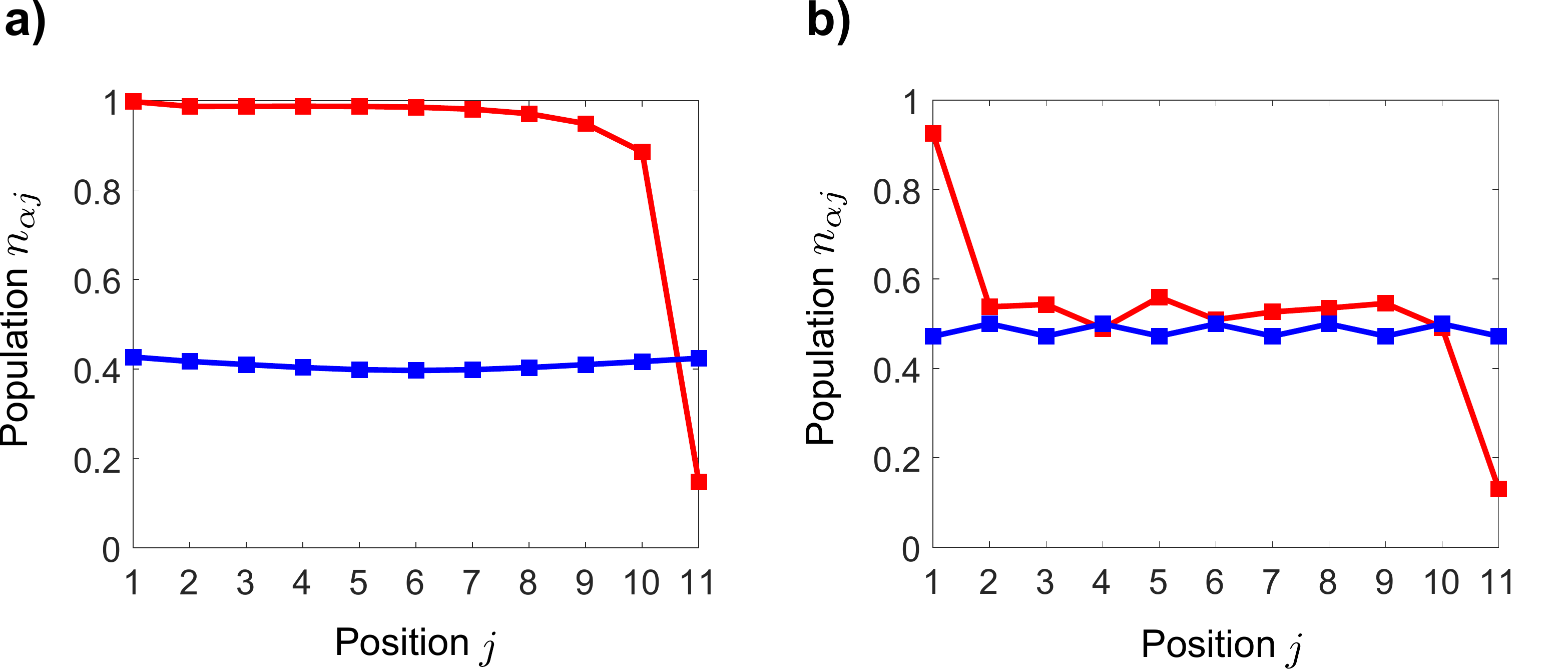}}
\caption{\textbf{a)}-\textbf{b)} Spatial profile of the electron population $n_{\alpha j}$ for $g=2.2 \times 10^{-3}$ and $N=11$. Populations in the lower ($\alpha=1$) and upper ($\alpha=2$) orbitals are respectively depicted as red and blue squares. \textbf{a)} Dissipative regime with $t_2=5\times 10^{-3}$ and $\kappa=0.1$. \textbf{b)} Individual dressing regime with $t_2=0.1$ and $\kappa=10^{-4}$. Other parameters are identical to that of Fig.~\ref{fig9}.}
\label{fig13}
\end{figure}

The particular band hybridization occuring in the individual dressing regime can be further investigated by computing the electron populations along the chain (NGFs method), and compare it to the population profile in the dissipative regime. This is represented on Fig.~\ref{fig13} for $g=2.2 \times 10^{-3}$ and $N=11$. 

First, in the dissipative regime [Fig.~\ref{fig13} \textbf{a)}], the current enhancement associated with the new transmission channel in the vicinity of $\omega_{1}$ can be interpreted as a transfer of population from the upper to the lower band [see Sec.~\ref{anal_app_part}]. On Fig.~\ref{fig13} \textbf{a)}, we observe that the lower orbital populations strongly increase when $g\neq 0$, while the upper band populations slightly decrease. In this regime, large photonic losses are responsible for a global (collective) transfer of populations down to the lower band. On the other hand, the population $n_{1 N}$ in the lower level of the last site is depopulated due to the coupling to the drain. Importantly, for $g\neq 0$, $n_{\alpha,N} \neq 1-n_{\alpha,1}$ (as it was the case for $g=0$), and the partial currents $J_1$ and $J_2$ resulting from the integration of $T_{1}(\omega)$ and $T_{2}(\omega)$ do not correspond to the currents $e\Gamma n_{1,N}$ and $e\Gamma n_{2,N}$ as one could have naively expected from Eq.~(\ref{current_ss_pop}). This implies that for $g\neq 0$, $J_1$ and $J_2$ can not be interpreted as two independent currents respectively flowing through the lower and the upper orbitals, as a result of band hybridization.

In the individual dressing regime [Fig.~\ref{fig13} \textbf{b)}], however, the density profiles exhibits small oscillations with a period of two sites consistent with the resonant coupling of the central Bloch state ($k_{0}=(N+1)/2$). Furthermore, the density profile in the lower band (red line) is reminiscent of the uncoupled case represented on Fig.~\ref{fig3} \textbf{d)}, but with a larger effective hopping $t_{1}'$ reducing (increasing) the population of the first (last) site. \textit{In this case, the current enhancement can be associated with a coherent hopping dynamics, sustained by the absorption and emission of cavity photons.}

\vspace{2mm}

\begin{figure}[ht]
\centerline{\includegraphics[width=1\columnwidth]{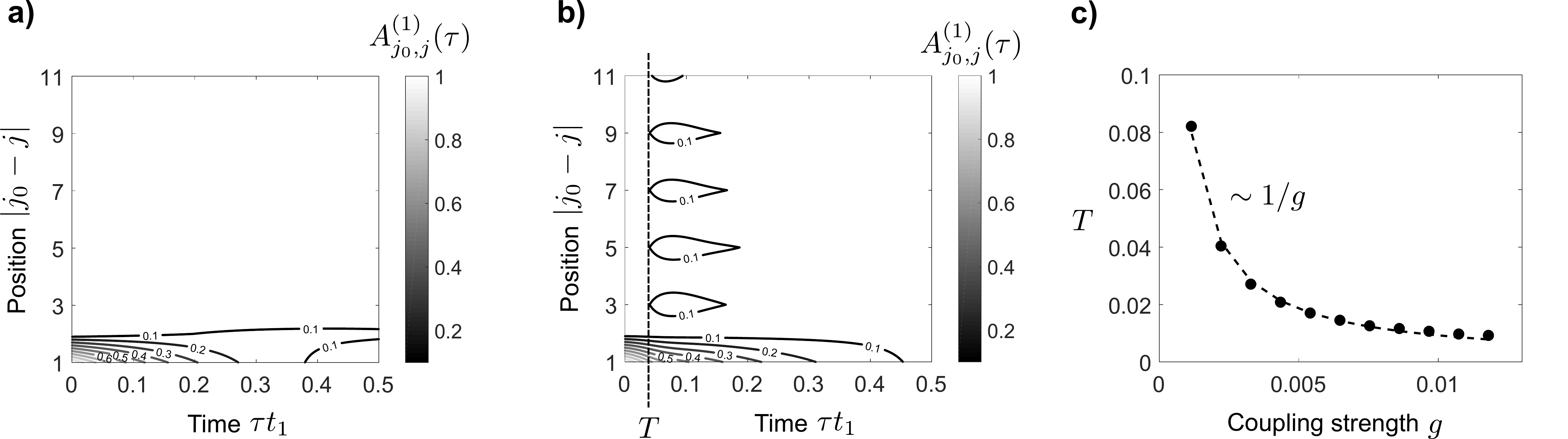}}
\caption{\textbf{a)}-\textbf{b)} Contour plot of the spectral function $A^{(1)}_{j_0,j} (\tau)$ of the lower band as a function of position and time. An electron is injected in the lower level at site $j_{0}=1$ and time $\tau=0$. The chain length is $N=11$ and the coupling strength $g=2.2 \times 10^{-3}$. \textbf{a)} Dissipative regime with $t_{2}=5 \times 10^{-3}$, and $\kappa=0.1$. \textbf{b)} Individual dressing regime with $t_2=0.1$ and $\kappa=10^{-4}$. A transfer of spectral weight occurs at a time $T$ represented as a vertical line. \textbf{c)} The time $T$ is represented as a function of the coupling strength. Other parameters are identical to that of Fig.~\ref{fig9}.}
\label{fig14}
\end{figure}

To further evidence the existence of a coherent dynamics in the individual dressing regime, we compare the spectral function $A^{(1)}_{j_0,j} (\tau)$ introduced in Sec.~\ref{no_coupl_part} in the collective (dissipative) and the individual dressing regimes. This function is computed using NGFs, and shown on Fig.~\ref{fig14}, with $j_0=1$, $g=2.2 \times 10^{-3}$, $N=11$, and the same other parameters as in Fig.~\ref{fig9}. In the dissipative regime [Fig.~\ref{fig14} \textbf{a)}], a particle injected at the first site for $g\neq 0$ stays essentially localized, and no propagation occurs through the lower band whatsoever, not even with the small hopping rate $t_1$ as in the case $g=0$ [Fig.~\ref{fig3} \textbf{e)}]. In this case, the dynamics consists of a collective damping of populations from the upper to the lower band, involving localized states (superpositions of different Bloch states). Pictorially, the large photon damping rate constantly projects the system onto its initial state (similarly to the quantum Zeno effect~\cite{zeno,PhysRevA.92.023825}), thereby preventing the hopping through the chain to occur.   

In the individual dressing regime [Fig.~\ref{fig14} \textbf{b)}], however, we observe a small transfer of spectral weight $\approx 10\%$ occuring after a time $T$ with a period of two sites, which shows that the properties of the resonant upper band Bloch state are transfered in the lower band. This results in a new state with energy $\sim \omega_1$ delocalized across the whole chain. In this sense, this corresponds to an effective hopping mechanism restoring propagation in the quasi-blocked lower band. However, one can not \textit{a priori} write an Hamiltonian term which reproduces this single-Bloch-state dynamics, as nearest neighbors hopping in a 1D chain involves the complete set of the chain Bloch states. Finally, we find that the spectral weight transfer induced by the coupling to the cavity mode occurs at a time $T\sim 1/g$, which corresponds to the time to emit a photon concurrently with the transfer of populations to the lower band [Fig.~\ref{fig14} \textbf{c)}].

\vspace{2mm}

Similarly as in the dissipative regime, we find that two polariton peaks appear in the cavity photon DOS (outside the upper electronic bandwidth), when $g$ exceeds the typical energy separation $\delta \omega$ between two adjacent Bloch states in the upper band. In the individual dressing regime, however, $\Omega_{S}\neq \Omega_{n}$ indicating that the dynamics does not involve a collective response of the Bloch states~\cite{hagenmuller}. Ultimately, for $\kappa \ll W_2$ and $g \gg \delta \omega$, all Bloch states are coupled to the cavity mode, and we expect to recover the physics of the collective dressing regime [see Fig.~\ref{fig1} \textbf{c)}]. We point out, however, that a quantitative study is difficult as neither the effective QME nor the NGFs method are valid in this strongly non-perturbative regime. The full QME is the only suitable method, but identifying collective effects is hard since this method is in any case limited to small $N$.

\subsection{Photons and scaling with $N$}
\label{comp_part}

As concluding remarks, we have checked that the mean cavity photon number in the steady-state $\bar{n}=\langle a^{\dagger} a \rangle$ (see end of Sec.~\ref{gf_self_ener_part}) remains small even for large coupling strength ($\bar{n}\lesssim 10^{-2}$ in the dissipative regime, and $\bar{n}\lesssim 1$ in the individual dressing regime), showing that the cavity operates in the quantum regime close to the vacuum state. Our methods can be easily generalized to consider a finite mean photon population $N_{P}$ in the bath, in which case we find that the current enhancement depends on the rescaled coupling strength $g\sqrt{N_{P}}$. On the other hand, order-of-magnitude current enhancements can occur when considering different injection/extraction rates $\Gamma_{1}\neq \Gamma_{2}$ for the two bands (e.g. for $\Gamma_{1}\gg \Gamma_{2}$), as well as a small photon population $N_{P} \lesssim 1$. In this case, still considering $t_{1}\ll \Gamma_{1}$ and $t_{2}\gg \Gamma_{2}$, the small injection/extraction rate in the upper band provides a strong reduction of the bare current $\approx e\Gamma_{2}/2$ obtained for $g=0$, leading to current enhancements only limited by the ratio $\Gamma_{1}/\Gamma_{2}$ when $g\neq 0$~\cite{hagenmuller}.      

\vspace{2mm}

For given $g$ and $\kappa$, we find that the saturation value for the steady-state current decreases sublinearly when increasing the chain length $N$, restricting the scope of our study to mesoscopic systems. In addition, the current typically exhibits small oscillations between odd and even values of $N$, with slightly larger values for $N$ odd. The existence of a Bloch state resonant with the cavity mode for $N$ odd leads to a slightly larger current enhancement than for $N$ even, where the two closest states to the upper band center $\omega_2$ are only quasi-resonant with the cavity mode. In the limit $N \gg 1$, the separation between adjacent Bloch states close to the upper band center is $\sim 2 \pi t_{2}/N$. Still considering a cavity mode resonant with the transition between the lower flat band and the upper band center, the number of interband transitions within the linewidth $\kappa$ is thus $\sim N \kappa/t_{2}$, and \textit{one can conclude that the individual band dressing in the large $N$ regime is only limited by the cavity quality factor. Ultimately, when $N \kappa/t_{2} \gg 1$, the system always enters the collective dressing regime.} 

\section{Conclusion}
\label{conclll}

In conclusion, we have studied in detail the interplay between the transport of fermions through a 1D mesoscopic chain of two-orbital systems and light-matter coupling to a single cavity mode close to its vacuum state. We have derived both analytical and numerical results using complementary methods based on Keldysh and QME techniques, providing new perspectives for the investigation of many-body fermionic systems coupled to confined photons. We have compared the steady-state current obtained with these different methods, and shown that light-matter coupling leads to a current enhancement. Depending on the ratio between the cavity photon decay rate and the upper electronic bandwidth, different regimes have been identified and discussed. In the dissipative regime, we have derived an analytical formula for the current enhancement valid for small coupling strengths, showing that the current enhancement scales with the cooperativity. We have characterized the presence of a collective coupling of all the Bloch states to the cavity mode, when the coupling strength is larger than the typical energy separation between two adjacent Bloch states in the upper band. In this case, the current enhancement is shown to stem from a global transfer of populations from the upper to the lower band, with only marginal propagation through the latter. In the coherent regime, however, we have shown that when the coupling strength is smaller than the typical energy separation between two adjacent Bloch states in the upper band, only the resonant Bloch state is ``individually'' coupled to the cavity mode. Moreover, a small transfer of spectral weight occurs from the upper to the lower band, resulting in a new state with energy $\sim \omega_1$ delocalized across the whole chain. In this case, the current enhancement has been interpreted as stemming from a coherent hopping dynamics sustained by the absorption and emission of cavity photons. Ultimately, when the coupling strength becomes larger than the upper electronic bandwidth, or when the system size becomes large, we expect to recover the collective dressing regime.  

In a realistic situation, additional random potentials due to disorder and impurities will affect transport properties through the chain. In the presence of light-matter coupling at optical frequencies, orbitals are separated by a large gap $\sim 1 {\rm eV}$, and since low-energy valence states are typically less affected by short-range random potentials than the upper delocalized orbitals, we expect that the effective transmission channel provided by the coupling to the cavity should be more robust to disorder than the standard channel involving the upper orbitals for $g=0$. Possible extensions of this model include considering a frequency-dependent leads coupling and/or cavity decay rate to study how non-Markovian (memory) effects affect our results. Further investigations could be also devoted to the symmetric case with equal lower and upper electronic bandwiths for $g\neq 0$. In this situation, charge transport can be reduced as the system exhibits interference between the different quantum paths connecting the same orbital at two distant sites for some specific coupling $g$. It would thus be interesting to study how this competes with the time-reversed loop trajectories leading to Anderson localization in random lattices~\cite{anderson}. Our model might find direct applications in several fields, such as transport in organic semiconductors~\cite{orgiu2015conductivity} and quantum dot arrays~\cite{kagan_charge_2015,wang_fabrication_2004,gudmundsson,moldoveanu}, which have recently been coupled to surface plasmon resonators~\cite{ebbesen,orgiu2015conductivity,jino} and microwave cavities~\cite{frey_dipole_2012,viennot_out_equilibrium_2014,liu_semiconductor_2015}.

\section*{Acknowledgements}

We are grateful to Stefano Azzini, Thibault Chervy, Roberta Citro, Thomas Ebbesen, Cyriaque Genet, Emanuele Orgiu, and Paolo Samor\`{i} for fruitful discussions. Work in Strasbourg was supported by the ERC St-Grant ColDSIM (No. 307688), with additional funding from Rysq and ANR-FWF grant BLUESHIELD. C.G. acknowledges support from the Max Planck Society and from the COST action NQO 1403 (Nano-scale Quantum Optics). This work is supported by IdEx Unistra (project STEMQuS) with funding managed by the French National Research Agency as part of the ``Investments for the future program''.

\appendix

\section{Keldysh formalism}
\label{negfs_form}

In this appendix, we propose a detailed derivation of the results presented in Sec.~\ref{gf_self_ener_part}. We first write the steady-state current in terms of electron GFs, and then show that electron and photon GFs can be computed by solving a closed set of equations involving electron and photon SEs. We consider $\hbar=1$, and use the short-hand notations $\partial_{\tau}\equiv \frac{\partial}{\partial \tau}$ and $\delta_{f (\tau)}\equiv \frac{\delta}{\delta f(\tau)}$, for function and functional derivatives, respectively. 

\vspace{2mm} 

\textbf{\textit{Steady-state current.}} As seen in Sec.~\ref{gf_self_ener_part}, the steady-state current $J_{\eta}$ flowing through the lead $\eta$ is proportional to the commutator between the total Hamiltonian $H$ and the number of electrons in the lead $\eta$. A direct calculation of this commutator allows us to express $J_{\eta}$ in terms of a GF which describes the correlations between the leads and the chain:

\begin{equation}
J_{\eta} = -2e\sum_{\alpha,k} \sum_{\bf q} \varphi^{j_{\eta}}_{k} \lambda_{\alpha,{\bf q}} \int \!\! \frac{d\omega}{2\pi} \Re \left[ G^{<}_{\alpha,k,{\bf q},\eta} (\omega) \right], 
\label{current_2_re}
\end{equation} 
where $\Re$ stands for real part, $\lambda_{\alpha,{\bf q}}$ is defined in Eq.~(\ref{cou_le_ref}), and $G^{<}_{\alpha,k,{\bf q},\eta} (\omega)$ denotes the Fourier transform of the ``lesser'' mixed system-leads GF $G^{<}_{\alpha,k,{\bf q},\eta} (\tau-\tau')$, which can be obtained from the time-ordered GF:

\begin{equation}
G_{\alpha,k,{\bf q},\eta} (\tau-\tau')= -i \langle \mathcal{T} \tilde{c}_{\alpha,k} (\tau) b^{\dagger}_{\alpha,{\bf q},\eta} (\tau') \rangle.
\label{to_GF1}
\end{equation} 

$\mathcal{T}$ denotes the time-ordered product for fermions. Taking the time derivative $\partial_{\tau'}$ of Eq.~(\ref{to_GF1}), and computing the different commutators entering the Heisenberg equation $\partial_{\tau'} b^{\dagger}_{\alpha,{\bf q},\eta} (\tau')=i[H,b^{\dagger}_{\alpha,{\bf q},\eta}] (\tau')$, the equation of motion of $G_{\alpha,k,{\bf q},\eta}(\tau-\tau')$ is derived as:

\begin{align}
\left(- i \partial_{\tau'}  - \omega_{\bf q} \right) G_{\alpha,k,{\bf q},\eta} (\tau-\tau')= - \lambda_{\alpha,{\bf q}} \sum_{k'} \varphi^{j_{\eta}}_{k'}  G_{\alpha,k,k'}(\tau-\tau'),
\label{eq_vcvc}
\end{align}
where $G_{\alpha,k,k'} (\tau-\tau')= -i \langle \mathcal{T} \tilde{c}_{\alpha,k} (\tau) \tilde{c}^{\dagger}_{\alpha,k'} (\tau') \rangle$ is the time-ordered GF of the chain, refered to as the ``electron GF''. Equation (\ref{eq_vcvc}) can be formally solved in the frequency domain as:

\begin{equation}
G_{\alpha,k,{\bf q},\eta} (\omega)= - \lambda_{\alpha,{\bf q}} \sum_{k'} \varphi^{j_{\eta}}_{k'} G_{\alpha,k,k'} (\omega) \mathcal{G}_{{\bf q},\eta} (\omega),
\label{final_GF_mixed00}
\end{equation}
where $G_{\alpha,k,k'} (\omega)$ and $\mathcal{G}_{{\bf q},\eta} (\omega)$ denote the Fourier transforms of the electron GF and the lead GF $\mathcal{G}_{{\bf q},\eta} (\tau-\tau')=-i\langle \mathcal{T} b_{\alpha,{\bf q},\eta} (\tau) b^{\dagger}_{\alpha,{\bf q},\eta} (\tau') \rangle_{0}$, and $\langle \cdots \rangle_{0}$ refers to the quantum average in the ground state of the Hamiltonian $H$ without the interaction terms $H_{I}$, $H_{L}$, and $H_{P}$. One can then use the Langreth rules~\cite{haug} in Eq.~(\ref{final_GF_mixed00}) to compute the ``lesser'' GF:

\begin{align}
G^{<}_{\alpha,k,{\bf q},\eta} (\omega) =- \lambda_{\alpha,{\bf q}} \sum_{k'} \varphi^{j_{\eta}}_{k'} G^{r}_{\alpha,k,k'} (\omega) \mathcal{G}^{<}_{{\bf q},\eta} (\omega) - \lambda_{\alpha,{\bf q}} \sum_{k'} \varphi^{j_{\eta}}_{k'} G^{<}_{\alpha,k,k'} (\omega) \mathcal{G}^{a}_{{\bf q},\eta} (\omega),
\label{gren_lead_el}
\end{align}
with $r$ and $a$ for retarded and advanced GFs, respectively. Using the results: 

\begin{align}
\mathcal{G}^{<}_{{\bf q},\eta} (\omega)&= 2i\pi \delta (\omega - \omega_{\bf q} ) n_{\eta} (\omega) \nonumber \\
\mathcal{G}^{a}_{{\bf q},\eta} (\omega) &= \frac{1}{\omega - \omega_{\bf q} - i 0^{+}},
\label{gg_petites}
\end{align}
where $0^{+}$ denotes an infinitesimal positive quantity and $n_{\eta} (\omega)$ is the Fermi occupation number of the lead $\eta$, we substitute Eq.~(\ref{gren_lead_el}) in the expression of the current Eq.~(\ref{current_2_re}), and convert the summation over ${\bf q}$ into a frequency integral $\sum_{\bf q} \to \int_{0}^{\infty} d\omega \rho (\omega)$, where $\rho (\omega)$ represents the electron density of states in the leads. Introducing the tunnelling rate between the chain and the leads as $\Gamma_{\alpha} =2\pi \rho (\omega) \lambda^{2}_{\alpha} (\omega)$ (assumed to be energy independent), we finally recover Eqs.~(\ref{transsm}) and (\ref{trans_long_coco2}). Note that we have assumed $n_{s} (\omega)=1$ and $n_{d} (\omega)=0$ $\forall \, \omega$ (high-bias regime).

\vspace{2mm} 

\textbf{\textit{Dyson equation for electrons GFs.}} In order to compute the transmission spectrum Eq.~(\ref{trans_long_coco2}), we now need an equation of motion of the time-ordered electron GFs. As before, we compute the time derivative $\partial_{\tau} G_{\alpha,k,k'} (\tau-\tau')$, use the Heisenberg equation $\partial_{\tau} \tilde{c}_{\alpha,k} (\tau)=i[H,\tilde{c}_{\alpha,k}] (\tau)$, and obtain: 

\begin{align}
\left( i \partial_\tau - \omega_{\alpha,k} \right) G_{\alpha,k,k'} (\tau-\tau')&= \delta_{k,k'} \delta (\tau-\tau') + g \sum_{\alpha'} (1-\delta_{\alpha,\alpha'}) F_{\alpha',k,\alpha,k'}(\tau-\tau') \nonumber \\
&- \sum_{{\bf q},\eta} \lambda_{\alpha,{\bf q}} \varphi^{j_{\eta}}_{k} G_{{\bf q},\eta,\alpha,k'}(\tau-\tau'),
\label{eq_momo_long_nc}
\end{align} 
where $G_{{\bf q},\eta,\alpha,k} (\tau-\tau')= -i \langle \mathcal{T} b_{\alpha,{\bf q},\eta} (\tau) \tilde{c}^{\dagger}_{\alpha,k} (\tau') \rangle$ is a mixed system-leads GF similar to the one defined in Eq.~(\ref{to_GF1}), and $F_{\alpha',k,\alpha,k'} (\tau-\tau') = -i \langle \mathcal{T} \tilde{c}_{\alpha',k} (\tau) \tilde{c}^{\dagger}_{\alpha,k'} (\tau') A (\tau) \rangle$ is a higher-order correlation function mixing the electronic and photonic degrees of freedom. First, the equation of motion for the Fourier transform $G_{{\bf q},\eta,\alpha,k} (\omega)$ is derived similarly as before and reads:

\begin{align}
G_{{\bf q},\eta,\alpha,k} (\omega) &= - \lambda_{\alpha,{\bf q}} \sum_{k'} \varphi^{j_{\eta}}_{k'} \mathcal{G}_{{\bf q},\eta} (\omega) G_{\alpha,k',k}(\omega).
\label{eq_of_motion2121} 
\end{align}

Secondly, the correlation function $F_{\alpha',k,\alpha,k'} (\tau-\tau')$ can be written in terms of single-particle GFs by considering a term $H'= \mathcal{J} A$ in the Hamiltonian, where $\mathcal{J}$ denotes a vanishing current source~\cite{engelsberg1963coupled}. Taking the functional derivative $\delta_{\mathcal{J} (\tau)} G_{\alpha',k,\alpha,k'} (\tau-\tau')$, where $G_{\alpha',k,\alpha,k'} (\tau-\tau')$ is given by:

\begin{equation}
G_{\alpha',k,\alpha,k'} (\tau-\tau') = -i \frac{\langle \mathcal{T} \tilde{c}_{\alpha',k} (\tau) \tilde{c}^{\dagger}_{\alpha,k'} (\tau') e^{-i \int \! d \tau_{1} H (\tau_{1})}\rangle_{0}}{\langle e^{-i \int \! d \tau_{1} H (\tau_{1})} \rangle_{0}}, 
\label{GF_def_ufc22}
\end{equation} 
we obtain:

\begin{align}
F_{\alpha',k,\alpha,k'} (\tau-\tau') = i g \sum_{k_{1},k_{2}} \int \!\! \{d\tau\} G_{\alpha',k,k_{1}} (\tau-\tau_{1}) \Lambda_{\alpha',k_{1},\alpha,k_{2}} (\{\tau\}) D (\tau_{3} - \tau) G_{\alpha,k_{2},k'} (\tau_{2}-\tau'),
\label{ff_fun_app}
\end{align} 
where $\{\tau\}\equiv \tau_1,\tau_2,\tau_3$, $\int \{d\tau\} \equiv \int \! d\tau_{1} \! \int \! d\tau_{2} \! \int \! d\tau_{3}$. The time-ordered photon GF is defined as:

\begin{align*}
D (\tau_{3} - \tau)=\delta_{\mathcal{J} (\tau)} \langle A (\tau_{3})\rangle = -i \langle \mathcal{T} A ( \tau_{3}) A (\tau) \rangle,
\end{align*} 
and the so-called vertex function as:

\begin{align*}
\Lambda_{\alpha',k_{1},\alpha,k_{2}} (\tau_{1},\tau_{2},\tau_{3}) = -\frac{1}{g} \delta_{\langle A (\tau_{3})\rangle} G^{-1}_{\alpha',k_{1},\alpha,k_{2}} (\tau_{1}-\tau_{2}).
\end{align*} 

It can be shown that this vertex function satisfies a self-consistent equation~\cite{engelsberg1963coupled}. The SCBA consists in considering only the leading term (undressed vertex) of this self-consistent equation, which provides: 

\begin{align}
\Lambda_{\alpha',k',\alpha,k} (\tau,\tau',\tau'') &= \left(1-\delta_{\alpha',\alpha} \right) \delta_{k,k'} \delta( \tau-\tau') \delta (\tau - \tau''). 
\label{vert_corr1}
\end{align}

Higher order corrections in $\Lambda$ correspond to the so-called vertex corrections associated with crossed diagrams~\cite{engelsberg1963coupled} such as the one sketched on Fig.~\ref{fig_self} \textbf{b)}, which are neglected in the SCBA. Using Eqs.~(\ref{eq_of_motion2121}), (\ref{ff_fun_app}), and (\ref{vert_corr1}), the equation of motion (\ref{eq_momo_long_nc}) written in the frequency domain takes the form:

\begin{align*}
\sum_{k_{1}} \left((G^{0}_{\alpha,k,k_{1}} (\omega))^{-1} - \Sigma_{\alpha,k,k_{1}} (\omega) \right) G_{\alpha,k_{1},k'} (\omega) =\delta_{k,k'},
\end{align*}
with the SCBA self-energy: 

\begin{align}
\Sigma_{\alpha,k,k'} (\omega) = i g^{2} \left(1-\delta_{\alpha,\alpha'}\right) \int \!\! \frac{d\omega'}{2\pi} G_{\alpha',k,k'} (\omega+\omega') D (\omega') + \sum_{{\bf q},\eta} \lambda^{2}_{\alpha,{\bf q}} \varphi^{j_{\eta}}_{k} \varphi^{j_{\eta}}_{k'} \mathcal{G}_{{\bf q},\eta} (\omega),
\label{lm_se_lala36}
\end{align}  
and the non-interacting time-ordered GF $G^{0}_{\alpha,k,k'} (\omega)$. Still considering the high-bias regime, we now use the Langreth rules together with Eq.~(\ref{gg_petites}), and convert the summation over ${\bf q}$ in Eq.~(\ref{lm_se_lala36}) into a frequency integral. This leads to the expressions of the ``lesser'' and ``greater'' electron SEs given by Eqs.~(\ref{lm_se_lala}) and (\ref{leads_se_true}). 

\vspace{2mm} 

\textbf{\textit{Dyson equation for photons GFs.}} The equation of motion for the time-ordered photon GF $D(\omega)$ can be derived by taking the second time derivative of the cavity vector potential $A(t)$, and then use the Heisenberg equation $\partial_{\tau} A (\tau)=i [H,A ] (\tau)$ two times in a row. As in the previous section, we consider a vanishing source term $H'= \mathcal{J} A$ in the Hamiltonian $H$. The functional derivative of the ground-state expectation of the obtained equation with respect to $\mathcal{J} (\tau')$ yields the following equation of motion for $D(\tau-\tau')$:

\begin{align}
\left(- \frac{\partial^{2}_{\tau}}{2\omega_{c}} - \frac{\omega_{c}}{2} \right) D (\tau-\tau') &=  \delta (\tau - \tau') - i g \sum_{\alpha,\alpha'} \sum_{k} \left(1-\delta_{\alpha,\alpha'} \right) \delta_{\mathcal{J} (\tau')} G_{\alpha,k,\alpha',k} (\tau,\tau^{+}) \nonumber \\
&+ \sum_{\bf p} \mu_{\bf p} D_{\bf p} (\tau-\tau'),
\label{eq_momo_ph}
\end{align} 
where the time $\tau^{+}=\tau + 0^{+}$, and the mixed GF $D_{\bf p} (\tau-\tau')= -i \langle \mathcal{T} A_{\bf p} (\tau) A (\tau') \rangle$ describes correlations between the cavity mode and the electromagnetic environment. The equation of motion for $D_{\bf p}$ can be derived similarly as before (by calculating its second time derivative): 

\begin{align*}
\left(-\partial^{2}_{\tau} - \omega^{2}_{\bf p}\right) D_{\bf p} (\tau-\tau') = 2 \omega_{\bf p} \mu_{\bf p}  D (\tau-\tau'),
\end{align*} 
which is solved in the frequency domain as $D_{\bf p} (\omega) = \mu_{\bf p}\mathcal{D}_{\bf p} (\omega) D (\omega)$. Here, $\mathcal{D}_{\bf p} (\omega)$ is the Fourier transform of the (time-ordered) extra-cavity photon GF $-i\langle \mathcal{T} A_{\bf p} (\tau) A_{\bf -p} (\tau') \rangle_{0}$. Using Eq.~(\ref{GF_def_ufc22}), the second term in the right-hand side of Eq.~(\ref{eq_momo_ph}) can be written in the form:

\begin{align}
\frac{\delta G_{\alpha,k,\alpha',k} (\tau,\tau^{+})}{\delta \mathcal{J} (\tau')} =  g \sum_{k_{1},k_{2}} \int \!\! \{d\tau\} G_{\alpha,k,k_{1}} (\tau-\tau_{1}) \Lambda_{\alpha,k_{1},\alpha',k_{2}} (\{\tau\})  D (\tau_{3} - \tau') G_{\alpha',k_{2},k} (\tau_{2}-\tau^{+}),
\label{green_12}
\end{align} 
where the vertex function is given by Eq.~(\ref{vert_corr1}). In the SCBA, we only consider the leading order $\Lambda_{\alpha,k,\alpha',k'} (\tau,\tau',\tau'')=\left(1-\delta_{\alpha,\alpha'} \right) \delta_{k,k'} \delta(\tau-\tau') \delta (\tau - \tau'')$, which we substitute in Eq.~(\ref{green_12}) to put the equation of motion (\ref{eq_momo_ph}) into the form:

\begin{align*}
\left(D^{-1}_{0} (\omega) - \Pi (\omega) \right) D (\omega) = 1
\end{align*} 
with the cavity photon SE:

\begin{align}
\Pi (\omega) = -i g^{2} \sum_{\alpha,\alpha'} \textbf{Tr} \left(1-\delta_{\alpha,\alpha'} \right) \int \!\! \frac{d\omega'}{2\pi} \underline{G}_{\alpha} (\omega+\omega') \underline{G}_{\alpha'} (\omega') + \sum_{\bf p} \mu^{2}_{\bf p} \mathcal{D}_{\bf p} (\omega),
\label{ph_se_2}
\end{align}
and the bare cavity photon GF $D_{0}(\omega)$. The summation over the continuous index ${\bf p}$ can again be converted into a frequency integral, namely $\sum_{\bf p} \to \int_{0}^{\infty} d\omega \rho_{0} (\omega)$, where $\rho_{0} (\omega)$ denotes the extra-cavity photon density of states. We introduce the cavity photon decay rate as $\kappa = 2\pi \rho_{0} (\omega) \mu^{2} (\omega)$ (assumed to be frequency-independent), and use the Langreth rules in Eq.~(\ref{ph_se_2}). Assuming a vanishing mean population in the photon bath, i.e. $\langle a^{\dagger}_{\bf p} a_{\bf p}\rangle =0$, one can compute the (non-interacting) extra-cavity photon GFs as $\mathcal{D}^{>}_{\bf p} (\omega)=-2i\pi \delta(\omega-\omega_{\bf p})$ and $\mathcal{D}^{<}_{\bf p} (\omega)=-2i\pi \delta(\omega+\omega_{\bf p})$, and show that the ``lesser'' and ``greater'' photon SEs correspond to Eqs.~(\ref{popo_ph_less}) and (\ref{popo_ph_less22}). 

\section{Elimination of the cavity field}
\label{master_form1}

In this appendix, we show that $\hat{\rho}$ -- the projection of the density operator $\tilde{\rho}$ (in the rotating frame) onto the cavity vacuum state -- evolves according to Eq.~(\ref{eff_master_gene}) in the dissipative regime.

\vspace{2mm} 

\textbf{\textit{Dissipative regime.}} We consider the case when the cavity decay rate $\kappa$ is large compared to the other rates. In particular, $\kappa$ is larger than the injection/extraction rates $\Gamma_{\alpha}$ and tunneling rates $t_{\alpha}$ governing the uncoupled evolution of the electronic degrees of freedom, and larger than the coupling strength $g$ between electronic and bosonic variables. This choice has two main consequences:

\begin{itemize}
\item We expect the strongly damped cavity field to stay close to its vacuum state (steady state for $g=0$).
\item The electrons' observables evolve on a much longer time-scale than the one associated with the cavity field.
\end{itemize} 

The last point allows us to adiabatically eliminate the light field from the overall dynamics. For this purpose, we first define the electron reduced density operator:

\begin{align*}
\tilde{\rho}_{el} = \text{Tr}_{F} [\tilde{\rho}] = \sum_{n} \bra{n} \tilde{\rho} \ket{n} = \sum_{n} \tilde{\rho}_{nn}, 
\end{align*}
which is a density matrix for the electronic degrees of freedom only. $\text{Tr}_{F} [A] = \sum_n \bra{n} A \ket{n}$ denotes the trace of the observable $A$ over the cavity field, and $\ket{n}$ with $n=0,1,2,\cdots$ is the photonic part of the (Fock) state containing $n$ photons. As already mentioned, we assume that the light field is close to its vacuum state, i.e. $\tilde{\rho}_{el} \simeq \tilde{\rho}_{00}$. In the following, we derive a closed time-evolution for the relevant part $\tilde{\rho}_{00}$ of the reduced density operator ~\cite{Bonifacio1971,Schuetz2013}. 

\vspace{2mm} 

\textbf{\textit{Projectors and coupled differential equations.}} We introduce the projectors $P$ and $Q$ with

\begin{align*}
P\tilde{\rho}&=\bra{0} \tilde{\rho} \ket{0} \ket{0}\bra{0} = \tilde{\rho}_{00} \ket{0}\bra{0} \equiv \hat{\rho}, \\
Q\tilde{\rho}&=\sum_{\substack{n,m \\ n,m\neq 0}} \bra{n} \tilde{\rho} \ket{m} \ket{n} \bra{m} = \sum_{\substack{n,m \\ n,m\neq 0}} \tilde{\rho}_{nm} \ket{n}\bra{m} \equiv \check{\rho}. 
\end{align*}

Using the decomposition (see Sec.~\ref{sec:eff_master}):

\begin{align*}
 \partial_{\tau}\tilde{\rho}= (\mathcal{L}_{e} + \mathcal{L}_{c} + \mathcal{L}_{I} + \mathcal{I}_{c})\tilde{\rho}, 
\end{align*}
with $\mathcal{I}_c = \kappa a \tilde{\rho} a^{\dagger}$, together with the property $P+Q=1$,
the coupled differential equations for $\hat{\rho}$ and $\check{\rho}$ can be written as:

\begin{align}
\label{hatrho_eq}
\partial_\tau \hat{\rho} &=  P \mathcal{L}_{e} \hat{\rho} + P \left(\mathcal{L}_{I}+ \mathcal{I}_{c} \right) \check{\rho}, \\
\partial_\tau \check{\rho} &= Q \mathcal{L}_{I} \hat{\rho} + Q \left(\mathcal{L}_{e} + \mathcal{L}_{c} + \mathcal{L}_{I} +\mathcal{I}_{c} \right)\check{\rho}. 
\label{checkrho_eq}
\end{align}

The formal solution of Eq.~(\ref{checkrho_eq}) is given by:

\begin{align}
\check{\rho}(\tau) = e^{Q (\mathcal{L}_c + \mathcal{L}_{e}) \delta \tau} \check{\rho}(\tau_{0}) + \int_{\tau_{0}}^{\tau} d \tau' e^{Q (\mathcal{L}_c + \mathcal{L}_{e}) (\tau-\tau')} V(\tau'),  
\label{eq_Qtorepl}
\end{align}
with 

\begin{align*}
V(\tau') = Q \mathcal{L}_{I} \hat{\rho}(\tau') + Q(\mathcal{L}_{I}+ \mathcal{I}_{c}) \check{\rho}(\tau'),
\end{align*}
and $\delta \tau = \tau - \tau_0$. The formal solution Eq.~(\ref{eq_Qtorepl}) can be plugged into Eq.~(\ref{hatrho_eq}), and keeping terms up to second order in $\mathcal{L}_{I}$, we obtain:

\begin{align}
\partial_\tau \hat{\rho}  
\simeq P \mathcal{L}_{e} \hat{\rho} 
+ &P \mathcal{L}_{I} \int_{\tau_{0}}^{\tau} \!\! d \tau' e^{Q (\mathcal{L}_c + \mathcal{L}_{e}) (\tau-\tau')} Q \mathcal{L}_{I} \hat{\rho}(\tau') \nonumber \\
+&P \mathcal{I}_{c} \int_{\tau_{0}}^{\tau} \!\! d \tau' e^{Q (\mathcal{L}_c + \mathcal{L}_{e}) (\tau-\tau')} Q \mathcal{L}_{I} \int_{\tau_{0}}^{\tau'} \!\! d \tau'' e^{Q (\mathcal{L}_c + \mathcal{L}_{e}) (\tau'-\tau'')} Q \mathcal{L}_{I} \hat{\rho}(\tau''), 
\label{eq_tot_PQ} 
\end{align}
with the initial condition $\check{\rho}(\tau_0)=0$ (cavity intially prepared in its vacuum state).

\vspace{2mm} 

\textbf{\textit{Time-scale separation and integration.}} We first focus on the second term in the right-hand side of Eq.~(\ref{eq_tot_PQ}), which, after change of variables, reads:

\begin{align}
P \mathcal{L}_{I} \int_{0}^{ \delta \tau} \!\! d \tau' e^{Q (\mathcal{L}_c + \mathcal{L}_{e}) \tau'} Q \mathcal{L}_{I} \hat{\rho}(\tau-\tau').
\label{eq:term1}
\end{align}

Letting the operator $Q \mathcal{L}_{I}$ act on $\hat{\rho}$, one obtains:

\begin{align}
Q \mathcal{L}_{I} \hat{\rho} (\tau - \tau') = - i g S^{-} \tilde{\rho}_{00}(\tau-\tau') \ket{1}\bra{0} + \textrm{h.c.},
\label{propo_inter}
\end{align}
with the collective lowering operator $S^{-}=\sum_{j} c^{\dagger}_{1,j} c_{2,j}$, ($S^{+} = (S^{-})^{\dagger}$). Subsequently, according to Eq.~(\ref{eq:term1}), we apply the free evolution $\exp( Q (\mathcal{L}_c + \mathcal{L}_{e}) \tau')$ to the previous expression Eq.~(\ref{propo_inter}):

\begin{align}
\int_{0}^{ \delta \tau} \!\! d \tau' e^{Q (\mathcal{L}_c + \mathcal{L}_{e}) \tau'} Q \mathcal{L}_{I} \hat{\rho}(\tau-\tau') \simeq \int_{0}^{ \delta \tau} \!\! d \tau'(- i g) S^{-} \tilde{\rho}_{00}(\tau-\tau') \ket{1}\bra{0} e^{\left(i \Delta - \frac{\kappa}{2}\right) \tau'} + \textrm{h.c.}, 
\label{eq:first_integ}
\end{align}
where we have used $e^{Q (\mathcal{L}_c + \mathcal{L}_{e}) \tau'} \approx e^{Q \mathcal{L}_c \tau'}$ in the integrand. This approximation is justified in the dissipative regime where $\vert i\Delta -\kappa/2\vert \gg \Gamma_{\alpha}, t_{\alpha}$. Corrections to the previous approximation could be taken into account, e.g. by using partial integration. They are expected to scale with $t_{\alpha}/|i \Delta - \kappa/2|$ and $\Gamma_{\alpha}/|i \Delta - \kappa/2|$ and are small whenever the light-field evolves on a much shorter time-scale than the electronic degrees of freedom. The time-scale separation allows us to further neglect the variation of $\tilde{\rho}_{00}$ during the relaxation time $\sim 1/\kappa$ of the cavity, namely:  

\begin{align}
\tilde{\rho}_{00}(\tau - \tau') e^{\left(i \Delta - \frac{\kappa}{2}\right) \tau'} \approx \tilde{\rho}_{00}(\tau) e^{\left(i \Delta - \frac{\kappa}{2}\right) \tau'}.
\label{eq:rho_00}
\end{align}

We point out that since the evolution of $\tilde{\rho}_{00}$ is governed by both the electronic term $\mathcal{L}_{e} \tilde{\rho}_{00}$ in Eq.~(\ref{eq_tot_PQ}), and the photon-mediated effective dynamics which we aim at calculating, checking the assumption Eq.~(\ref{eq:rho_00}) will be required (for consistency) at the end of the calculation. Under these approximations, the second term in the right-hand side of Eq.~(\ref{eq_tot_PQ}) takes the form:

\begin{align}
P \mathcal{L}_{I} \int_{0}^{\delta \tau} \!\! d \tau' e^{Q \mathcal{L}_c \tau'} Q \mathcal{L}_{I} \hat{\rho} (\tau)&= - 2 i \Gamma_{\Delta} \left( S^+ S^- \hat{\rho} (\tau) \left( 1 - e^{\left(i \Delta - \frac{\kappa}{2}\right)\delta \tau} \right) - \textrm{h.c.} \right) \nonumber \\
 &- 2\Gamma_{\kappa} \left( S^+ S^- \hat{\rho} (\tau) \left( 1 - e^{\left(i \Delta - \frac{\kappa}{2} \right)\delta \tau} \right) + \textrm{h.c.} \right),
\label{double_int_00}
\end{align}
where $\Gamma_{\Delta}= \frac{g^{2}\Delta}{2\Delta^{2}+\kappa^{2}/2}$ and $\Gamma_{\kappa}=\frac{g^{2}\kappa}{4\Delta^{2}+\kappa^{2}}$. We now turn to the third term in the right-hand side of Eq.~(\ref{eq_tot_PQ}). After change of variables, integration provides: 

\begin{align}
P \mathcal{I}_{c} \!\! \int_{0}^{\delta \tau} \!\!\! d \tau' e^{Q \mathcal{L}_c \tau'} Q \mathcal{L}_{I} \!\!\! \int_{0}^{\delta \tau-\tau'} \!\!\!\!\!\!\!\!\!\! d \tau'' e^{Q \mathcal{L}_c \tau''} Q \mathcal{L}_{I} \hat{\rho}(\tau) 
= 4 \Gamma_{\kappa} S^- \hat{\rho}(\tau) S^+ \left( 1 + e^{- \kappa \delta \tau} - 2 e^{-\frac{\kappa\delta\tau}{2}} \cos(\Delta \delta \tau) \right),
\label{double_int}
\end{align} 
where we have used $\tilde{\rho}_{00} (\tau - \tau' - \tau'') \simeq \tilde{\rho}_{00} (\tau)$ and made use of similar considerations as for the second term in the right-hand side of Eq.~(\ref{eq_tot_PQ}). Collecting the two contributions Eqs.~(\ref{double_int_00}) and (\ref{double_int}), and substituting them in Eq.~(\ref{eq_tot_PQ}), we finally obtain:

\begin{align}
\partial_\tau \hat{\rho} &= P \mathcal{L}_{e} \hat{\rho} - 2 i \Gamma_{\Delta} \left( S^+ S^- \hat{\rho} (\tau) \left(1 - e^{\left(i \Delta - \frac{\kappa}{2}\right) \delta \tau}\right) - \textrm{h.c.} \right) \nonumber \\
&- 2\Gamma_{\kappa} \left( S^+ S^- \hat{\rho} (\tau) \left(1 - e^{\left(i \Delta - \frac{\kappa}{2}\right) \delta \tau}\right) + \textrm{h.c.} \right) \nonumber\\
&+ 4 \Gamma_{\kappa} S^- \hat{\rho} (\tau) S^+ \left(1 + e^{- \kappa \delta \tau} - 2 e^{-\frac{\kappa\delta \tau}{2}} \cos(\Delta \delta \tau)\right).
\label{eq:integrand}
\end{align}

This result shows that the photon-mediated dynamics of the electrons scales with $\Gamma_{\Delta}$ and $\Gamma_{\kappa}$. Those rates should be small compared to $\kappa$ in order to use the time-scale separation, and in particular the approximation Eq.~(\ref{eq:rho_00}). This provides an upper bound for the coupling strength $g$. We point out that neglecting further corrections scaling with $\Gamma_{\Delta}$ and $\Gamma_{\kappa}$ in Eq.~(\ref{eq:rho_00}) is consistent with the approximation of keeping terms only up to second order in $\mathcal{L}_I$ in Eq.~(\ref{eq_tot_PQ}). In the regime $\kappa \delta \tau \gg 1$, the effective master equation (\ref{eff_master_gene}) can be finally derived from Eq.~(\ref{eq:integrand}):

\begin{align}
\partial_{\tau} \hat{\rho} \equiv \mathcal{L}_{\text{red}} \hat{\rho}
&= \mathcal{L}_{e} \hat{\rho} - 2 i \Gamma_{\Delta} [S^+ S^- , \hat{\rho}] - 2\Gamma_{\kappa} \big( S^+ S^- \hat{\rho} + \hat{\rho} S^+ S^- - 2 S^- \hat{\rho} S^+ \big). 
\label{eq:master_eff}
\end{align}

\textbf{\textit{Coarse graining and discussion.}} In the regime of parameters considered here, a coarse-grained time scale $\Delta \tau$ satisfying: 

\begin{align}
\kappa^{-1} \ll \Delta \tau \ll t_{\alpha}^{-1}, \Gamma_{\alpha}^{-1},
\label{time_scale_sepparat}
\end{align} 
can be introduced. Using Eqs.~(\ref{eq:integrand}) and (\ref{time_scale_sepparat}), one can show that the evolution of the reduced density operator on the time scale $\Delta \tau$ is:

\begin{align}
\frac{\hat{\rho}(\tau + \Delta \tau) - \hat{\rho}(\tau)}{\Delta \tau}  
= \int\limits_{\tau}^{\tau + \Delta \tau} \!\! d\tau' \frac{\partial_{\tau'} \hat{\rho} (\tau')}{\Delta \tau} 
\simeq \mathcal{L}_{\text{red}} \hat{\rho}(\tau),
\label{eq:coarse_evolv}
\end{align}
with $\tau \geq \tau_0$, and where consistently with the approximations used in Eqs.~(\ref{eq:first_integ}) and (\ref{eq:rho_00}), contributions $\sim (\kappa \Delta \tau)^{-1}$ and $\sim \Gamma_{\alpha} \Delta \tau, t_{\alpha} \Delta \tau$ have been neglected. The effective master equation (\ref{eff_master_gene}) is well-established on such a footing, and it is not suitable to describe the dynamics occuring on time-scales smaller than $1/\kappa$. In addition, $\Delta \tau$ has to be small compared to the time-scale associated with the photon-mediated dynamics, such that $\Delta \tau (\mathcal{L}_{\rm red} - \mathcal{L}_{e}) \hat{\rho} (\tau)$ is negligible. Together with the condition $\Delta \tau \gg \kappa^{-1}$, this provides an upper bound for the coupling strength $g$, whose exact form depends on the states that are involved in the dynamics, and whether collective effects play a role or not. In coupled spin-cavity systems (with $\Delta=0$), the condition: 

\begin{align}
\sqrt{N} g \ll \kappa
\label{eq:coll_g}
\end{align}
has been considered sufficient, or even required~\cite{Bonifacio1971}. We expect that the condition Eq.~(\ref{eq:coll_g}) is also sufficient in our fermionic model to use the time-scale separation. As a matter of fact, since quantum states with both orbitals either empty or fully occupied are not coupled to light, the collective coupling constant in our open fermionic model is $<g\sqrt{N}$ (see Sec.~\ref{coll_dress_part}), which places us on the safe side regarding Eq.~(\ref{eq:coll_g}). Note that the equations of motion (\ref{eq:master_eff}) and (\ref{eq:coarse_evolv}) are given in the adiabatic limit, and retardation effects between cavity and electronic dynamics~\cite{Jaeger2017} are neglected. We conclude this appendix by a short discussion on how to compute the mean photon number of the cavity mode, when the latter can be considered as close to its vacuum state. 

\vspace{2mm} 

\textbf{\textit{Photon number.}} In the adiabatic limit considered above, and for $\Delta = 0$, it can be shown that the mean photon number is well-approximated by the formula~\cite{Bonifacio1971,Meiser2010}:  

\begin{align}
\langle \hat{a}^{\dagger} \hat{a} \rangle \simeq \frac{g^2}{(\kappa/2)^2} \langle S^+ S^- \rangle.
\label{eq:ph_num}
\end{align} 

Drawing the conjecture

\begin{align*}
\langle S^+ S^- \rangle = \sum_i \langle s_i^+ s_i^-\rangle + \sum_{i \neq j} \langle s_i^+ s_j^- \rangle \leq \sum_i \langle s_i^+ s_i^-\rangle
\end{align*}
from numerical simulations, and using Eq.~(\ref{eq:ph_num}) together with

\begin{align*}
\langle s_i^+ s_i^- \rangle = \langle \hat{n}_{2i} (1-\hat{n}_{1i}) \rangle \leq 1,
\end{align*}
we obtain $\langle \hat{a}^{\dagger} \hat{a} \rangle \leqslant \frac{N g^2}{(\kappa/2)^2}$. Restricting the light-matter coupling strength to values such that the condition Eq.~(\ref{eq:coll_g}) is fulfilled, one can thus reasonably expect the cavity mode to stay close to its vacuum state, consistently with the time-scale separation argument discussed before. 

\newpage

\section*{References}

\bibliography{charge_long}

\providecommand{\newblock}{}
\begin{thebibliography}{100}
\expandafter\ifx\csname url\endcsname\relax
  \def\url#1{{\tt #1}}\fi
\expandafter\ifx\csname urlprefix\endcsname\relax\def\urlprefix{URL }\fi
\providecommand{\eprint}[2][]{\url{#2}}

\bibitem{plenio1}
Plenio M~B and Huelga S~F 2008 {\em New Journal of Physics\/} {\bf 10} 113019

\bibitem{reben}
Rebentrost P, Mohseni M, Kassal I, Lloyd S and Aspuru-Guzik A 2009 {\em New
  Journal of Physics\/} {\bf 11} 033003

\bibitem{marais}
Marais A, Sinayskiy I, Kay A, Petruccione F and Ekert A 2013 {\em New Journal
  of Physics\/} {\bf 15} 013038

\bibitem{levi}
Levi F, Mostarda S, Rao F and Mintert F 2015 {\em Reports on Progress in
  Physics\/} {\bf 78} 082001

\bibitem{schachenmayer2015cavity}
Schachenmayer J, Genes C, Tignone E and Pupillo G 2015 {\em Phys. Rev. Lett.\/}
  {\bf 114}(19) 196403

\bibitem{feist2015extraordinary}
Feist J and Garcia-Vidal F~J 2015 {\em Phys. Rev. Lett.\/} {\bf 114}(19) 196402

\bibitem{tavis}
Tavis M and Cummings F~W 1968 {\em Phys. Rev.\/} {\bf 170}(2) 379--384

\bibitem{taptap}
Zhong X, Chervy T, Wang S, George J, Thomas A, Hutchison J~A, Devaux E, Genet C
  and Ebbesen T~W 2016 {\em Angewandte Chemie International Edition\/} {\bf 55}
  6202--6206 ISSN 1521-3773

\bibitem{mahan}
Mahan G~D 1993 {\em {Many-Particle Physics}\/} 2nd ed (New York, N.Y.: Plenum)

\bibitem{0038-5670-12-1-A09}
Landau L 1933 {\em Phys. Z. Sowjetunion\/} {\bf 3} 884

\bibitem{pekar}
Pekar S 1963 {\em US AEC Report AEC-tr-5575\/}

\bibitem{doi:10.1080/00018735400101213}
Fr\"{o}hlich H 1954 {\em Advances in Physics\/} {\bf 3} 325--361

\bibitem{fro}
Fr\"{o}hlich H 1952 {\em Proceedings of the Royal Society of London A:
  Mathematical, Physical and Engineering Sciences\/} {\bf 215} 291--298 ISSN
  0080-4630

\bibitem{PhysRev.104.1189}
Cooper L~N 1956 {\em Phys. Rev.\/} {\bf 104}(4) 1189--1190

\bibitem{PhysRev.108.1175}
Bardeen J, Cooper L~N and Schrieffer J~R 1957 {\em Phys. Rev.\/} {\bf 108}(5)
  1175--1204

\bibitem{peierls1996quantum}
Peierls R 1996 {\em Quantum Theory of Solids\/} International Series of
  Monographs on Physics (Clarendon Press) ISBN 9780192670175

\bibitem{afre}
Fr\"{o}hlich H 1954 {\em Proceedings of the Royal Society of London A:
  Mathematical, Physical and Engineering Sciences\/} {\bf 223} 296--305 ISSN
  0080-4630

\bibitem{liberato}
De~Liberato S and Ciuti C 2009 {\em Phys. Rev. B\/} {\bf 79}(7) 075317

\bibitem{lindner}
Lindner N~H, Refael G and Galitski V 2011 {\em Nature Physics\/} {\bf 7} 490

\bibitem{yamamoto}
Masayuki S, Kato R and Yamamoto H~M 2015 {\em Science\/} {\bf 347}(6223)
  743--746

\bibitem{mitrano}
Mitrano M, Cantaluppi A, Nicoletti D, Kaiser S, Perucchi A, Lupi S, Di~Pietro
  P, Pontiroli D, Riccò M, Clark S~R, Jaksch D and Cavalleri A 2016 {\em
  Nature\/} {\bf 530}(7591) 461--464

\bibitem{sri}
Rajasekaran S, Casandruc E, Laplace Y, Nicoletti D, Gu G~D, Clark S~R, Jaksch D
  and Cavalleri A 2016 {\em Nature Physics\/} {\bf 12} 1012

\bibitem{theory_mitrano}
Babadi M, Knap M, Martin I, Refael G and Demler E 2017 {\em Phys. Rev. B\/}
  {\bf 96}(1) 014512

\bibitem{theory_mitrano2}
Sentef M~A, Tokuno A, Georges A and Kollath C 2017 {\em Phys. Rev. Lett.\/}
  {\bf 118}(8) 087002

\bibitem{scalp}
Schlawin F, Dietrich A~S~D, Kiffner M, Cavalleri A and Jaksch D 2017 {\em Phys.
  Rev. B\/} {\bf 96}(6) 064526

\bibitem{theory_mitrano3}
Kennes D~M, Wilner E~Y, Reichman D~R and Millis A~J 2017 {\em Phys. Rev. B\/}
  {\bf 96}(5) 054506

\bibitem{theory_mitrano4}
Mazza G and Georges A 2017 {\em Phys. Rev. B\/} {\bf 96}(6) 064515

\bibitem{mani}
Mani R~G, Smet J~H, von Klitzing K, Narayanamurti V, Johnson W~B and Umansky V
  2002 {\em Nature\/} {\bf 420} 646

\bibitem{zudov}
Zudov M~A, Du R~R, Pfeiffer L~N and West K~W 2003 {\em Phys. Rev. Lett.\/} {\bf
  90}(4) 046807

\bibitem{theory_mani}
Durst A~C, Sachdev S, Read N and Girvin S~M 2003 {\em Phys. Rev. Lett.\/} {\bf
  91}(8) 086803

\bibitem{theory_mani2}
Dmitriev I~A, Vavilov M~G, Aleiner I~L, Mirlin A~D and Polyakov D~G 2005 {\em
  Phys. Rev. B\/} {\bf 71}(11) 115316

\bibitem{tsintzos}
Tsintzos S~I, Pelekanos N~T, Konstantinidis G, Hatzopoulos Z and Savvidis P~G
  2008 {\em Nature\/} {\bf 453}(7193) 372--375

\bibitem{lagoudakis}
Lagoudakis K~G, Wouters M, Richard M, Baas A, Carusotto I, Andre R, Dang L~S
  and Deveaud-Pledran B 2008 {\em Nat. Phys.\/} {\bf 4}(9) 706--710

\bibitem{kenacohen}
Kena-Cohen S and Forrest S~R 2010 {\em Nat. Photon.\/} {\bf 4}(6) 371--375

\bibitem{cristofolini}
Cristofolini P, Christmann G, Tsintzos S~I, Deligeorgis G, Konstantinidis G,
  Hatzopoulos Z, Savvidis P~G and Baumberg J~J 2012 {\em Science\/} {\bf 336}
  704--707 ISSN 0036-8075

\bibitem{schneider}
Schneider C, Rahimi-Iman A, Kim N~Y, Fischer J, Savenko I~G, Amthor M, Lermer
  M, Wolf A, Worschech L, Kulakovskii V~D, Shelykh I~A, Kamp M, Reitzenstein S,
  Forchel A, Yamamoto Y and Hofling S 2013 {\em Nature\/} {\bf 497}(7449)
  348--352

\bibitem{javadi}
Javadi A, S\"{o}llner I, Arcari M, Hansen S~L, Midolo L, Mahmoodian S,
  Kir\v{s}ansk\.{e} G, Pregnolato T, Lee E~H, Song J~D, Stobbe S and Lodahl P
  2015 {\em Nature Communications\/} {\bf 6} 8655

\bibitem{sipahigil}
Sipahigil A, Evans R~E, Sukachev D~D, Burek M~J, Borregaard J, Bhaskar M~K,
  Nguyen C~T, Pacheco J~L, Atikian H~A, Meuwly C, Camacho R~M, Jelezko F,
  Bielejec E, Park H, Lon{\v c}ar M and Lukin M~D 2016 {\em Science\/} {\bf
  354} 847--850 ISSN 0036-8075

\bibitem{cao}
Cao B, Mahmud K~W and Hafezi M 2016 {\em Phys. Rev. A\/} {\bf 94}(6) 063805

\bibitem{manzoni}
Manzoni M~T, Chang D~E and Douglas J~S 2017 {\em Nature Communications\/} {\bf
  8}(1) 1743

\bibitem{ebbesen}
Shalabney A, George J, Hutchison J~A, Pupillo G, Genet C and Ebbesen T~W 2015
  {\em Nature Communications\/} {\bf 6} 5981

\bibitem{jino}
George J, Shalabney A, Hutchison J~A, Genet C and Ebbesen T~W 2015 {\em The
  Journal of Physical Chemistry Letters\/} {\bf 6} 1027--1031

\bibitem{herrera}
Herrera F and Spano F~C 2016 {\em Phys. Rev. Lett.\/} {\bf 116}(23) 238301

\bibitem{baumberg}
Chikkaraddy R, de~Nijs B, Benz F, Barrow S~J, Scherman O~A, Rosta E,
  Demetriadou A, Fox P, Hess O and Baumberg J~J 2016 {\em Nature\/} {\bf
  535}(7610) 127--130

\bibitem{galego}
Galego J, Garcia-Vidal F~J and Feist J 2016 {\em Nature Communications\/} {\bf
  7} 13841

\bibitem{tredicucci}
Tredicucci A, Chen Y, Pellegrini V, B\"orger M, Sorba L, Beltram F and Bassani
  F 1995 {\em Phys. Rev. Lett.\/} {\bf 75}(21) 3906--3909

\bibitem{kasprzak}
Kasprzak J, Richard M, Kundermann S, Baas A, Jeambrun P, Keeling J~M~J,
  Marchetti F~M, Szymanska M~H, Andre R, Staehli J~L, Savona V, Littlewood P~B,
  Deveaud B and Dang L~S 2006 {\em Nature\/} {\bf 443}(7110) 409--414

\bibitem{amo}
Amo A, Sanvitto D, Laussy F~P, Ballarini D, Valle E~d, Martin M~D, Lemaitre A,
  Bloch J, Krizhanovskii D~N, Skolnick M~S, Tejedor C and Vina L 2009 {\em
  Nature\/} {\bf 457}(7227) 291--295

\bibitem{deng}
Deng H, Haug H and Yamamoto Y 2010 {\em Rev. Mod. Phys.\/} {\bf 82}(2)
  1489--1537

\bibitem{carusotto}
Carusotto I and Ciuti C 2013 {\em Rev. Mod. Phys.\/} {\bf 85}(1) 299--366

\bibitem{kavokin}
Kavokin A and Lagoudakis P 2016 {\em Nat. Mater.\/} {\bf 15}(6) 599--600

\bibitem{ouss}
Maghrebi M~F, Yao N~Y, Hafezi M, Pohl T, Firstenberg O and Gorshkov A~V 2015
  {\em Phys. Rev. A\/} {\bf 91}(3) 033838

\bibitem{Zhu}
Zhu G, Suba\c{s}{\i} Y, Whitfield J~D and Hafezi M 2017 {\em ArXiv e-prints\/}
  (\textit{Preprint} \eprint{1707.04760v1})

\bibitem{imam1}
Smolka S, Wuester W, Haupt F, Faelt S, Wegscheider W and Imamoglu A 2014 {\em
  Science\/} {\bf 346}(6207) 332--335

\bibitem{imam2}
Cotle\c{t} O, Zeytino\v{g}lu S, Sigrist M, Demler E and Imamo\v{g}lu A 2016
  {\em Phys. Rev. B\/} {\bf 93}(5) 054510

\bibitem{orgiu2015conductivity}
Orgiu E, George J, Hutchison J, Devaux E, Dayen J~F, Doudin B, F~Stellacci F,
  Genet C, Schachenmayer J, Genes C, Pupillo G, Samori P and Ebbesen T~W 2015
  {\em Nature Materials\/} {\bf 14} 1123 -- 1129

\bibitem{hagenmuller}
Hagenm\"uller D, Schachenmayer J, Sch\"utz S, Genes C and Pupillo G 2017 {\em
  Phys. Rev. Lett.\/} {\bf 119}(22) 223601

\bibitem{kagan_charge_2015}
Kagan C~R and Murray C~B 2015 {\em Nat. Nano.\/} {\bf 10} 1013--1026 ISSN
  1748-3387

\bibitem{wang_fabrication_2004}
Wang Z~M, Holmes K, Mazur Y~I and Salamo G~J 2004 {\em Appl. Phys. Lett.\/}
  {\bf 84} 1931--1933 ISSN 0003-6951

\bibitem{frey_dipole_2012}
Frey T, Leek P~J, Beck M, Blais A, Ihn T, Ensslin K and Wallraff A 2012 {\em
  Phys. Rev. Lett.\/} {\bf 108} 046807

\bibitem{viennot_out_equilibrium_2014}
Viennot J~J, Delbecq M~R, Dartiailh M~C, Cottet A and Kontos T 2014 {\em
  Physical Review B\/} {\bf 89} 165404

\bibitem{liu_semiconductor_2015}
Liu Y~Y, Stehlik J, Eichler C, Gullans M~J, Taylor J~M and Petta J~R 2015 {\em
  Science\/} {\bf 347} 285--287

\bibitem{gudmundsson}
Gudmundsson V, Abdullah N~R, Sitek A, Goan H~S, Tang C~S and Manolescu A 2017
  {\em Phys. Rev. B\/} {\bf 95}(19) 195307

\bibitem{moldoveanu}
Moldoveanu V, Gudmundsson V and Manolescu A 2007 {\em Phys. Rev. B\/} {\bf
  76}(8) 085330

\bibitem{rurali_colloquium_2010}
Rurali R 2010 {\em Rev. Mod. Phys.\/} {\bf 82} 427--449

\bibitem{laird_quantum_2015}
Laird E~A, Kuemmeth F, Steele G~A, Grove-Rasmussen K, Nyg\r{a}rd J, Flensberg K
  and Kouwenhoven L~P 2015 {\em Rev. Mod. Phys.\/} {\bf 87} 703--764

\bibitem{vidar2}
Gudmundsson V, Gainar C, Tang C~S, Moldoveanu V and Manolescu A 2009 {\em New
  Journal of Physics\/} {\bf 11} 113007

\bibitem{Brantut1069}
Brantut J~P, Meineke J, Stadler D, Krinner S and Esslinger T 2012 {\em
  Science\/} {\bf 337} 1069--1071 ISSN 0036-8075 (\textit{Preprint}
  \eprint{http://science.sciencemag.org/content/337/6098/1069.full.pdf})

\bibitem{laflamme}
Laflamme C, Yang D and Zoller P 2017 {\em Phys. Rev. A\/} {\bf 95}(4) 043843

\bibitem{devoret2004}
Devoret M~H and Martinis J~M 2004 {\em Quantum Information Processing\/} {\bf
  3} 163--203 ISSN 1573-1332

\bibitem{clarke}
Clarke J and Wilhelm F~K 2008 {\em Nature\/} {\bf 453} 1031

\bibitem{koch}
Houck A~A, T\''{u}reci H~E and Koch J 2012 {\em Nature Physics\/} {\bf 8} 292

\bibitem{klein}
Klein A and Marshalek E~R 1991 {\em Rev. Mod. Phys.\/} {\bf 63}(2) 375--558

\bibitem{raimond}
Raimond J~M, Brune M and Haroche S 2001 {\em Rev. Mod. Phys.\/} {\bf 73}(3)
  565--582

\bibitem{scully}
Scully M~O and Zubairy M~S 1997 {\em Quantum Optics:\/} (Cambridge: Cambridge
  University Press) ISBN 9780511813993

\bibitem{landauer}
Landauer R 1987 {\em Zeitschrift f{\"u}r Physik B Condensed Matter\/} {\bf 68}
  217--228 ISSN 1431-584X

\bibitem{haug}
Haug H and Jauho A~P 2008 {\em Quantum Kinetics in Transport and Optics of
  Semiconductors\/} (Springer-Verlag Berlin Heidelberg) ISBN 9783540735649

\bibitem{engelsberg1963coupled}
Engelsberg S and Schrieffer J~R 1963 {\em Phys. Rev.\/} {\bf 131}(3) 993--1008

\bibitem{pourfath}
Pourfath M 2014 {\em The Non-Equilibrium Green's Function Method for Nanoscale
  Device Simulation\/} (Springer-Verlag Wien) ISBN 9783709117996

\bibitem{PhysRevB.72.115303}
Ciuti C, Bastard G and Carusotto I 2005 {\em Phys. Rev. B\/} {\bf 72}(11)
  115303

\bibitem{pol1}
Rza\ifmmode~\dot{z}\else \.{z}\fi{}ewski K, W\'odkiewicz K and
  \ifmmode~\dot{Z}\else \.{Z}\fi{}akowicz W 1975 {\em Phys. Rev. Lett.\/} {\bf
  35}(7) 432--434

\bibitem{yamanoi}
Yamanoi M 1976 {\em Physics Letters A\/} {\bf 58} 437 -- 439 ISSN 0375-9601

\bibitem{knight}
Knight J~M, Aharonov Y and Hsieh G~T~C 1978 {\em Phys. Rev. A\/} {\bf 17}(4)
  1454--1462

\bibitem{lagaf}
Bialynicki-Birula I and Rza\ifmmode \mbox{\c{}}\else
  \c{}\fi{}\ifmmode~\dot{z}\else \.{z}\fi{}ewski K 1979 {\em Phys. Rev. A\/}
  {\bf 19}(1) 301--303

\bibitem{gaston}
Gaw\ifmmode~\mbox{\c{e}}\else \c{e}\fi{}dzki K and Rza\ifmmode \mbox{\c{}}\else
  \c{}\fi{}\ifmmode~\acute{z}\else \'{z}\fi{}ewski K 1981 {\em Phys. Rev. A\/}
  {\bf 23}(5) 2134--2136

\bibitem{keeling1}
Keeling J 2007 {\em Journal of Physics: Condensed Matter\/} {\bf 19} 295213

\bibitem{nataf}
Nataf P and Ciuti C 2010 {\em Nature Communication\/} {\bf 72}

\bibitem{viehmann}
Viehmann O, von Delft J and Marquardt F 2011 {\em Phys. Rev. Lett.\/} {\bf
  107}(11) 113602

\bibitem{hagenmuller2}
Hagenm\"uller D and Ciuti C 2012 {\em Phys. Rev. Lett.\/} {\bf 109}(26) 267403

\bibitem{PhysRevLett.110.133603}
De~Liberato S and Ciuti C 2013 {\em Phys. Rev. Lett.\/} {\bf 110}(13) 133603

\bibitem{vukics2}
Vukics A, Grie\ss{}er T and Domokos P 2014 {\em Phys. Rev. Lett.\/} {\bf
  112}(7) 073601

\bibitem{bamba}
Bamba M and Ogawa T 2014 {\em Phys. Rev. A\/} {\bf 90}(6) 063825

\bibitem{Medvedyeva2013}
Medvedyeva M~V and Kehrein S 2013 {\em ArXiv e-prints\/} (\textit{Preprint}
  \eprint{1310.4997})

\bibitem{PhysRevB.80.035110}
Benenti G, Casati G, Prosen T, Rossini D and \ifmmode \check{Z}\else
  \v{Z}\fi{}nidari\ifmmode~\check{c}\else \v{c}\fi{} M 2009 {\em Phys. Rev.
  B\/} {\bf 80}(3) 035110

\bibitem{PhysRevB.91.205416}
Medvedyeva M~V, \ifmmode \check{C}\else \v{C}\fi{}ubrovi\ifmmode~\acute{c}\else
  \'{c}\fi{} M~T and Kehrein S 2015 {\em Phys. Rev. B\/} {\bf 91}(20) 205416

\bibitem{carlos}
Navarrete-Benlloch C 2015 {\em ArXiv e-prints\/} (\textit{Preprint}
  \eprint{1504.05266})

\bibitem{moritz}
Moritz G 2007 {\em On a New Solution to the Electron Correlation Problem in
  Quantum Chemistry: The Density Matrix Renormalization Group Algorithm\/}
  Ph.D. thesis ETH Zurich

\bibitem{Bonifacio1971}
Bonifacio R, Schwendimann P and Haake F 1971 {\em Phys. Rev. A\/} {\bf 4}(1)
  302--313

\bibitem{Schuetz2013}
Sch\"utz S, Habibian H and Morigi G 2013 {\em Phys. Rev. A\/} {\bf 88}(3)
  033427

\bibitem{Bullough1987}
Bullough R~K 1987 {\em Hyperfine Interactions\/} {\bf 37} 71--108 ISSN
  1572-9540

\bibitem{Garraway2011}
Garraway B~M 2011 {\em Philosophical Transactions of the Royal Society of
  London A: Mathematical, Physical and Engineering Sciences\/} {\bf 369}
  1137--1155 ISSN 1364-503X

\bibitem{Meiser2010}
Meiser D and Holland M~J 2010 {\em Phys. Rev. A\/} {\bf 81}(6) 063827

\bibitem{Jaeger2017}
J\"ager S~B, Xu M, Sch\"utz S, Holland M~J and Morigi G 2017 {\em Phys. Rev.
  A\/} {\bf 95}(6) 063852

\bibitem{wendler}
Wendler L and Kraft T 1996 {\em Phys. Rev. B\/} {\bf 54}(16) 11436--11456

\bibitem{lee}
Lee S~C and Galbraith I 1999 {\em Phys. Rev. B\/} {\bf 59}(24) 15796--15805

\bibitem{todorov}
Todorov Y, Andrews A~M, Colombelli R, De~Liberato S, Ciuti C, Klang P, Strasser
  G and Sirtori C 2010 {\em Phys. Rev. Lett.\/} {\bf 105}(19) 196402

\bibitem{hopfield}
Hopfield J~J 1958 {\em Phys. Rev.\/} {\bf 112}(5) 1555--1567

\bibitem{holstein}
Holstein T and Primakoff H 1940 {\em Phys. Rev.\/} {\bf 58}(12) 1098--1113

\bibitem{bruus}
Bruus H and Flensberg K 2004 {\em Many-body quantum theory in condensed matter
  physics - an introduction\/} (United States: Oxford University Press)

\bibitem{zeno}
Misra B and Sudarshan E~C~G 1977 {\em Journal of Mathematical Physics\/} {\bf
  18} 756--763

\bibitem{PhysRevA.92.023825}
Debierre V, Goessens I, Brainis E and Durt T 2015 {\em Phys. Rev. A\/} {\bf
  92}(2) 023825

\bibitem{anderson}
Anderson P~W 1958 {\em Phys. Rev.\/} {\bf 109}(5) 1492--1505

\end{thebibliography}

\end{document}